\documentclass[twoside,11pt]{article}

\usepackage{jmlr2e}

 \usepackage{color}
 \usepackage{hyperref}
 \usepackage{colortbl}
 \usepackage{breakcites}
 
\usepackage{amsmath}

 \usepackage{mathtools}
\mathtoolsset{centercolon}

\newif\ifdiff
 \difffalse

\ifdiff
  \newcommand{\removed}[1]{{\color{red}{#1}}{}}
  \newcommand{\added}[1]{{\color{green!80!black}{#1}}{}}
\else
  \newcommand{\removed}[1]{} 
  \newcommand{\added}[1]{#1}
\fi
\newcommand{\changed}[2]{\added{#1}\removed{#2}}

\newcommand\transpose{^{\top}}
\newcommand\var{ \mathrm{Var}  }
\newcommand\cov{ \mathrm{Cov}  }

\newcommand\varh{ \widehat{\mathrm{Var}}  }
\newcommand\covh{ \widehat{\mathrm{Cov} } }

\newcommand\blambda{ { \boldsymbol{\lambda} } }
\newcommand\hblambda{ \widehat{ \boldsymbol{\lambda} } }

\newcommand\bTheta{ \boldsymbol{\theta} } 
\newcommand\hbTheta{ {\widehat{ \boldsymbol \theta}} } 
\newcommand\hTheta{\hat{ \theta} } 

\newcommand\bSigma{ \boldsymbol{\Sigma} } 
\newcommand\hDelta{ \widehat{\Delta} } 

\newcommand\hbT{\widehat{ \mathbf T} }
\newcommand\bT{ \mathbf{ T} }
\newcommand\hT{\widehat{  T}} 

\newcommand\bbE{\mathbb{ E } }

\newcommand \hbA{\widehat{ \mathbf{A}}}
\newcommand \bA{\mathbf{A}}
\newcommand \hA{\widehat{A}}

\newcommand \hbb{\widehat{ \mathbf{b}}}
\newcommand \hb{\hat{b}}
\newcommand \bb{\mathbf{b}}

\newcommand \hbC{\widehat{ \mathbf{C}}}
\newcommand \bC{{ \mathbf{C}}}

\newcommand \bR{{ \mathbf{R}}}

\newcommand \bS{{ \mathbf{S}}}

\newcommand \bX{{ \mathbf{X}}}
\newcommand \bx{{ \mathbf{x}}}

\newcommand \bY{{ \mathbf{Y}}}

\newcommand \bZ{{ \mathbf{Z}}}

\newcommand \bmu{\boldsymbol \mu}
\newcommand \hbmu{{\widehat{ \boldsymbol \mu} }}
\newcommand \hmu{{\hat{ \mu} }}

\newcommand \balpha{\boldsymbol{\alpha}}

\usepackage{environ}
\makeatletter
\NewEnviron{Lalign}{\tagsleft@true\begin{align}\BODY\end{align}}
\makeatother

\DeclareMathOperator*{\argmin}{arg\,min}
\DeclareMathOperator*{\argmax}{arg\,max}

\jmlrheading{?}{2014}{?-?}{11/14}{??/??}{Daniel Bartz, Johannes H\"ohne and Klaus-Robert M\"uller}

\ShortHeadings{Multi-Target Shrinkage}{Bartz et al.}
\firstpageno{1}

\begin{document}

\title{Multi-Target Shrinkage}

\author{\name  Daniel Bartz\ thanks{corresponding authors.}
\email daniel.bartz@tu-berlin.de \\
\addr  Department of Computer Science, TU Berlin \\
Marchstra\ss e 23, 10587 Berlin, Germany 
   \AND
\name   Johannes H\"ohne
\email  j.hoehne@tu-berlin.de \\
\addr  Department of Computer Science, TU Berlin \\
Marchstra\ss e 23, 10587 Berlin, Germany 
   \AND
\name   Klaus-Robert M\"uller$^*$
\email klaus-robert.mueller@tu-berlin.de \\
\addr  Department of Computer Science, TU Berlin \\
Marchstra\ss e 23, 10587 Berlin, Germany \\
Korea University, Korea, Seoul \\
}

\editor{???}

\maketitle

\begin{abstract}
Stein showed that the multivariate sample mean is outperformed by ``shrinking''  to a constant \emph{target} vector. Ledoit and Wolf extended this approach to the sample covariance matrix and proposed a multiple of the identity as shrinkage target.
In a general framework, independent of a specific estimator, we extend the shrinkage concept by allowing simultaneous shrinkage to a set of targets. Application  scenarios include settings with (A) additional data sets from potentially similar distributions, (B) non-stationarity, (C) a natural grouping of the data or (D) multiple alternative estimators which could serve as targets.

We show that this \emph{Multi-Target Shrinkage} can be translated into a quadratic program and derive conditions under which the estimation of the shrinkage intensities yields optimal expected squared error in the limit.
For the sample mean and the sample covariance as specific instances, we derive conditions under which the optimality of MTS is applicable. 
We consider two asymptotic settings: the large dimensional limit (LDL), where the dimensionality and the number of observations go to infinity at the same rate, and the finite observations large dimensional limit (FOLDL), where only the dimensionality goes to infinity while the number of observations remains constant.
We then show the effectiveness  in extensive simulations and on real world data.
\end{abstract}

\begin{keywords}
  Covariance estimation, Shrinkage, Large Dimensional Limit, Linear Discriminant Analysis, Transfer Learning
\end{keywords}

\section{Introduction and Motivation}
Shrinkage is a widely applied estimation technique  dating back to Charles Stein \citep{Ste56,Ste61}. Stein showed that the sample mean is not admissible, e.g.\ that the shrinkage mean estimator is always better. The performance gain is achieved by optimizing the bias-variance-trade-off between the unbiased, high variance sample estimate and a biased, low variance target.

Over the last years, shrinkage has become very popular for the estimation of covariance matrices. Ledoit and Wolf proposed an analytic formula for covariance shrinkage which allows to calculate the optimal shrinkage intensity w.r.t.\ expected squared error (ESE) with low computational cost \citep{LedWol04} and serves as an alternative to time-consuming cross-validation.
Shrinkage has further been applied to wavelets \citep{Don95} and density estimators \citep{San13}.

\begin{figure} 
\begin{center}
\includegraphics[width= 0.95 \linewidth]{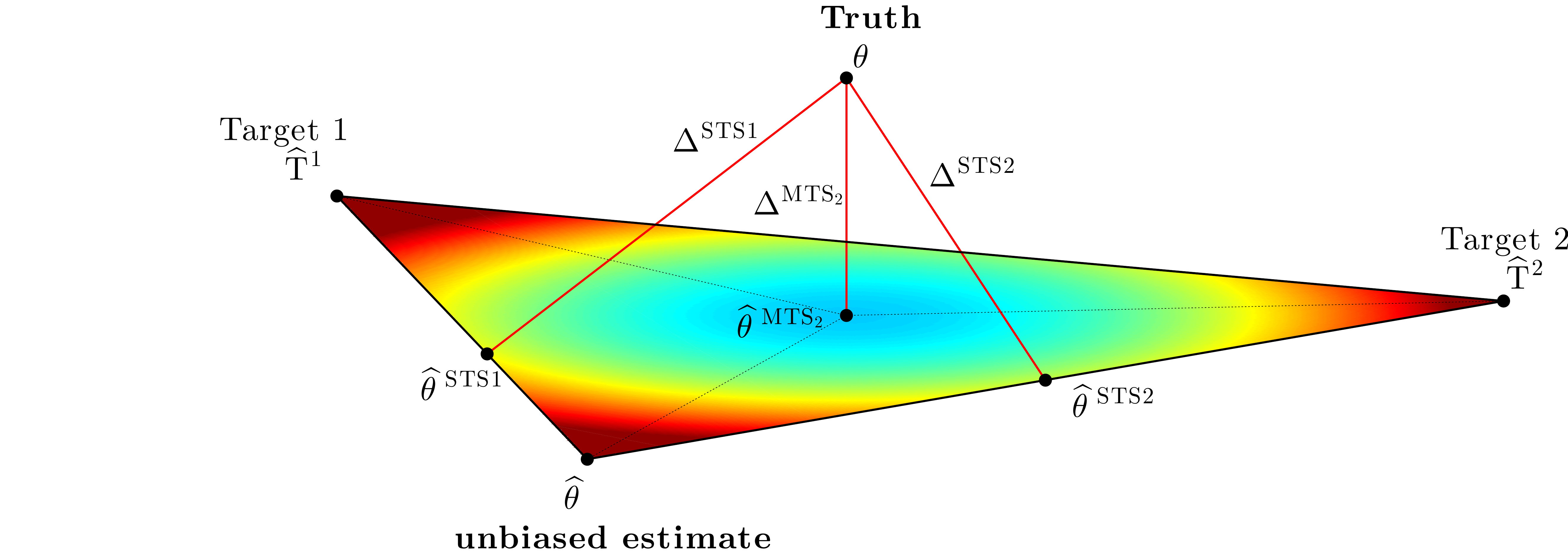}
\caption{Geometric illustration of Multi-Target Shrinkage. The unbiased estimate and the two targets span a convex set. The optimal MTS estimate is the estimate in the convex set with minimum squared distance to the truth.}
\label{fig:illustration}
\end{center}
\end{figure}

In the following, we will propose a generalization of the analytic shrinkage approach, in the following called Single-Target Shrinkage (STS), to multiple shrinkage targets. Figure~\ref{fig:illustration} illustrates Single- and Multi-Target Shrinkage (MTS) of an unbiased estimator\footnote{Note that we do not use different symbols for the estimator (a random variable) and the estimate (a realization of the random variable). It will be clear from the context to which we refer.}  $\hbTheta$ of a parameter $\bTheta$ for the case of two available shrinkage targets $\hbT^1$ and $\hbT^2$.
The convex combinations of the  three estimators span a triangle whose color coding visualizes the  squared error  of each combination\footnote{The optimum can lie on the border of the triangle if one of the targets is completely useless. Otherwise it will lie within the triangle.
}.
The two standard Single-Target Shrinkage estimators 
\begin{align*}
\hbTheta^{\mathrm{STS1}} (\lambda) & = (1-\lambda) \hbTheta + \lambda \hbT^1 \\
\hbTheta^{\mathrm{STS2}} (\lambda) & = (1-\lambda) \hbTheta + \lambda \hbT^2 
\end{align*}
are restricted to the lines connecting $\hbTheta$ with $\hbT^1$ and $\hbT^2$, respectively.  For the optimal shrinkage  intensities $\lambda_\text{STS1}^{\star}$ and $\lambda_\text{STS2}^\star$, both estimators improve over $\hbTheta$. 
Further improvement can be achieved by the Multi-Target Shrinkage estimator
$$\hbTheta^{\mathrm{MTS}_2} (\lambda_1,\lambda_2)
= (1-\lambda_1 - \lambda_2) \hbTheta + \lambda_1 \hbT^1 +\lambda_2 \hbT^2, $$
the optimal convex combination of the sample estimate and the two targets. This is nicely seen in Figure \ref{fig:illustration} where we have
\begin{align*}
\Delta^{\mathrm{MTS}_2 }
:= \| \bTheta  - \hbTheta^{\mathrm{MTS}_2} \| 
< \| \bTheta  - \hbTheta^{\mathrm{STS1/STS2}} \| 
:= \Delta^{\mathrm{STS1/STS2}}.
\end{align*}

\begin{figure} 
\begin{center}
\includegraphics[width= 0.95 \linewidth]{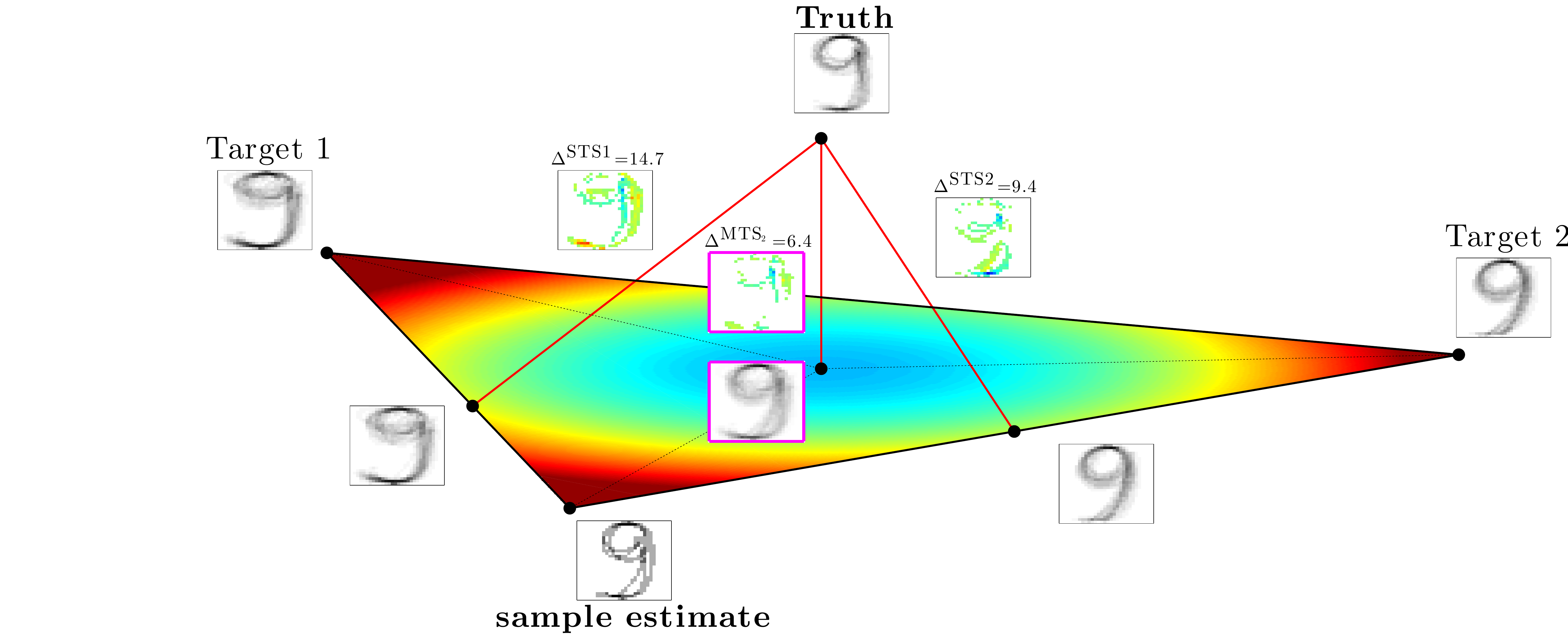}
\caption{Geometric illustration of Multi-Target Shrinkage for handwritten digits. The targets are the mean images of digit 9 for two different subjects.}
\label{fig:illustration_digits}
\end{center}
\end{figure}

As an illustration we consider MTS for the estimation of subject-specific mean images on a data set of handwritten digits\footnote{The data set consists of 10992 traces, approximately equally distributed over 44 subjects and the 10 digits $0,1,\dots 9$. We converted the traces into images of size $30\times30$.}
 \citep{Ali97,BacLic13}. Assume we want to estimate the mean image of digit $9$ of person \emph{A} from a small number of observations. In this case MTS improves over the sample mean image and STS by shrinking towards the mean images of two other subjects \emph{T1} and \emph{T2}. This can be seen in Figure  \ref{fig:illustration_digits}: for MTS, the differences to the truth\footnote{The mean of the hold-out data for subject $A$  serves as a proxy to the truth.} are less pronounced than in STS and the squared error is smaller.

\begin{figure} 
\begin{center}
\includegraphics[width= 0.75 \linewidth]{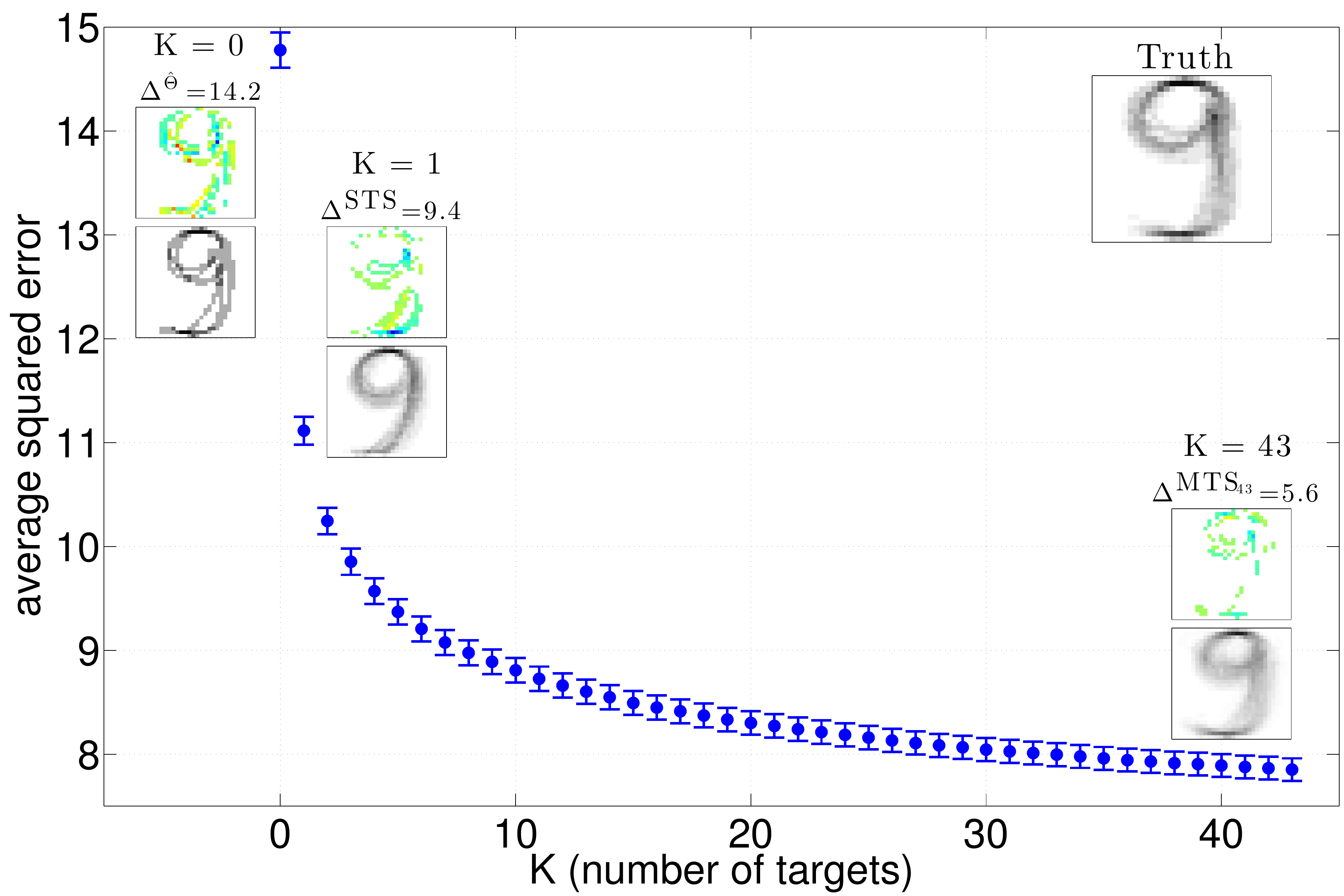}
\caption{Decay of the squared error for increasing number of shrinkage targets. Average over $R = 10000$ random choices of digits and subjects.} 
\label{fig:illustration_digits_numerics}
\end{center}
\end{figure}

The illustrations Figure  \ref{fig:illustration} and   \ref{fig:illustration_digits}  are limited to the case of simultaneous shrinkage to two shrinkage targets. MTS can handle an arbitrary number of shrinkage targets $\hbT^1, \hbT^2, \dots, \hbT^k$.  Figure \ref{fig:illustration_digits_numerics} shows this for the handwritten digits: incorporating more and more targets, the squared error decreases.

There are a many application scenarios for Multi-Target Shrinkage:
\begin{itemize}
\item similar data sets: assume that $K$ additional data sets from similar distributions exist. Then, we can calculate a target $\hbT^k$ on each additional data set and use MTS to decide  how useful  the other data sets are for the estimation task. This is a special case of \emph{transfer learning} (see  \citep{pan2010survey} for a recent review). The handwritten digits example (Figure \ref{fig:illustration_digits}) falls into this category.
\item data with group structure: if there is a natural group structure in a data set, one can estimate $\bTheta$ either (A) on the whole data set or (B) on each group separately.  
\begin{itemize}
\item When $\bTheta$ is independent of  group membership,  (A) is optimal and MTS yields approximately equal weights.
\item When $\bTheta$ is very different for each group,  (B) is optimal and MTS puts approximately no weight on the targets.
\item When $\bTheta$ is dependent of group membership, but similar,  MTS provides an optimal weighting of each group which is superior to both (A) and (B).
\end{itemize}

\item non-stationarity: assume that  the parameter $\bTheta$ is non-stationary. MTS can yield a superior estimate of the current value of $\bTheta$ by treating older segments of the data as shrinkage targets. 
\item multiple available targets: for covariance shrinkage, a set of biased estimators has been proposed as shrinkage targets: the identity, a multiple of the identity, a diagonal matrix, constant and perfect correlation matrices or, in a finance context, a factor model (see \citep{SchStr05,LedWol03}. Which one of these structured estimators constitutes the  best target depends on the structure of the true covariance matrix. The choice is based on expert knowledge or cross-validation. 
In contrast, MTS does not make a choice but yields an  an optimal weighting of all targets which is equal  or superior to the optimal choice.
\end{itemize}

We have stated above that the optimal  STS can be estimated by minimizing the ESE or by a slower cross-validation approach. 
For MTS, the computational cost to cross-validate $K$ parameters grows with the power of $K$ which is not feasible. We therefore extend the approach of minimizing the ESE to multiple shrinkage targets.

In Section~\ref{sec:MTS} we will introduce the MTS approach independently of a specific estimator and derive a quadratic program for the optimal shrinkage intensities. We then prove conditions under which the MTS estimate on a sequence of statistical model converges to the optimum. 

For the sample mean (section~\ref{sec:MTSmean}) and the sample covariance matrix (section~\ref{sec:MTScov}) we show when these conditions are fulfilled. We consider two asymptotic settings: the large dimensional limit (LDL), where the dimensionality and the number of observations go to infinity at the same rate, and the finite observations large dimensional limit (FOLDL), where only the number of dimensions goes to infinity while the number of observations remains constant. In both settings MTS is consistent, although we will show that the FOLDL requires stronger restrictions on the covariance structure.

Section~\ref{sec:simulations} presents simulations which illustrate the theorems and demonstrate the capabilities of MTS. Section \ref{sec:realworld} shows applications on real world data.





\section{Notation, distributional assumptions and asymptotic framework}
\label{sec:notationNassumptions}
\paragraph{General notation} Our notation adheres to the following conventions:
\begin{itemize}
\item 
Matrices $\mathbf{M}$ and vectors $\mathbf{v}$ are written in upper case and lower case bold letters, respectively, their entries are given by $M_{ij}$ and $v_i$. $\mathbf{m}_j$ denotes the $j^{th}$ column of the matrix $\mathbf{M}$ with entries $m_{ij} \equiv M_{ij}$.  
\item
Quantities with a hat, $\widehat{\mathbf{M}}$ and $\widehat{ \mathbf{v}}$ always denote estimators. 
\item 
$\var(a)$ and $\cov(a,b)$ denote the variance of $a$ and the covariance between $a$ and $b$, respectively. 
\item 
$\varh(a)$ and $\covh(a,b)$ denote estimators of variance and covariance which have to be specified for each set of parameters $a$ and $b$.
\item For asymptotic behaviour, we make use of the Bachmann-Landau symbols $\mathcal O$, $o$ and $\Theta$. We here only define the less frequently used $\Theta$, which denotes \emph{asymptotically bounded from above and below}:
\begin{align*}
f = \Theta(g)  
\; \Longleftrightarrow \; 
\exists c > 0 \; \exists C > 0 \; \exists x_0 > 0 \; \forall x > x_0: c \cdot | g(x) | \leq  | f(x) | \leq C \cdot | g(x) |.
\end{align*}

\end{itemize}

\paragraph{Notation for MTS} In section~\ref{sec:MTS} the  general case is analysed:
\begin{itemize}
 \item we consider the estimation of a set of parameters $\bTheta = (\theta_1, \theta_2, \dots, \theta_{q}) \in \mathbb{R}^{q}$ for which we assume the existence of an unbiased estimator $\hbTheta$. 
 \item optimality is defined w.r.t.\ expected squared error (ESE) which we denote by $\Delta$. For example, the ESE of the unbiased  estimator $\hbTheta$ is denoted by  
 \begin{align*}
\Delta^{\hbTheta} := \bbE \| \hbTheta - \bTheta \|^2.
 \end{align*}
  We always consider the $2$-norm (the Frobenius norm for multivariate parameters).
\item to study the behaviour in the limit, we will consider the estimation on a general sequence of models indexed by $p$.

\end{itemize}

\paragraph{Notation for MTS of the mean and the covariance}
In sections~\ref{sec:MTSmean} and \ref{sec:MTScov}, we consider the estimation of the mean and the covariance matrix, respectively. There, 
\begin{itemize}
 \item the sequence index $p$ also denotes the dimensionality of $n_p$ i.i.d.\ observations  with mean $\bmu_p$ and covariance $\bC_p$, given by the $(p\times n_p)$-matrix $\mathbf{X}_p$. 
  \item  We consider $K_1$ additional data sets  with mean $\bmu^k_p$ and covariance $\bC_p^k$, their $n_p^k$ i.i.d.\ observations are given by the $(p
  \times n_p^k)$-matrices $\mathbf{X}_p^k$.
 \item $\gamma_{p,1}^{(k)},\gamma_{p,2}^{(k)}, \dots,\gamma_{p,p}^{(k)}$ denote the eigenvalues of $\bC^{(k)}_p$.
\item $\bY^{(k)}_p = \bR^{(k)}_p{} \transpose \bX^{(k)}_p$ denote the observations in their respective eigenbasis, where the covariance matrices $\bSigma^{(k)}_p = \bR^{(k)}_p{} \transpose \bC_p \bR_p^{(k)}$ are diagonal. The mean in the eigenbasis is denoted by $\bmu^{Y(k)}_p$.
\item For two datasets $\bX_p^{(k)}$ and $\bX_p^{(l)}$, we denote $\bZ_p^{(k)} = \bR_p^{(l)} {}\transpose \bX_p^{(k)} $. From the context, it will be clear which $l$ was used to obtain $\bZ_p^{(k)}$.
\item in the following, we will always omit the sequence index $p$ to obtain a less cluttered notation.
\end{itemize}
\begin{table}
\centering
\begin{tabular}{l*{2}{l}l}
\hline
Setting
& set of parameters	
&  unbiased est. 		
& \#parameters \\
\hline
\hline
\arrayrulecolor{white}
\hline
\hline
general 			
& $\bTheta$ 	
& $\hbTheta$ 					
& $q$\\
mean                 		
& $ \bmu := {\mathbb E} [ \bx_i ] $				
& $\hbmu  := n^{-1}\sum_i \bx_i$		
& $q=p$ \\
covariance              	
& $ \bC  := {\mathbb  E}  [ (\bx_i - \bmu) (\bx_i - \bmu) \transpose] $ 
& $\hbC := n^{-1}\sum_i (\bx_i - \hbmu) (\bx_i - \hbmu) \transpose$
& $ q=p^2$ \\
\hline
\hline
\arrayrulecolor{black}\hline
\end{tabular}
\caption{general, mean and covariance MTS.}
\label{tab:settings}
\end{table}
Table~\ref{tab:settings} gives an overview of the different  MTS scenarios considered in this paper.


\paragraph{Distributional assumptions}
We assume
\begin{Lalign}
\tag{A1}
\label{ass:sum_sigma}
(\forall k: )
\frac{1}{p} \sum_{i=1}^p  \gamma_i^{(k)}
&  = \Theta( 1 ). \\
\tag{A2}
\label{ass:sum_sigma2}
(\forall k) \; \exists \tau_\gamma^{(k)}: 
\frac{1}{p} \sum_{i=1}^{p}  {\gamma_i^{(k)}}^2
& = \Theta \left( p^{\tau^{(k)}_\gamma} \right). 
\\
\tag{A3}
\label{ass:fourth_moms}
\exists \alpha_4, \beta_4: \qquad
 \quad  (1 + \beta_4) \mathbb{E}^2[y_i^2]   \quad
\leq 
& \quad \mathbb{E}[y_i^4]  \quad 
\leq \quad (1 + \alpha_4) \mathbb{E}^2[y_i^2]  
\\
\tag{A4}
\label{ass:eighth_moms}
\exists \alpha_8, \beta_8: \qquad
\quad  (1 + \beta_8) \mathbb{E}^2[y_i^4]   \quad
\leq 
& \quad \mathbb{E}[y_i^8]  \quad 
\leq \quad (1 + \alpha_8) \mathbb{E}^2[y_i^4] 
\end{Lalign}

The assumption \eqref{ass:sum_sigma} states, for each data set, that for an increasing number of dimensions the variance per dimension is bounded from above and below.

The assumption \eqref{ass:sum_sigma2} restricts the dispersion of the eigenvalues: for increasing dimensionality, the dispersion is assumed to have a well-defined limit behaviour. Note that \eqref{ass:sum_sigma} implies  $ 0 \leq \tau_\gamma^{(k)} \leq 1$.

The assumptions \eqref{ass:fourth_moms} and \eqref{ass:eighth_moms} have two functions: first they guarantee the existence of fourth and eighth moments, respectively. Second, they impose an (arbitrary) upper bound on the heaviness of the tails in the sequence $p$.

\paragraph{Asymptotic settings}
We consider two different asymptotic settings: 
\begin{itemize}
\item LDL: 
the standard setting in Random Matrix Theory and for the analysis of covariance shrinkage is the \emph{large dimensional limit} ($n, p \rightarrow \infty$, $n/p \rightarrow c$) \citep{LedWol04}. In the LDL,  the sample mean remains a consistent estimator, this does not hold for the sample covariance matrix. We assume that for the additional data sets $n^k/p \rightarrow c^k$ holds.
\item FOLDL: 
in addition we consider the \emph{finite observations large dimensional limit} ($p \rightarrow \infty$, $n = c$, $n^k = c^k$). In the FOLDL, neither sample covariance nor sample mean are consistent.
\end{itemize}
\begin{table}
\centering
\begin{tabular}{ll}
\hline
symbol
& meaning	 \\
\hline
\hline
\arrayrulecolor{white}
\hline
\hline
$n $ 					& number of observations 	\\
$p$  					& dimensionality / index of the sequence of models \\
$q$ 					& number of parameters \\
$f = \Theta(g)$ 			& $f$ asymptotically bounded from above and below by $g$ \\
$f = {\mathcal O}(g)$ 	& $f$ asymptotically  bounded from above by $g$ \\
$f = o(g)$ 				& $f$ asymptotically dominated by $g$ \\
$\bTheta$			& set of parameters \\
$\hbTheta$			& unbiased estimate of the set of parameters \\
$\tau_\hbTheta$		& limit behaviour of the unbiased estimator \eqref{eq:G1} \\	
$\Delta^{\hbTheta}$	& expected squared error, here of the unbiased estimator\\
$\bmu$,
 $\hbmu$				& mean and sample mean\\
$\bC$, 	
$\bS$				& covariance and sample covariance \\ 	
$\gamma_{1}^{(k)},
\dots,
\gamma_{p}^{(k)}$		& eigenvalues of $\bC$\\
$\widehat{
\text{symbol}}$		& estimate claculated on the data \\
$\tau_\gamma$		& limit behaviour of the average squared eigenvalue \eqref{ass:sum_sigma2} \\
$\mathbf{X}$			& observations ($p \times n$ matrix) \\
$\mathbf{Y}$			& observations in the eigenbasis($p \times n$ matrix) \\
$\mathbf{R}$			& rotation into the eigenbasis ($p \times p$ matrix) \\
$\mathbf{Z}$			& observations in the eigenbasis of a different data set ($p \times n$ matrix) \\
(symbol)$^k$			& for each symbol, $k$ stands for the data set $k$ \\
$\alpha_4, \beta_4$		& bounds on the ratio between second and fourth moments \eqref{ass:fourth_moms} \\
$\alpha_8, \beta_8$		& bounds on the ratio between fourth and eighth moments \eqref{ass:eighth_moms} \\
$c$					& ratio between number of observations and dimensionality $n/p$\\
$K$					& number of shrinkage targets \\
$\bT^k$				& $k^{th}$ shrinkage target \\
$\lambda^k$			& shrinkage intensity of the $k^{th}$ shrinkage target \\
$\bA$				& matrix containing estimates of the quality of the targets \\
$\bb$				& vector containing  variance of sample estimate and correlation to targets \\
$\tau_A^k$			& limit behaviour of the quality of target $k$ \eqref{eq:G3} \\	
$\tau_\mu^k$			& limit behaviour of the quality of the mean of data set $k$ \eqref{eq:LDLweakness} \\	
$\tau_\bC^k$			& limit behaviour of the quality of the covariance of data set $k$ \eqref{eq:covsimilarity} \\	
$Q_p$				& set of all quadruples consisting of distinct integers between 1 and $p$ \\	
$|Q_p|$				& cardinality of $Q_p$ \\	
\hline
\hline
\arrayrulecolor{black}\hline
\end{tabular}
\caption{overview of the notation.}
\label{tab:notation}
\end{table}
Table~\ref{tab:notation} gives an overview of the notation in the paper.

\section{Multi-Target Shrinkage}
\label{sec:MTS}
In Single-Target Shrinkage, the linear combination of an unbiased estimator $\hbTheta$ with another estimator $\hbT$ (called the shrinkage target)
is optimized. 
In most cases, the linear combination 
 is restricted to be convex \citep{LedWol04,SchStr05}:
\begin{align*}
\hbTheta^{\mathrm{STS}}( \lambda ) := (1-\lambda) \hbTheta + \lambda \hbT.
\end{align*}
In this manuscript, we generalize to optimizing the convex combination\footnote{Setting
 $\hbT^{K+1} = 0$ and allowing for $\blambda  \in \mathbb{R}^{K+1}$, this turns into an arbitrary linear combination which can deal with arbitrarily rescaled targets. Theoretical results can be extended at the cost of clarity and accesibility.} 
with a set of $K$ targets
\begin{align}
\label{eq:MTS}
\hbTheta^{\mathrm{MTS}}( \boldsymbol{\lambda} ) := \left(1-\sum_{k=1}^K \lambda_k \right) \hbTheta + \sum_{k=1}^K \lambda_k \hbT^k, 
\end{align}
where $\blambda = ( \lambda_1,\lambda_2, \dots, \lambda_K ) \in \mathbb{R}^K_{\geq0}$ is subject to $\sum_k \lambda_k \leq 1$. 
The MTS objective is given by
\begin{align}
 \Delta^{\mathrm{MTS}} ( \blambda )
:= {\mathbb E} \left\|  \bTheta - \hbTheta^{\mathrm{MTS}}(  \blambda ) \right\|^2.
\label{eq:lw-loss}
\end{align}
From the MTS objective we derive a quadratic program for the optimal value of $\blambda$:


\begin{theorem}[MTS quadratic program]
\label{th:solutionMTS}
Let  the MTS quadratic program be defined by 
\begin{align}
 \Delta_\mathrm{qp}^{\mathrm{MTS}} ( \blambda )
 :=  \frac{1}{2} \boldsymbol{\lambda}\transpose \mathbf{A} \boldsymbol{\lambda} 
- \bb\transpose \boldsymbol{\lambda} 
\label{eq:quadprog}
\end{align}
with
\begin{align*}
A_{kl}  := {\sum_{i=1}^{q} {\mathbb E} \left[ \left( \hT^k_i - \hTheta_{i} \right) \left( \hT^l_i - \hTheta_{i} \right)  \right] }, 
\qquad
b_k  :=   \sum_{i=1}^{q} \left\{  \var (\hTheta_{i})  - \cov ( \hT^k_i,\hTheta_{i} )  \right\},
\end{align*}
Then it is equivalent to optimize $\Delta^{\mathrm{MTS}} ( \blambda )$ and $\Delta_\mathrm{qp}^{\mathrm{MTS}} ( \blambda )$:
\begin{align}
\blambda^\star 
:= \argmin_{\substack{%
        \blambda \in \mathbb{R}^K_{\geq0} \\
        \sum_k \lambda_k \leq 1}}
         \Delta^{\mathrm{MTS}} ( \blambda )
= \argmin_{\substack{%
        \blambda \in \mathbb{R}^K_{\geq0} \\
        \sum_k \lambda_k \leq 1}}  \;  \Delta_\mathrm{qp}^{\mathrm{MTS}} ( \blambda ).
\label{eq:quadprog}
\end{align}
\begin{proof}
see appendix.
\end{proof}
\end{theorem}

The quadratic program is governed by the parameters $\bA$ and $\bb$, quantifying the quality of the targets and the unbiased estimator, respectively. 
The vector $\bb$ contains the variance of the unbiased estimator, adjusted for correlation with the targets.
The diagonal elements in the matrix $\bA$ contain information on the variance and bias of the targets and the correlation with the unbiased estimator. 
A target $\bT^k$ is useful if the entry in $A_{kk}$ is small relative to the variance of the unbiased estimator. 
The off-diagonal elements in the matrix $\bA$ contain information on the correlation between  targets. 

\subsection{Estimation of Multi-Target Shrinkage}
The optimal shrinkage intensities $\blambda^\star$ depend on the unknown parameters $\mathbf{A}$ and $\mathbf{b}$ of the quadratic program eq.~\eqref{eq:quadprog}.  We propose the following estimators:
\begin{align}
\hblambda
& := \argmin_{\substack{%
        \blambda \in \mathbb{R}^K_{\geq0} \\
        \sum_k \lambda_k \leq 1}}
        \hDelta^{\mathrm{MTS}}_\mathrm{qp} ( \blambda),
\qquad 
\hDelta^{\mathrm{MTS}}_\mathrm{qp} ( \blambda)
 := \frac{1}{2} \blambda \transpose \hbA \blambda - \hbb \transpose \blambda
        \label{eq:empquadprog} \qquad \mathrm{with} 
        \\
\widehat  A_{kl} 
 := \sum_{i=1}^{q} &  \left( \hT^k_i - \hTheta_{i} \right) \left( \hT^l_i - \hTheta_{i} \right)  
 ,\qquad
 \widehat  b_k 
 :=   \sum_{i=1}^{q} \left\{  \varh (\hTheta_{i})  - \covh ( \hT^k_i,\hTheta_{i} )  \right\},
\end{align}
where the unbiased estimator $\hbTheta$, the targets $\hbT^k$ and the estimators of variance and covariance appearing in $\hbb$ depend on the application scenario.

For a general parameter set $\bTheta$, the following theorem relates the limit behaviour of  the estimators  in $\hbb$ and of linear combinations of the estimators in $\hbA$  to to the limit behaviour of  $\Delta^{\mathrm{MTS}}(\hblambda)$ and $\widehat \blambda$:


\begin{theorem}(consistency of MTS)
\label{the:qpconsistency}
 Let us assume  a sequence of models indexed by $p$ such that 
\begin{Lalign}
\tag{G1}
\label{eq:G1}
\exists \tau_{\hTheta}:
\Delta^{\hTheta}
& = \Theta \left (p^{\tau_{\hTheta}} \right), 
\\
\tag{G2}
\label{eq:G2}
\quad 
\forall k \; \exists \tau_A^k:
 A_{kk} 
 = \Theta \left (p^{\tau_A^k} \right),
& \qquad 
\forall k :
 b_{k}  
 = \Theta \left (p^{\tau_{\hTheta}} \right) 
 \\
\tag{G3}
\label{eq:G3}
\left\| \hA_{kl}   - A_{kl} \right \|  
= o \left( p^{0.5(\tau_A^k + \tau_A^l)}\right)
, & 
   \qquad 
\left\| \hb_k   - b_{k} \right \|
= o \left(   p^{ \tau_{\hTheta}}   \right)
\\
\tag{G4} 
\label{eq:G3'}
\forall k: 
\min_{\substack{%
        \balpha \in \mathbb{R}^K_{\geq0} \\
        \alpha_k = 1}}
\sum_{i=1}^{q}  \bbE \Bigg[
 \Bigg(   \sum_{l=1}^K  & \alpha_l    (\hT^l_i - \hTheta_{i} ) \Bigg) ^2 \Bigg]
 = \Theta \left( p^{\tau_A^k} \right) 
\end{Lalign}
We then have
\begin{Lalign}
\tag{i}    \forall k : \lambda^\star_k , \hat \lambda_k  = \mathcal O \left( p^{(\tau_{\hTheta} - \tau_A^k)/2} \right) . \\
\tag{ii}
\frac{\Delta^{\mathrm{MTS}}(\hblambda) - \Delta^{\mathrm{MTS}}(\blambda^\star)}
{  \Delta^\hbTheta }
= o(1) 
\end{Lalign}
If one strenghtens \eqref{eq:G3'} to hold $\forall \balpha  \in \mathbb R^K$,
we also have
\begin{Lalign}
\tag{iii} 
\| \blambda^{\star} - \hblambda     \| = o(1)
\end{Lalign}
 \begin{proof}
 see appendix.
 \end{proof}
\end{theorem}

The assumptions (G1) and (G2) state that all estimators have a well-defined limit behaviour w.r.t.\ ESE. In addititon, $\Delta^{\hTheta}$ and $b_k^p$ having the same limit behaviour implies that none of the targets is identical to the unbiased estimator.

Assumption (G3) states that the relative errors\footnote{for an off-diagonal element $A_{kl}$, we consider the error relative to $\sqrt{A_{kk} A_{ll}}$.} in the entries of the estimators $\hA_{kl}$ and $\hb_k$ 
go to zero
in the limit. 
We call this property \emph{consistency of $\hbA$ and $\hbb$}.

Assumption (G4) states that the linear combination of  a set of targets cannot have better  limit behaviour w.r.t.\ ESE than the best single target in the set. 
This is needed because linear dependence of targets can result in $\bA$ having small eigenvalues  for which the relative error does not go to zero.

To illustrate the assumptions consider the handwritten digits example. A possible sequence of models consists of images with increasing resolution ($p\times p$ pixels) and an increasing number of observations for each subject. Then the sequence of ESE of the sample estimator for subject $A$ would have a clear limit behaviour and hence fullfil (G1). 
The similarity between the digits of subjects $A$ and e.g.\ \emph{T1} defines the similarity of the images. Hence a clear limit behaviour of $\bA$ (G2) is to be expected. 
With increasing $p$ and $n$, we can better estimate the variance of the sample mean and the similarity between subjects and hence the relative errors in  $\bb$ and $\bA$ would go to zero (G3).
Two subjects \emph{T1} and \emph{T2} whose differences to subject $A$ exactly cancel out in a linear combination would violate Assumption (G4). This is highly unlikely.

Part (i) of Theorem \ref{the:qpconsistency} states that a target $\bT^k$ which has worse limit behaviour w.r.t.\  ESE than the sample estimator $\hTheta$ does not contribute in the limit. 

Part (ii) is the most important result. It states that the  expected  squared error of the MTS estimator  $\hblambda$ (normalized by the error of the sample estimator) converges to the ESE of the optimal $\lambda^\star$\footnote{Note that
$
\frac{\Delta^{\mathrm{MTS}}(\hblambda) - \Delta^{\mathrm{MTS}}(\blambda^\star)}
{\Delta^{\mathrm{MTS}}(\blambda^\star)}
= o(1) 
$
does not hold in general.}. We call this property \emph{consistency of MTS}.

Part (iii) shows  that $\blambda^\star$ is, under a restriction on the linear dependency of the targets, identifiable and that the estimator $\hblambda$ converges to $\blambda^\star$. We call this \emph{consistency of the estimator $\hblambda$}.

\section{Multi-Target Shrinkage of the  mean}
\label{sec:MTSmean}
In this section we apply the MTS approach on the $p$-dimensional sample mean:
\begin{align*}
\bTheta = \bmu, \qquad \hbTheta = \hbmu = (\hmu_1, \hmu_2, \dots, \hmu_{q=p}).
\end{align*}
As shrinkage targets, we take a set of sample means  $\hbmu^{1}, \hbmu^{2}, \dots,  \hbmu^{K}$ of additional data sets $\mathbf{X}^1,\mathbf{X}^2, \dots \mathbf{X}^K$, drawn from potentially different distributions. We obtain
\begin{align}
\label{eq:Ab_mu}
A_{kl}  = {\sum_{i=1}^p {\mathbb E} \left[ \left( \hat \mu^k_i - \hat \mu_{i} \right) \left( \hat  \mu^l_i - \hat \mu_{i} \right)  \right] }
\qquad 
\qquad 
b_k  =   \sum_{i=1}^p \left\{  \var (\hat \mu_{i})  - \cov ( \hat \mu^k_i,\hmu_{i} )  \right\}.
\end{align}
$\cov ( \hat \mu^k_i,\hat  \mu_{i} )  = 0$ holds and for the sample estimates $\hbA$ and $\hbb$ we propose
\begin{align}
\widehat  A_{kl} 
 := \sum_{i=1}^p  \left( \hat \mu^{k}_i - \hat \mu_{i} \right) \left( \hat  \mu^{l}_i - \hat \mu_{i} \right)
\qquad 
\qquad 
\hb_k 
 :=
 \hat  b :=    \sum_{i=1}^p  \varh (\hat \mu_{i}) , \label{eq:Abhat_mu}
\end{align}
where the estimator of the variance of the sample mean is given by
\begin{align*}
 \varh (\hat \mu_{i})  := \frac{1}{n(n-1)} \sum_{t=1}^n{ (x_{it} - \hat \mu_i})^2.
\end{align*}

\paragraph{Remark} MTS of the mean can be seen as a weighting of each data point. Data points in $\mathbf{X}$ are weighted by $(1-\sum_{l=1}^K \lambda_l^\star) n ^{-1}$ and data points in $\mathbf{X}^k$ are weighted by $\lambda_k^\star n^{-1}_k$. 
Assuming that the distributions of the data sets only differ with respect to their means, the optimal weight of each original data point is larger than or equal to the weight of the data points from the additional data sets. 

This translates into a constraint on the quadratic program:
\begin{align*}
\forall k: \quad \lambda_k^\star n^{-1}_k \leq (1-\sum_{l=1}^K \lambda_l^\star) n ^{-1}.
\end{align*}
The constraint is reasonable to impose in many applications and increases numerical stability.

\subsection{Consistency of MTS of the mean}
In this section we will establish the conditions under which MTS of the mean is consistent by showing when the estimators eq.\,\eqref{eq:Abhat_mu} fulfill the assumptions of Theorem~\ref{the:qpconsistency}. We will show this for both asmptotic settings.



\paragraph{LDL consistency of MTS of the mean} We first consider the LDL.

\begin{theorem}[LDL consistency of MTS of the mean] 
\label{th:LDLconsistency}
Let us assume a sequence of statistical models indexed by $p$ for which 
\eqref{ass:sum_sigma},
\eqref{ass:sum_sigma2},
\eqref{ass:fourth_moms} and 
\begin{Lalign}
\tag{M1}
\label{eq:LDLweakness} 
\forall k \; 
\exists \tau_\mu^k \leq 1: &
\| \bmu^k - \bmu \|^2  = \Theta( p^{\tau_\mu^k} ), 
\\
\tag{M2}  
\label{eq:LDLmeancovrestrictions}
\forall k: \qquad
\tau^k_\gamma <  2 \max(0,\tau_\mu^k)   +1
\quad
& \text{and}
\quad
\tau_\gamma  <  2 \max(0,\min_k \tau_\mu^k) + 1
\\
\tag{M3}
\label{eq:LDLmean_lin_comb}
\forall k | \tau_\mu^k > 1 : 
\min_{\substack{%
        \balpha \in  \mathbb R^K_{\geq0} \\
        \alpha_k = 1}}
        &
 \left\| \sum_l \alpha_l (\bmu^l - \bmu) \right\|^2 = \Theta \left( p^{\tau_\mu^k} \right)
\end{Lalign}
hold. 

Then assumptions \eqref{eq:G1}, \eqref{eq:G2}, \eqref{eq:G3}, and \eqref{eq:G3'} of Theorem~\ref{the:qpconsistency} are fulfilled, MTS of the mean is consistent and 
\begin{align}
\notag
\forall k | \tau_\mu^k > 0 :  
\lambda^\star_k
 = \hat \lambda_k  
 = \mathcal O \left( p^{ - \tau_\mu^k/2} \right)
\end{align}
holds. If  \eqref{eq:LDLmean_lin_comb} holds for $\alpha \in \mathbb R^K$, $\blambda^\star$ is identifiable and $\hblambda$ is consistent.
\begin{proof}
see appendix.
\end{proof}
\end{theorem}

Assumption \eqref{eq:LDLweakness} states that the distance between data and target mean  needs to have a clear limit behaviour.  
We exclude unrealistic sequences of models $\tau_\mu^k > 1$ in which the distance between data and target mean grows faster than the dimensionality.

Assumption \eqref{eq:LDLmeancovrestrictions} limits the eigenvalue dispersion of the data sets in dependence of the distance between data and target mean. 
Intuitively, if there are strong directions whose contributions are at a constant level independent of $p$ and hence do not average out, small distances beweent data and target mean cannot be estimated reliably.

Assumption  \eqref{eq:LDLmean_lin_comb} states that there are no target means which, linearly combined, have better asymptotic behaviour than the single target means. 

 Theorem \ref{th:LDLconsistency} states conditions und which  MTS of the mean is consistent in the LDL.  
In addition it states that data sets with increasing mean distance \eqref{eq:LDLweakness} do not contribute to the MTS estimate in the LDL limit: for $n\rightarrow \infty$, these data sets do not remain useful because the sample mean is consistent.


\paragraph{FOLDL consistency of MTS of the mean}
We now consider the case where only the dimensionality $p$ goes to infinity, while $n$ remains constant. 

\begin{theorem}[FOLDL consistency of MTS of the mean]
\label{th:FOLDLconsistency}
Let us assume a sequence of statistical models indexed by $p$ for which
\eqref{ass:sum_sigma},
\eqref{ass:sum_sigma2},
\eqref{ass:fourth_moms},
 assumption~\eqref{eq:LDLweakness} 
from Theorem~\ref{th:LDLconsistency} and
\begin{Lalign}
\tag{M2$'$}  
\label{eq:FOLDLmeancovrestrictions}
\forall k: \quad 
& 
\tau^k_\gamma <  1
\quad
\text{and}
\quad
\tau_\gamma  <  1
\\
\tag{M4}
\label{eq:avdims}
( \forall k: )
  \quad \sum_{i,j\neq i}  &   \cov \left(  {y_{is}^{(k)}}^2, {y_{js}^{(k)}}^2 \right) = o \left( p^2 \right) 
\end{Lalign}
hold. Then assumptions \eqref{eq:G1}, \eqref{eq:G2}, \eqref{eq:G3}, and \eqref{eq:G3'} of Theorem~\ref{the:qpconsistency} are fulfilled and MTS of the mean is consistent and $\hblambda$ is a consistent estimator.
%
\end{theorem}

In the FOLDL, consistency results from averaging over dimensions. Therefore, consistency requires stronger restrictions on the correlation between dimensions. 
Assumption \eqref{eq:FOLDLmeancovrestrictions} states that the dispersion  of the eigenvalues \eqref{ass:sum_sigma2} has to grow slower than $\Theta(p)$. Otherwise, strong eigendirections exist whose influence on the MTS estimate remains at a constant level in the sequence of models.
Assumption \eqref{eq:avdims} states that the correlation between squared uncorrelated variables, on average, converges to zero.

Note that identifiability holds even without Assumption \eqref{eq:LDLmean_lin_comb}.

\section{Multi-Target Shrinkage of the covariance matrix}
\label{sec:MTScov}
In the second application of MTS we consider sample covariance matrices:
\begin{align*}
\bTheta = \bC, 
\qquad \hbTheta = \bS, 
\qquad S_{ij} = n^{-1} \sum_{s=1}^n (x_{is} - \hat \mu_i) (x_ {js} - \hat \mu_j).
\end{align*}
For the sample covariance matrix, we will consider two classes of targets:
\begin{itemize}
 \item as for the sample mean, it is possible to shrink to a set of  sample covariance matrices $\bS^1, \dots, \bS^{K_1}$ from additional data sets $\bX^1, \bX^2, \dots \bX^{K_1}$.
 \item a variety of biased estimators $\hbC^{1}, \hbC^{2}, \dots, \hbC^{K_2}$ of the sample covariance matrix exists which can be used as targets.  An overview is given in \citep{SchStr05}. Examples:
 \begin{itemize}
 \item $\bT^{\mathrm{id}} =  \text{trace}(\bS) \cdot \mathbf{I}$
 \item $\bT^{\mathrm{diag}} =  \bS \circ \mathbf{I}$ (elementwise product) 
 \item $\bT^{\mathrm{const.\,corr.}} =  \bS \circ \mathbf{I} + \mathbf{F} \circ (1 - \mathbf{I})$, \\
where  $F_{ij} = \sqrt{S_{ii} S_{jj}} \cdot \bar{r}$ and $\bar{r}$ is the average correlation between dimensions.
 \end{itemize}

 \end{itemize}
In total, we obtain a set of targets $\hbT^{1}, \hbT^{2}, \dots,  \hbT^{K}$ for which we have
\begin{align*}
A_{kl}  = {\sum_{i,j=1}^p \bbE \left[ \left( \hT^k_{ij} - S_{ij} \right) \left( \hT^l_{ij} -  S_{ij} \right)  \right] }
\qquad  \text{and} \qquad
b_k  =   \sum_{i,j=1}^p \left\{  \var (S_{ij})  - \cov ( \hT^k_{ij}, S_{ij} )  \right\}.
\end{align*}
For the sample estimates $\hbA$ and $\hbb$ we propose
\begin{align}
\widehat  A_{kl} 
= {\sum_{i,j=1}^p  \left( \hT^k_{ij} - S_{ij} \right) \left( \hT^l_{ij} -  S_{ij} \right)   }
 \qquad \text{and} \qquad
\hb_k 
 \equiv \hat  b =    \sum_{i,j=1}^p  \varh (S_{ij}) , \label{eq:Abhat_S}
\end{align}
where the estimator of the variance of the sample covariance is given by
\begin{align*}
 \varh (S_{ii'})  := \frac{1}{(n-1)n}   \sum_s \Big( x_{is} x_{js}  - \frac{1}{n} \sum_{t} x_{it} x_{jt} \Big)^2.
\end{align*}
To keep the notation simple, we assume $\forall \; k: \bmu  = \bmu^k =  0$. 
 
\subsection{Consistency of MTS of the covariance}

In this section we will establish the conditions under which MTS of the mean is consistent by showing when the estimators eq.\,\eqref{eq:Abhat_S} fulfill the assumptions of Theorem~\ref{the:qpconsistency}. We will consider both asmptotic settings.

\paragraph{LDL consistency of MTS of the covariance}
We first consider the LDL.
\begin{theorem}[LDL consistency of MTS of the covariance]
\label{th:LDLconsistencyCov}
Let us assume a sequence of statistical models indexed by $p$ for which
\eqref{ass:sum_sigma},
\eqref{ass:sum_sigma2},
\eqref{ass:fourth_moms},
\eqref{ass:eighth_moms} and
\begin{Lalign}
\tag{C1} 
\label{eq:covsimilarity}
& \qquad \qquad \quad 
\forall k \; 
\exists \tau_C^k \leq 2:
\| \bC^k - \bC \|^2 = \Theta( p^{\tau_\bC^k} ), 
\\
\tag{C2} 
\label{eq:LDLcovcons}
 \qquad  
& \qquad   \qquad      
\frac{  \sum_{i,j,k,l \in Q_p} \big(\cov \left[ y_{i1}y_{j1},y_{k1}y_{l1} \right] \big)^2 } { | Q_p |} = o(1) \\
& \text{where $Q_p$ is the set of all quadruples consisting of distinct integers } \notag \\
& \text{between 1 and $p$,} \notag \\
\tag{C3}
\label{eq:EVdispgrowthrate}
& \qquad \qquad \qquad
1 + 2 \tau_\gamma^{(k)} < 2 \max(1, \min_k \tau_\bC^k), 
\\
\tag{C4}
\label{eq:LDL_cov_lin_combs}
& \qquad \quad
\forall k | \tau^k_C > 1  : 
\min_{\substack{ \balpha  \in \mathbb R^K_{\geq0} \\ \alpha_k = 1}} 
\left\| \sum_l \alpha_l (\bC^l - \bC) \right\|^2 = \Theta \left( p^{\tau_\bC^k} \right)
\end{Lalign}
hold. 
Then, for the set of targets in \citep{SchStr05} and targets given by additional data sets,
assumptions \eqref{eq:G1}, \eqref{eq:G2},  \eqref{eq:G3} and \eqref{eq:G3'} of Theorem~\ref{the:qpconsistency} are fulfilled. Hence MTS of the covariance is consistent and
\begin{align}
\notag
\forall k | \tau_C^k > 1:  
\lambda^\star_k , \hat \lambda_k  
 = \mathcal O \left( p^{(1 - \tau_C^k)/2} \right) 
\end{align}
holds.  If  \eqref{eq:LDL_cov_lin_combs} holds for $\alpha \in \mathbb R^K$, $\blambda^\star$ is identifiable and $\hblambda$ is consistent.
\begin{proof}
see appendix.
\end{proof}
\end{theorem}

Assumption \eqref{eq:covsimilarity} states that the distance of the data covariance matrices to each target covariance needs to have a clear limit behaviour.  
We exclude unrealistic sequences of models with $\tau_C^k > 2$ in which the distance between data and target grows faster than the number of entries in $\bC$.


Assumption \eqref{eq:LDLcovcons} restricts the average covariance between products of uncorrelated variables. This assumption is quite weak (compare to~\citep{LedWol04}).

Assumption \eqref{eq:EVdispgrowthrate}  limits the eigenvalue dispersion of the data sets in dependence of the distance between data and target covariance. 
This is analogue to Assumption~\eqref{eq:LDLmeancovrestrictions} for MTS of the mean.

Assumption  \eqref{eq:LDL_cov_lin_combs} states that there are no additional data sets which, linearly combined, have better limit behaviour than the single data sets.

Theorem \ref{th:LDLconsistencyCov} shows that MTS of the covariance is consistent in the LDL. 
We also see that data sets with  covariance distance \eqref{eq:covsimilarity} increasing faster than $\mathcal O(p)$ do not contribute to the MTS estimator in the LDL limit: for $n\rightarrow \infty$, these data sets do not remain useful.

\paragraph{FOLDL consistency of MTS of the covariance}
We now consider the case where only the dimensionality $p$ goes to infinity, while $n$ remains constant. 
\begin{theorem}[FOLDL consistency of MTS of the covariance]
\label{th:FOLDLconsistencyCov}
Let us assume a sequence of statistical models indexed by $p$ for which
\eqref{ass:sum_sigma},
\eqref{ass:sum_sigma2},
\eqref{ass:fourth_moms},
\eqref{ass:eighth_moms},
\eqref{eq:covsimilarity},
\eqref{eq:LDLcovcons}
 (see Theorem \ref{th:LDLconsistencyCov}) and 
\begin{Lalign}
\tag{C3$'$}
\label{eq:EVdispgrowthrateFOLDL}
& \qquad \qquad
\tau^k_\gamma <  1
\quad
\text{and}
\quad
\tau_\gamma  <  1
\\
\label{eq:FOLDLcovcons}
\tag{C5}
&
\frac{  \sum_{i,j,k,l \in Q_p}  \cov \left[ (y_{i1} y_{j1})^2,  (y_{k1} y_{l1} )^2 \right] } { | Q_p | } 
= o(1) 
\end{Lalign}
hold.
Then, for the set of targets in \citep{SchStr05} and targets given by additional data sets,
assumptions \eqref{eq:G1}, \eqref{eq:G2},  \eqref{eq:G3}, and \eqref{eq:G3'} of Theorem~\ref{the:qpconsistency} are fulfilled,
 and
 MTS of the covariance and $\hblambda$ are  consistent.
\begin{proof}
see appendix.
\end{proof}
\end{theorem}
As for the mean, consistency in the FOLDL  requires a restriction \eqref{eq:EVdispgrowthrateFOLDL} on the largest eigenvalue (compare to Theorem~\ref{th:FOLDLconsistency})
Assumption~\eqref{eq:FOLDLcovcons} further restricts covariances between uncorrelated random variables.
Note that identifiability holds even without Assumption~\eqref{eq:LDL_cov_lin_combs}.

\section{Simulations}
\label{sec:simulations}
Our proposed MTS has more free parameters than standard shrinkage and therefore the vector of shrinkage intensity estimates~$\hblambda$ has a higher variance than the single shrinkage intensity estimate $\hat \lambda$ in STS. In this section, we will provide simulations for both MTS of the mean and MTS of the covariance which show that already at moderate data set sizes, MTS accurately estimates $\blambda$. We will consider
\begin{itemize}
\item expected squared error: this quantity is optimized by MTS. We directly measure the \emph{percentage improvement in average loss} (PRIAL) with respect to the sample estimator~$\hbTheta$:
\begin{align*}
\text{PRIAL}\big( \hbTheta^{shr} \big) 
= 100 \cdot \frac{\bbE \| \hbTheta - \bTheta \| - \bbE \| \hbTheta^{shr} - \bTheta \| }
{ \bbE \| \hbTheta - \bTheta \|}.
\end{align*} 
The PRIAL is a measure relative to the ESE of the sample estimator. A PRIAL of 100 means that the shrinkage estimator has no error while a PRIAL of 0 means that it yields no improvement. Negative values indicate performance worse than the sample estimator.
\item classification accuracies: in classification tasks, the ESE of the covariance matrix is not the quantity of interest: it only serves as a proxy for classification accuracies. We measure accuracy relative to the unbiased estimator:
\begin{align*}
\text{accuracy\;gain}\big( \, \hbTheta^{shr} \big)  
= \text{ accuracy}\big( \, \hbTheta^{shr} \, \big) 
-  \text{ accuracy}\big( \, \hbTheta \, \big) 
\end{align*}
We use MTS to estimate
\begin{itemize}
\item means in \emph{Linear Discriminant Analysis} (LDA)
\item covariances for \emph{Common Spatial Patterns} as an LDA preprocessing.
\end{itemize}
\end{itemize}
\subsection{Simulations for MTS of the mean}

\subsubsection{Simulation 1: MTS of the mean to additional data sets}
In the first simulation we illustrate the behaviour of MTS of the mean in the large dimensional limit (LDL, $p,n \rightarrow \infty$). We generate $n$ standard normal data points of dimensionality $p = n$ with mean $\mu_i = 0$. For the shrinkage targets we generate $K=4$ standard normal data sets with $n^k = p$ data points and means $\mu^k_i  = (\pm 1)_i \eta_k$, where the sign is random and $\boldsymbol \eta =  ( \sqrt{p}^{-1} , 0.5, 1.0, 2.0)/5$ defines the quality of the four additional data sets\footnote{Drawing 
the means from normal distributions with different variances seems more straightforward. In particular for small dimensionalities it has the disadvantage that the quality of the additional data sets varies a lot and that often $\| \bmu - \bmu^1 \| > \| \bmu - \bmu^2 \|$.
}. 
In this setting, the first additional data set $\bX^1$ has $\tau_\mu^1 = 0$ and $\bX^{2/3/4}$ have $\tau_\mu^{2/3/4} = 1$.

This setting fulfills the assumptions of Theorem~\ref{th:LDLconsistency}: targets have a clear limit behaviour (M1), from standard normality follows $\tau_\gamma^{(k)} = 0$ (M2) and the means of the targets are independently sampled (M3).
The Theorem tells us that the MTS estimator will converge and that targets $\hbT^{2/3/4}$ will not receive any weight in the LDL.

\begin{figure} 
\begin{center}
\includegraphics[width= 0.49 \linewidth]{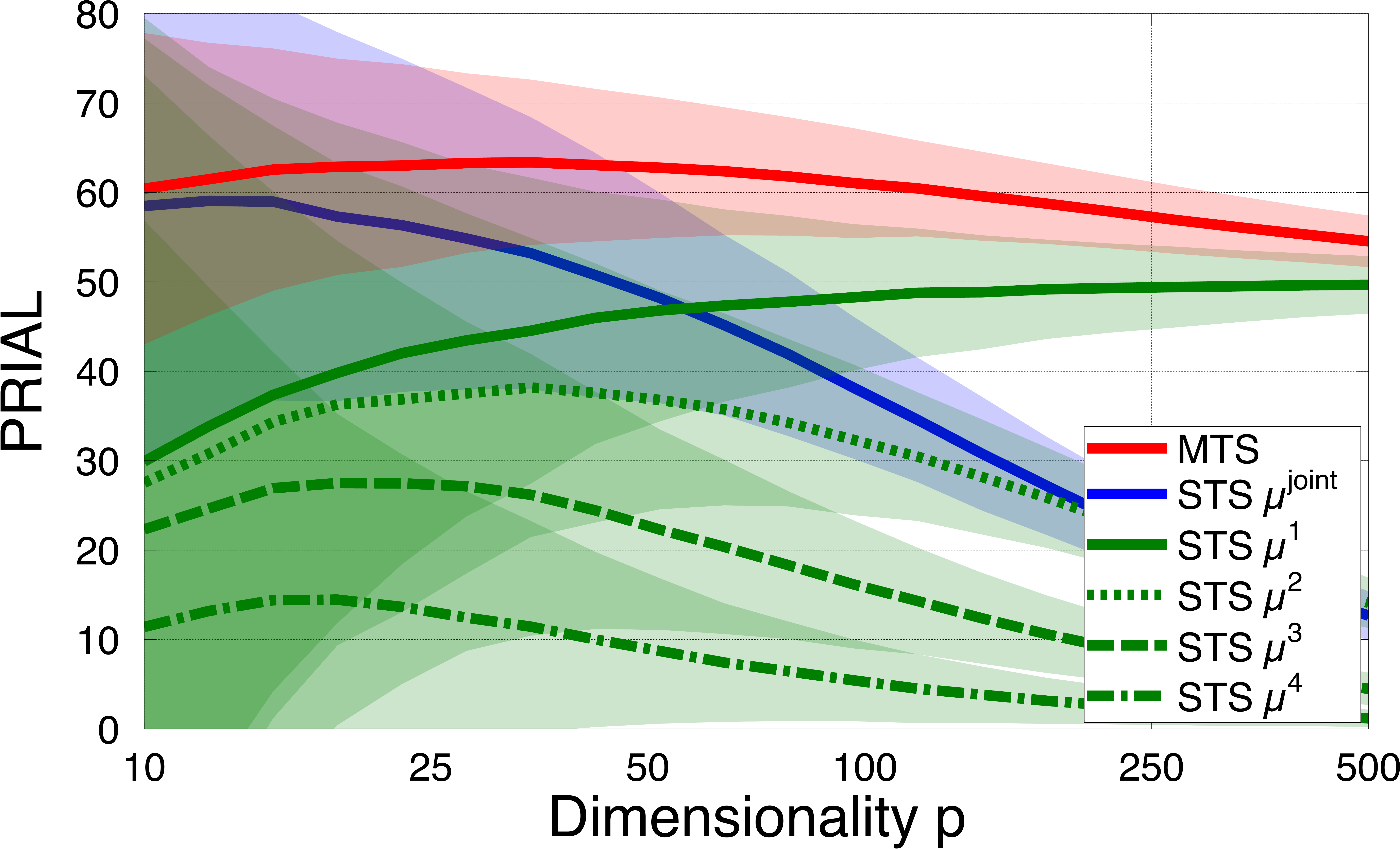}
\includegraphics[width= 0.49 \linewidth]{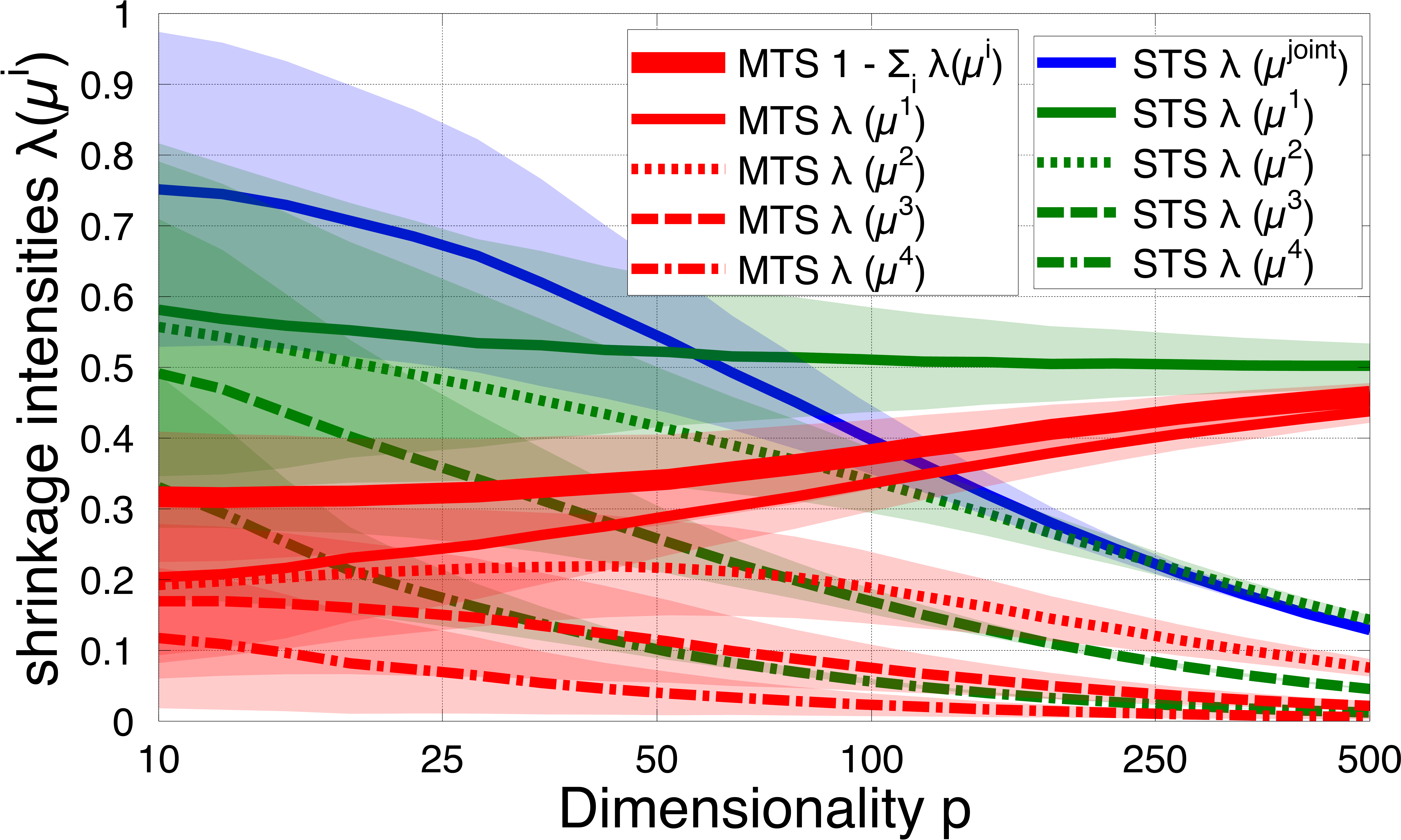}
\caption{Large dimensional limit (LDL) of MTS of the mean to additional data sets.  Average obtained over $R_r = 20$ repetitions for $R_m = 500$ models. Shaded areas show one standard deviation.} \label{fig:sotm_poc}
\end{center}
\end{figure}
We compare MTS to five versions of STS: STS to each of the targets $\hbT^k = \hbmu^k$ and STS to the joint target $\hbT^{joint} := \hbmu^{joint} := 0.25 \cdot \sum_k \hbmu^k$. Figure \ref{fig:sotm_poc} shows the dependency of the PRIAL (left) and the shrinkage intensities (right) on the dimensionality $p$.

As predicted for the LDL by Theorem \ref{th:LDLconsistency}, the STS and MTS shrinkage intensities for targets $\hbmu^2$, $\hbmu^3$, $\hbmu^4$ and $\hbmu^{joint}$ go to zero: these targets are not useful in the limit.
Only the target $\hbmu^1$ remains useful. As $n = n_1$ and the entries is $\bmu^1$ converge to the entries in $\bmu$, the shrinkage intensity $\hblambda^1$ goes to 0.5. 

The PRIALs reflect this picture: For the asymptotically useless targets, the improvement over the sample mean goes to zero, for $\hbmu^1$ it goes to a constant. 
For low $p$ and $n$, it is less relevant that $\bmu^2$, $\bmu^3$ and $\bmu^4$ are different from $\bmu$: as a consequence, the joint target is better than $\hbmu^1$. 
Over the whole range of $p$, $\hbmu^{\mathrm{MTS}}$ outperforms all STS estimators. 
For $p \rightarrow \infty$, MTS converges to STS to $\hbmu^1$.

\begin{figure} 
\begin{center}
\includegraphics[width= 0.49 \linewidth]{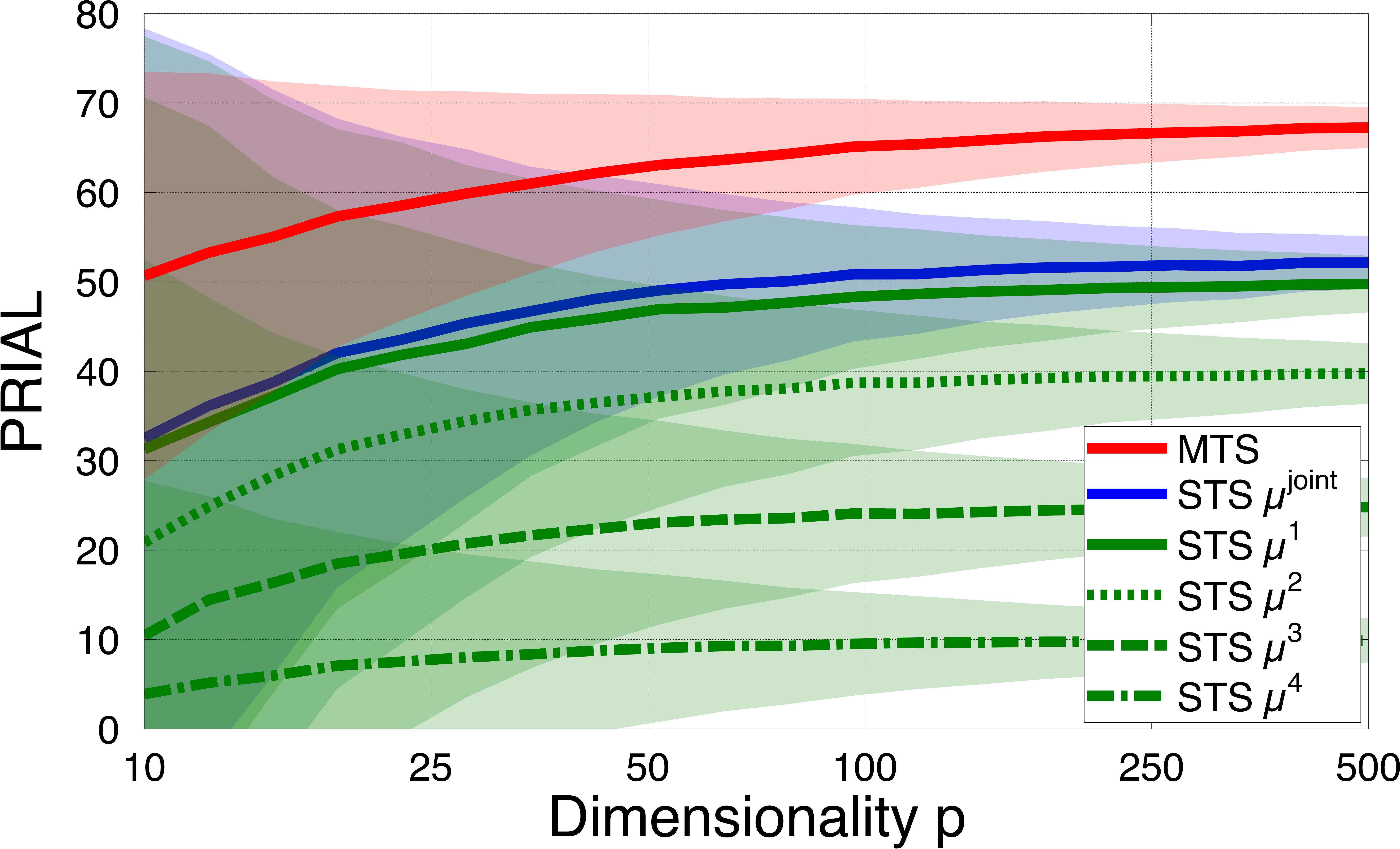}
\includegraphics[width= 0.49 \linewidth]{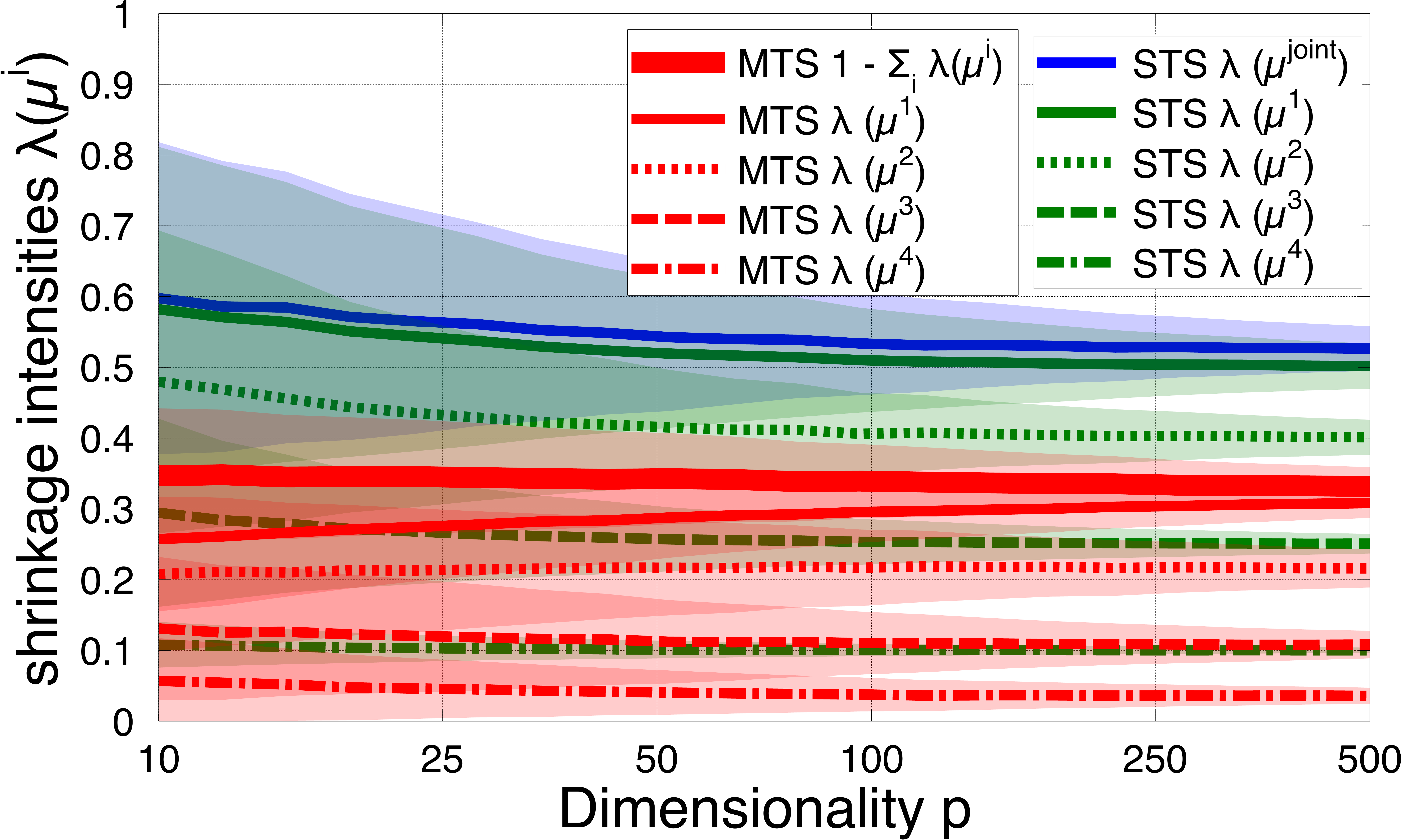}
\caption{Finite observations large dimensional limit (FOLDL) of MTS of the mean to additional data sets. Average obtained over $R_r = 20$ repetitions for $R_m = 500$ models. Shaded areas show one standard deviation.} \label{fig:sotm_FOLDL}
\end{center}
\end{figure}

Figure~\ref{fig:sotm_FOLDL} shows convergence for the finite observations large dimensional limit (FOLDL). The experiment is analogous to the one above, only 
$n = n^k = 50$ 
is kept fixed.  Contrary to the LDL, all shrinkage intensities remain finite. As above, over the whole range of~$p$, $\hbmu^{\mathrm{MTS}}$ outperforms all STS estimators.

\subsubsection{Simulation 2:  MTS for Linear Discriminant Analysis}
To test MTS in a classification setting we extended the above simulations to two class means $\bmu_{A/B}$ ($p = 50$, $n=50$). 
The difference of the class means is identical in each dimension,
chosen such that the Bayes optimal classifier achieves 80\% accuracy.
For both classes there are four additional data sets, $n^k=100$ with mean differences 
$$\Delta \mu_{A/B,i}^k = \mu_{A/B,i}^k - \mu_{A/B,i} =    (\pm 1)_i \eta_k, $$
  $\boldsymbol \eta = 10^\kappa \cdot (0.25, 0.5, 1, 2)$ where the parameter $\kappa$ governs the similarity of the additional data sets.  The covariance of each data set is $\bC_{A/B}^{(k)} = \mathbf{I}$. To make the setting slightly more realistic, we transform the data to have diagonal covariance with eigenvalues $\gamma_i= 10^{2(i-1)/(p-1)-1}$ (log-spaced  between $10^{\pm \alpha}$, $\alpha=1$). 
This is achieved by rescaling  all data points:
$$x_{A/B,it}^{(k),rescaled} = x_{A/B,it}^{(k)} \cdot \sqrt{\gamma_i}. $$
  
  We train Linear Discriminant Analysis using diffferent mean estimators: 
We compare MTS 
to 
(A) sample means $\hbmu_{A/B}$, where we ignore the additional data sets\footnote{to increase comparability, we use the sample covariance averaged over all data sets, independently of the estimator of the mean.}, 
(B) pooled means where we take $\hbmu_{A/B}^{pooled} := (K+1)^{-1} (\sum_k \hbmu_{A/B}^{k} + \hbmu_{A/B})$, 
and (C) STS where we shrink both sample means $\hbmu_{A/B}$ to the corresponding joint target $\hbmu_{A/B}^{joint} := K^{-1} \sum_k \hbmu_{A/B}^{k}$.

\begin{figure} 
\begin{center}
\includegraphics[width= 0.49 \linewidth]{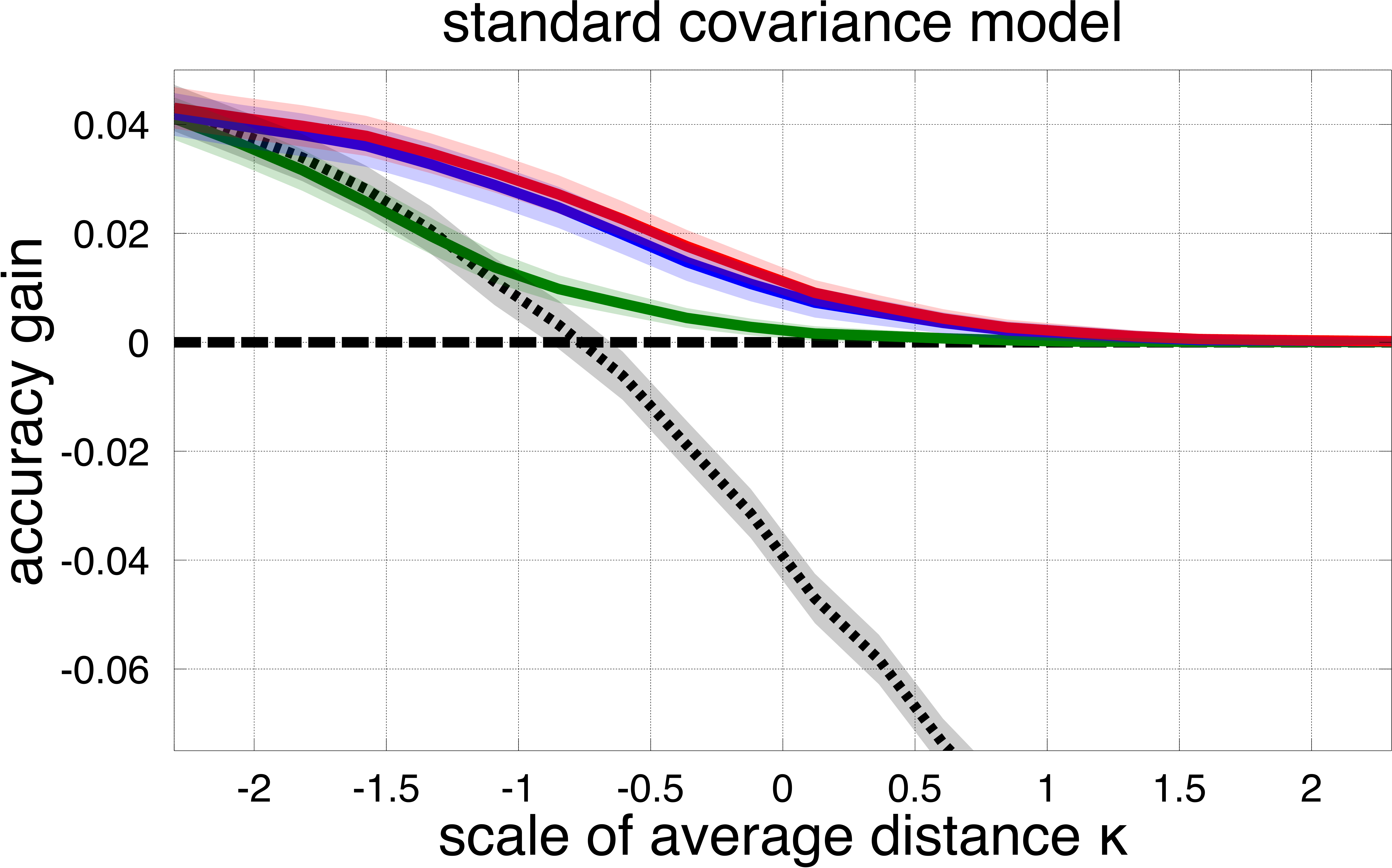}
\includegraphics[width= 0.49 \linewidth]{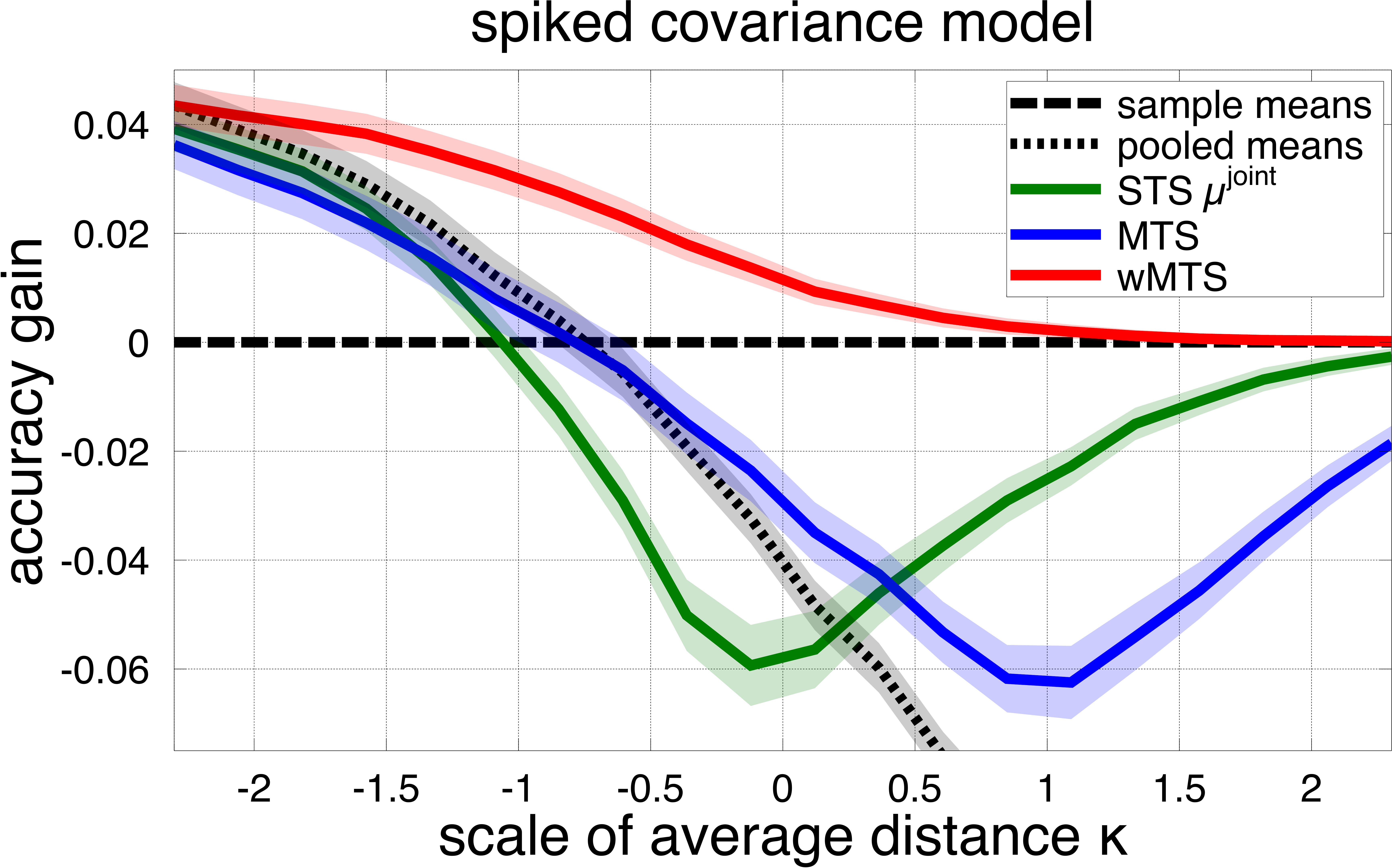}
\caption{accuracy gain for MTS for Linear Discriminant Analysis. Average obtained over $R_r = 20$ repetitions for $R_m = 500$ models. Shaded areas show one fourth standard deviation.} 
\label{fig:sotm_lda1}
%
\end{center}
\end{figure}

Figure~\ref{fig:sotm_lda1} (left) shows the gain in classification accuracy  relative to the baseline of sample means in dependence of the scale parameter~$\kappa$. 
When the target means are  very similar ($\kappa  \rightarrow -\infty$), pooled means is the optimal solution. For very different target distributions  ($\kappa \rightarrow \infty$) we cannot improve over the sample means $\hbmu_{A/B}$.
For  these extremes, STS to the pooled data performs as well as the superior method, in between it outperforms both. MTS improves on STS by finding a superior weighting of the target means. 

For Figure~\ref{fig:sotm_lda1} (right),  a spike has been added to the covariance model:  The largest eigenvalue has been multiplied by 100 and the corresponding direction has been made non-discriminative. The drop in performance indicates that STS and MTS now give too much weight to the targets, especially to the less useful targets $\bmu^{3/4}_{A/B}$. All targets are similar to the original data in the non-discriminative direction of the spike, but still vary in quality  in the discriminative directions.

\paragraph{Whitening -- a practical trick} Shrinkage puts too much weight on the  direction of highest variance. Whitening the data before MTS (wMTS) helps: wMTS gives equal importance to all directions, yields proper weights for the $\bmu^{k}_{A/B}$ and superior accuracies.

Interestingly, wMTS also performs better than standard MTS when there is no spike in the covariance~(left). In this case the estimation of the shrinkage intensities is dominated by the few directions of largest variance. This  causes high variance in the shrinkage intensity estimates $\hbmu$. Using wMTS, the estimation of the shrinkage intensities becomes an evenly weighted average over dimensions and hence gets more stable.

In general, whitening leads to large improvements if the discriminative information is not restricted to the subspace of highest variance.



\subsection{Simulations for MTS of the covariance}
\label{sec:simulation}
\subsubsection{Simulation 3: MTS of the covariance to additional data sets}
Here we illustrate the behaviour of MTS of the covariance in the large dimensional limit (LDL, $p,n \rightarrow \infty$). We generate $n$ normal data points of dimensionality $p = n$ with covariance $\mathbf{C}$ diagonal with  logarithmically spaced eigenvalues. For the shrinkage targets we generate $K=4$ standard normal data sets with $n^k = p$ data points. The covariance matrices $\mathbf{C}^k$ of the additional data sets only differ in the largest eigenvalue $\gamma_{max}^k =  \eta_k \cdot p$, with $\boldsymbol \eta = ( \sqrt{p}^{-1} , 1.0, 2.5, 5.0)/10$. Therefore the first additional data set $\bX^1$ has $\tau_C^1 = 1$ and $\bX^{2/3/4}$ have $\tau_C^{2/3/4} = 2$.

\begin{figure} 
\begin{center}
\includegraphics[width= 0.49 \linewidth]{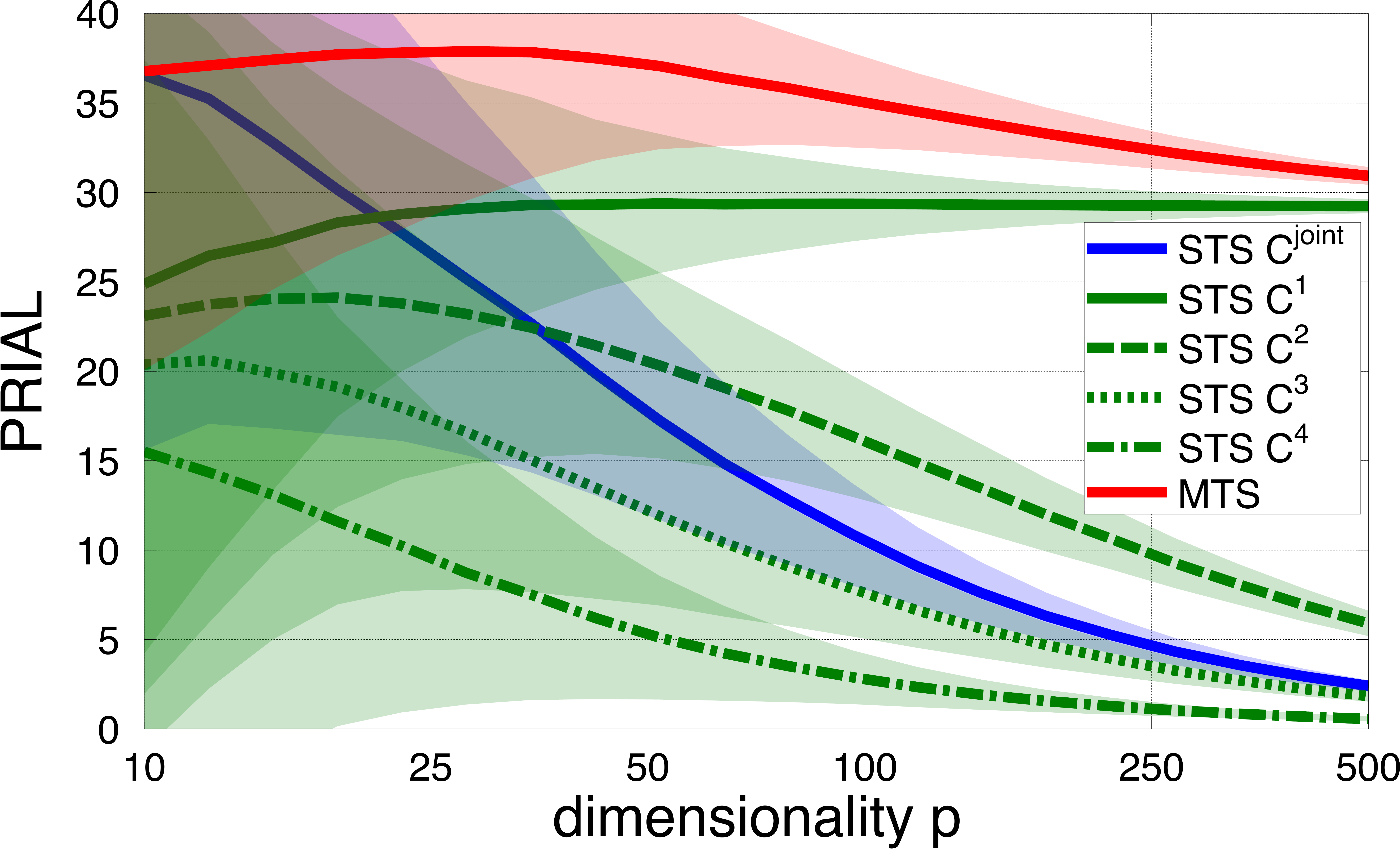}
\includegraphics[width= 0.49 \linewidth]{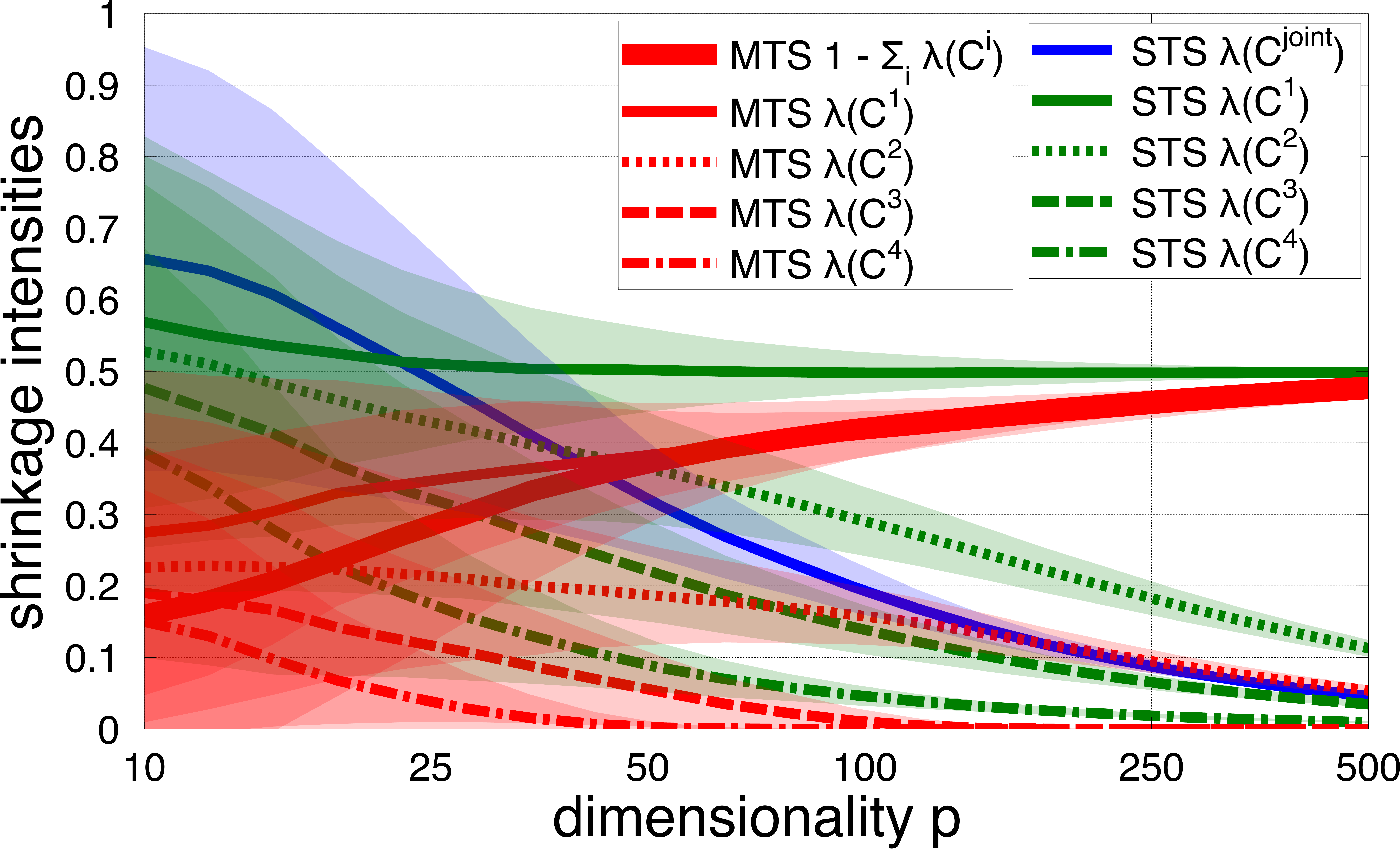}
\caption{Large dimensional limit (LDL) of MTS of the covariance to additional data sets.  Average obtained over $R_r = 20$ repetitions for $R_m = 500$ models. Shaded areas show one standard deviation.} \label{fig:sotc_poc}
\end{center}
\end{figure}


This makes the setting analog to simulation 1. Figure \ref{fig:sotc_poc} shows the dependency of the PRIAL (left) and the shrinkage intensities (right) on the dimensionality $p$:
the STS and MTS shrinkage intensities for targets $\hbC^{2/3/4}$ and $\hbC^{joint}$ go to zero, only the target $\hbC^1$ remains useful in the LDL. As $n = n^1$, the shrinkage intensity goes to 0.5. 
For the asymptotically useless targets, the PRIAL over the sample covariance goes to zero, for $\hbC^1$ it goes to a constant. 
For low $p$ and $n$, it is less relevant that $\bC^{2/3/4}$ are different from $\bC$: as a consequence, the joint target is better than $\hbC^1$. 
Over the whole range of $p$, $\hbC^{\mathrm{MTS}}$ outperforms all STS estimators.

\begin{figure} 
\begin{center}
\includegraphics[width= 0.49 \linewidth]{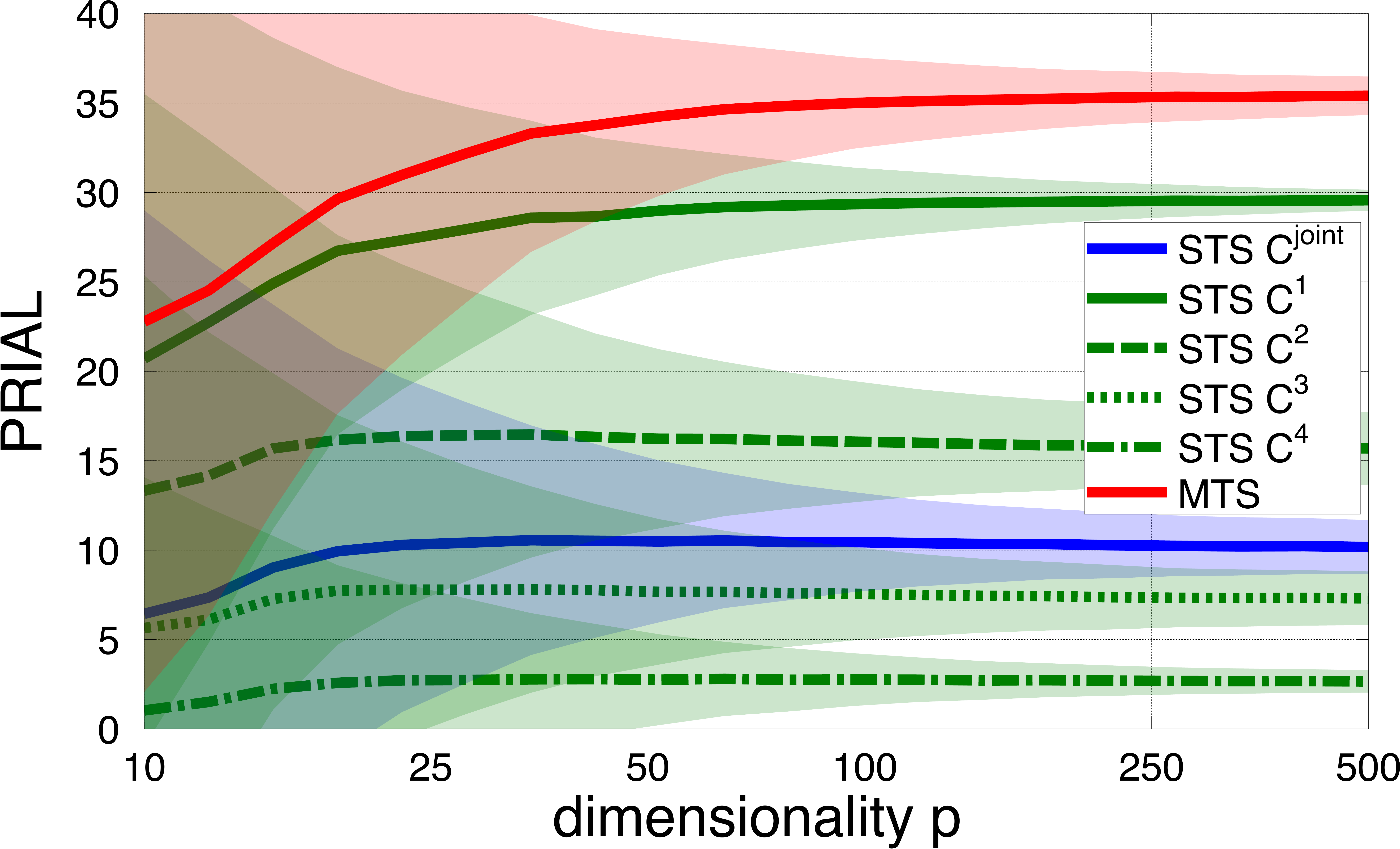}
\includegraphics[width= 0.49 \linewidth]{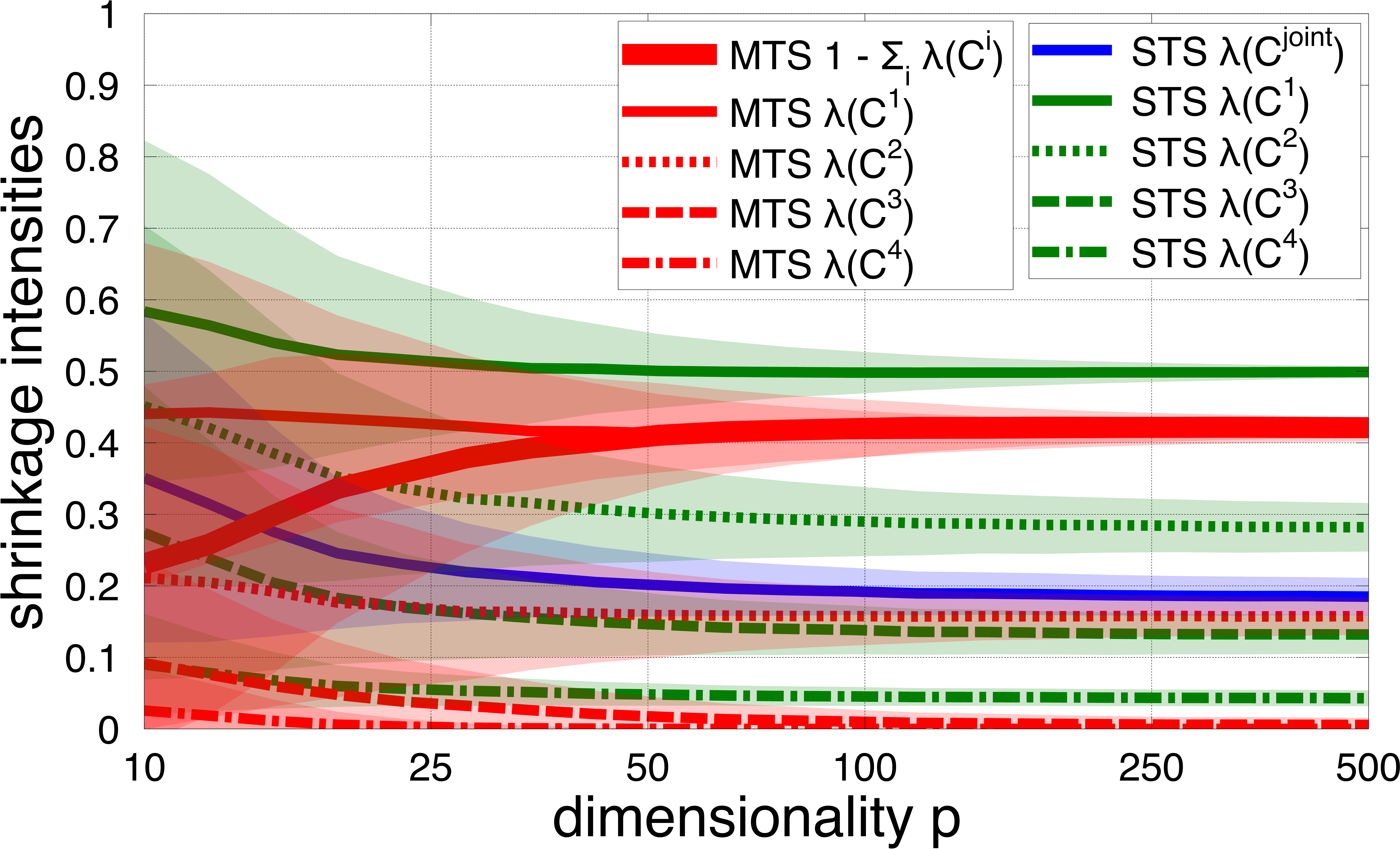}
\caption{Finite observations large dimensional limit (FOLDL) of MTS of the covariance to additional data sets.  Average obtained over $R_r = 20$ repetitions for $R_m = 500$ models. Shaded areas show one standard deviation.} \label{fig:sotc_FOLDL}
\end{center}
\end{figure}

Figure~\ref{fig:sotc_FOLDL} shows  results for the FOLDL, where $n = n^1 = n^2 = n^3 = n^4 = 50$ is kept fixed. As for the mean, all shrinkage intensities remain finite and over the whole range of~$p$, $\hbC^{\mathrm{MTS}}$ outperforms all STS estimators.

\subsubsection{Simulation 4: shrinkage to identity and additional data}
\begin{figure} 
\begin{center}
\includegraphics[width= 0.49 \linewidth]{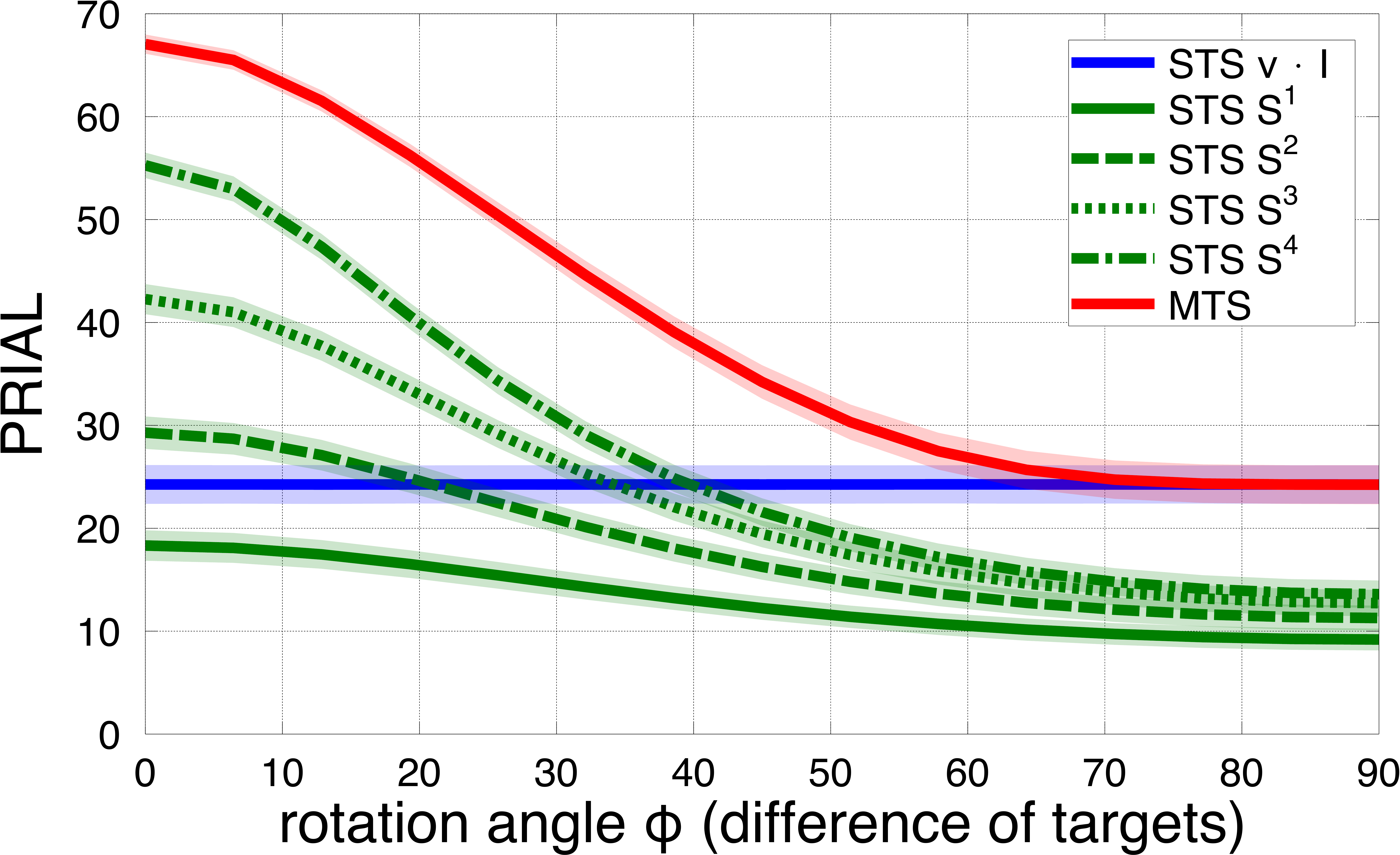}
\includegraphics[width= 0.49 \linewidth]{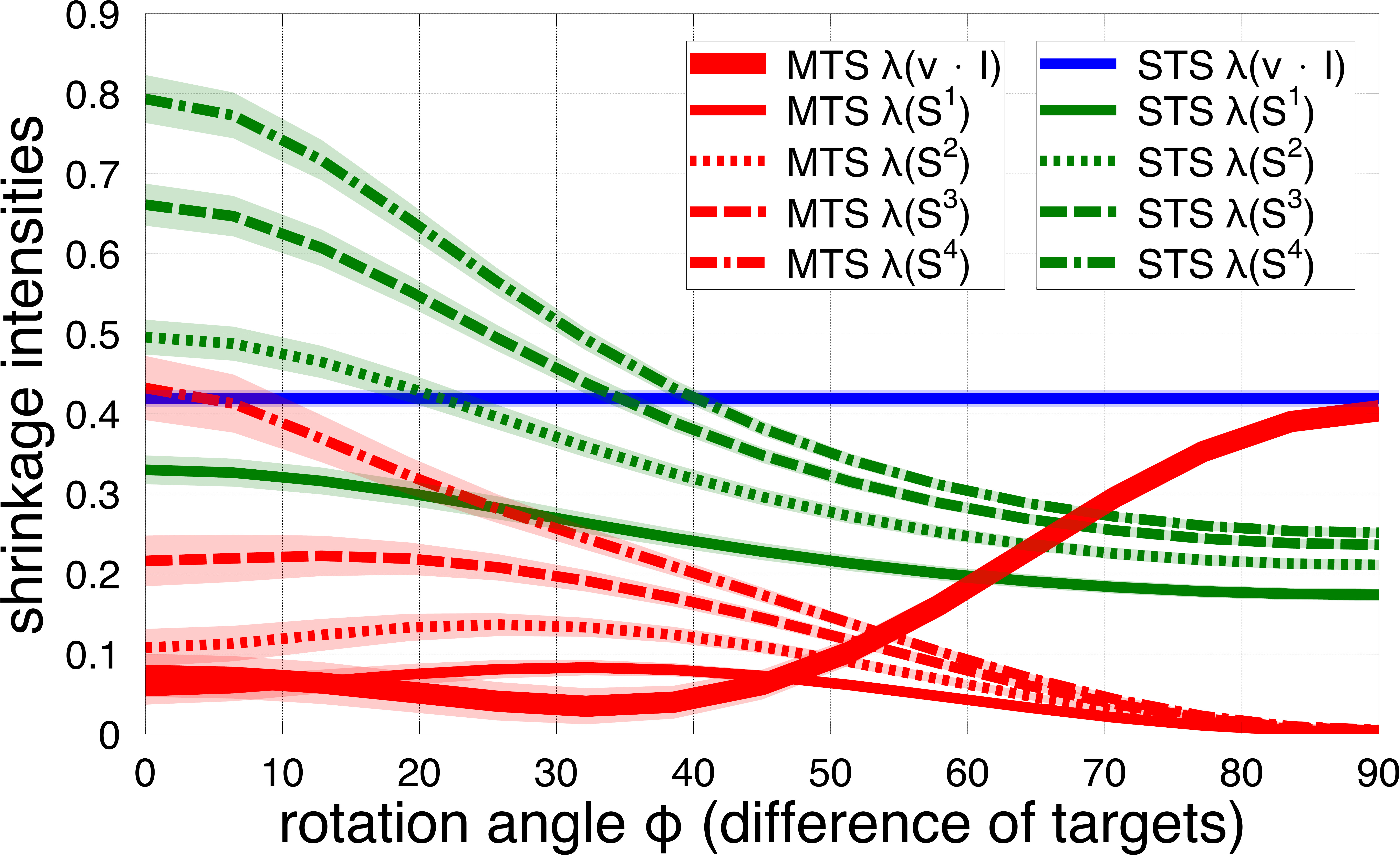}
\caption{MTS of the covariance to identity and additional data sets.  Average obtained over $R_r = 20$ repetitions for $R_m = 500$ models. Shaded areas show four standard deviations.} \label{fig:MTS_MT}
\end{center}
\end{figure} 

For MTS of the covariance there is also the possibility to include a biased estimator as a shrinkage target. The most widely used biased estimator is the identiy multiplied by the average sample eigenvalue: $\hbT^{id} := \nu \mathbf{I}$. In this simulation, we  shrink to  $\hbT^{id}$  and the covariance matrices of four additional sets of observations. We choose $\mathbf{C}$ and $\mathbf{C}^k$ diagonal with  logarithmically spaced eigenvalues between $10^{-1}$ and $10^1$. Each of the additional data sets is rotated randomly constrained to a rotation angle $ \phi$. 
 We generate multivariate normal random data sets $\mathbf{X}$ and $\mathbf{X}^{1/2/3/4}$ of size $p = n= 500$, $ n^1= p/2$, $ n^2= p$, $ n^3=2p$ and $ n^4=4p$.

Figure \ref{fig:MTS_MT} shows PRIAL and shrinkage intensities in dependence of the rotation angle~$\phi$. Shrinkage to $\hbT^{id}$ is independent of $\phi$, while STS to the other data is good when distributions are similar (small rotation angle) and  yields only small improvements for very different distributions (large rotation angle). The MTS shrinkage intensities   show that for large $\phi$ MTS yields approximately the same estimate as STS to $\bT^{id}$, while for small $\phi$ it yields a weighting of all five targets. This weighting yields superior PRIAL compared to each  STS estimator. 

\subsubsection{Simulation 5: MTS of  the covariance and CSP}

In this section we apply MTS to the preprocessing method \emph{Common Spatial Patterns} (CSP). CSP is  used for dimension reduction in classification settings where  (A) each datapoint is a time series of observations and (B) the discriminative information between two classes lies in the signal variance. Then CSP yields filters for the classes A and B which are defined by the directions where the ratio of the variances is maximal:
\begin{align*}
\mathbf{f}^{A/B}_i 
:= \argmax_{ \mathbf{f} : \mathbf{f} \perp \mathbf{f}^{A/B}_j \forall j<i}
\frac{ \mathbf{f} \transpose \hbC^{A/B} \mathbf{f}  }
{\mathbf{f} \transpose (\hbC^A + \hbC^B) \mathbf{f} }.
\end{align*}
As common in Brain-Computer Interfacing, an LDA classifier is trained on features $x_i^{CSP} = \log \left( \varh ( \bX  \, \mathbf{f}_i  ) \right)$.

For this simulation, a $p = 50$ dimensional diagonal covariance matrix $\bC$ with logarithmically spaced eigenvalues between $10^{-1}$ and $10^1$ is generated.  The covariances of the two classes $\bC^{A,B}$ and a set of different covariances  $\bC^{A/B,k}_{\mathrm{diff}}$ are each obtained by rescaling $P = 10$ random eigenvalues of $\bC$ by $p_i = (1 + i/P), i=1,2,\dots,P$. In addition, we rotate the $\bC^{A/B,k}_{\mathrm{diff}}$ randomly by an angle $\phi^k$, $\boldsymbol \phi = (0, 5, 10, 90)$.
 To study the dependency on the similarity of targets we set the covariance matrices of the additional data sets to
$$\bC^{A/B,k}(w) = (1 - w) \bC^{A/B,k}_{\mathrm{diff}} + w \bC^{A/B}.$$
For each class and each target we generate $n = n^k = 200$ data points. The classification accuracy is calculated for test trials of length $n^{test} = 20$. 
\begin{figure} 
\begin{center}
\includegraphics[width= 0.49 \linewidth]{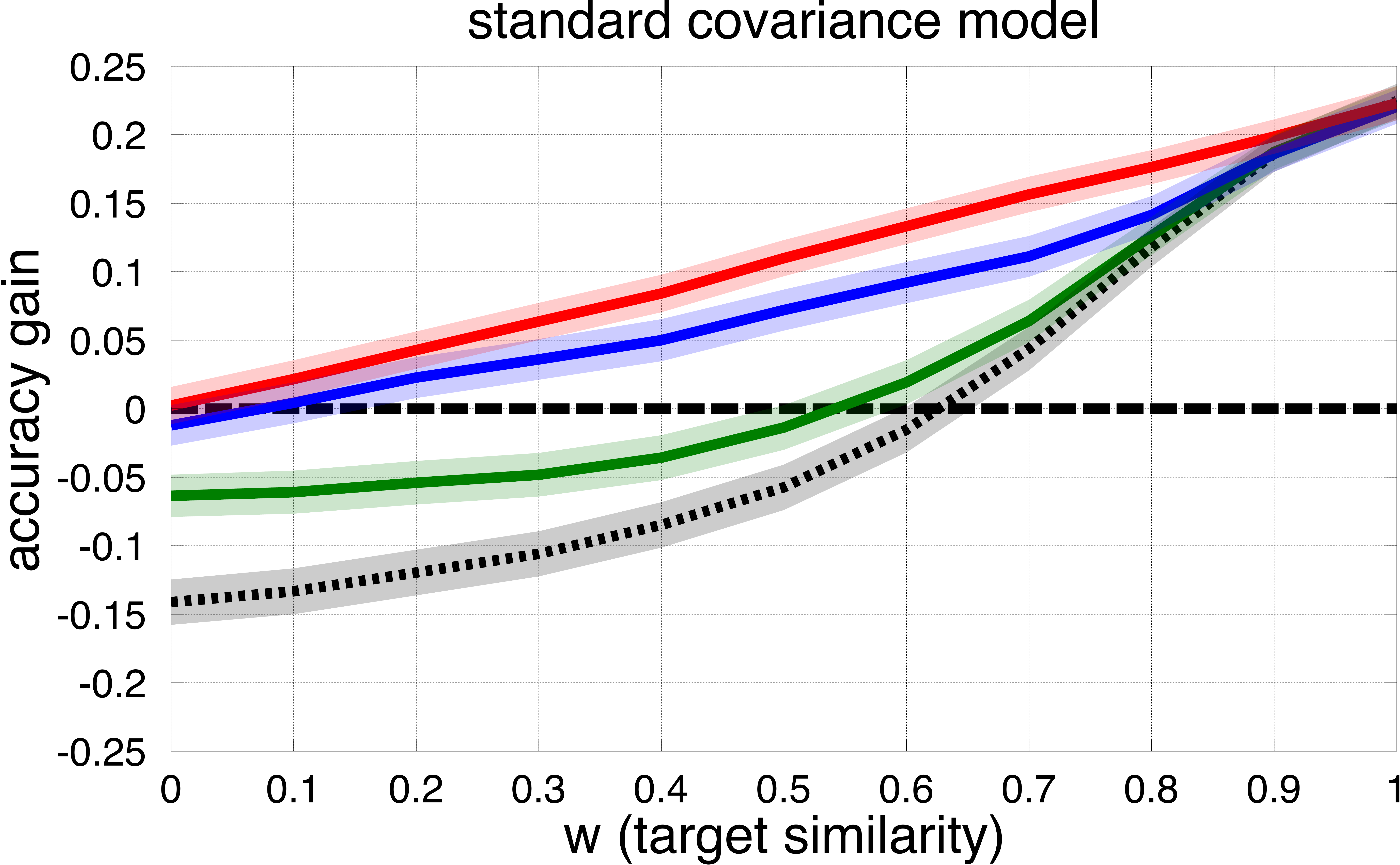}
\includegraphics[width= 0.49 \linewidth]{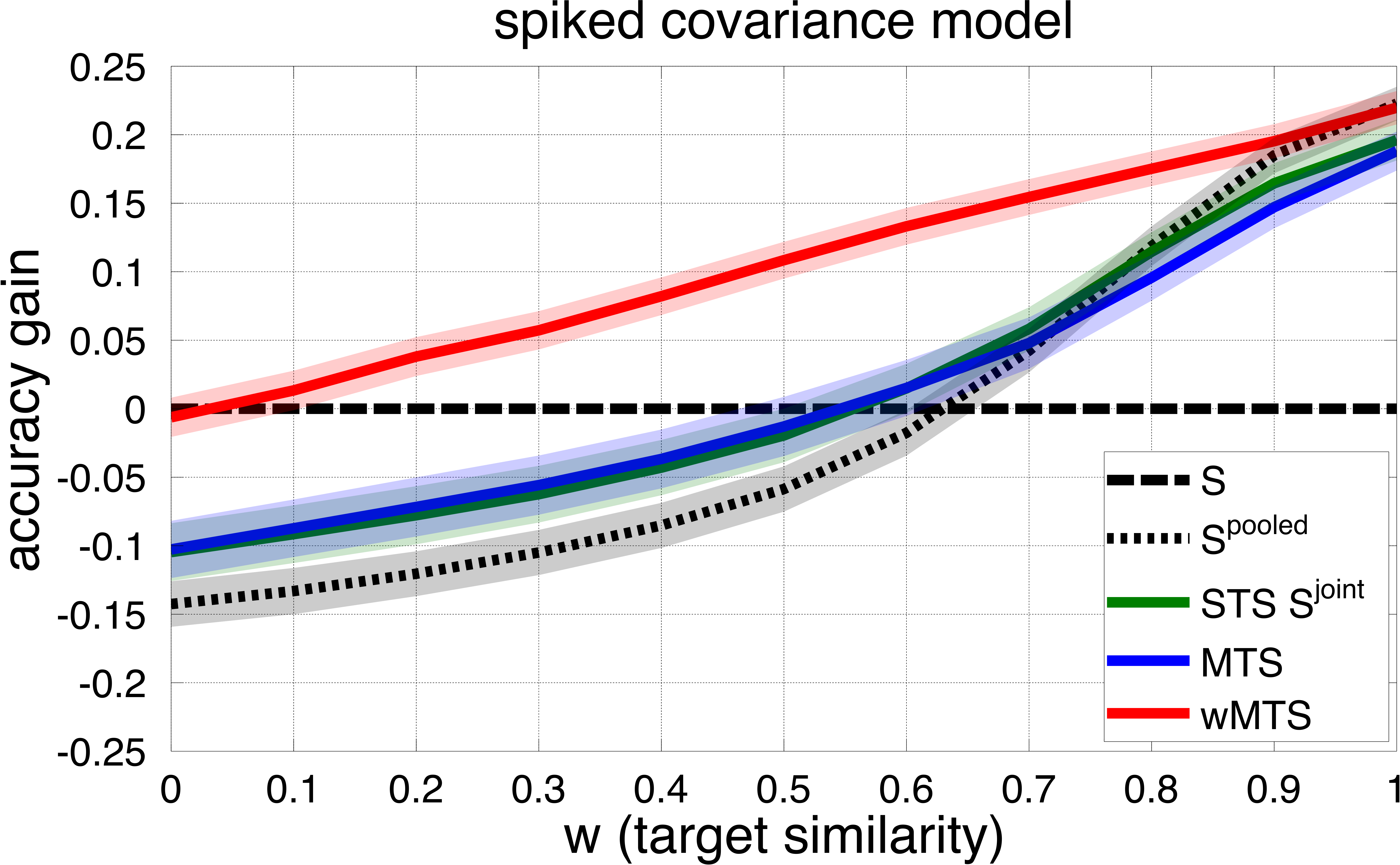}
\caption{accuracy gain for MTS of the covariance for CSP.  Average obtained over $R_r = 20$ repetitions for $R_m = 500$ models.} \label{fig:MTS_CSP}	
\end{center}
\end{figure}

Figure \ref{fig:MTS_CSP} (left) shows the relative classification accuracies of the different covariance estimation approaches. For $w = 1$, the target covariances are equal to the class covariances and $\bS^{pooled} = 1/(k+1)(\sum_k \bS^k + \bS)$ is optimal. For $w \rightarrow 0$, the targets do not contain discriminative information, hence the  sample covariance becomes optimal. STS to the joint covariance of the additional data sets performs better then the pooled covariance, but is clearly outperformed by MTS. Whitened MTS performs even better.

For Figure \ref{fig:MTS_CSP} (right) a spike has been added to all covariance matrices: The largest eigenvalue has been multiplied by 100 and the corresponding direction was excluded from the random rotations. This strong direction dominates the standard STS and MTS estimates and causes a strong degradation of performance. The performance of whitened MTS, on the other hand, is not affected.

\section{Multi-Target Shrinkage on Real World Data}
\label{sec:realworld}
In this section we will spotlight two application scenarios of MTS on real world data, one for MTS of the mean estimation and one for MTS of the covariance. Detailed articles on these applications are in preparation.

\subsection{MTS of the mean for Event Related Potentials}
In a Brain-Computer Interface (BCI) paradigm based on event related potentials (ERPs), Linear Discriminant Analysis (LDA) is commonly applied to a binary classification problem (targets vs.~nontargets). A detailed overview of the state-of-the-art approaches for feature extraction and classification for ERP data in BCI application is given in \citep{BlaLemTreHauMue10}. 

Generally, a sequence of $k$ different stimuli are presented repetitively in an random order. The user attends on only one stimulus (target\footnote{Note that despite having the same name,  there is no relation between the targets in an ERP experiment and Shrinkage targets.
}), while neglecting all others (non-targets). For each stimulus, the brain response is evaluated and it is assessed whether or not the user was attending. Then, a one-out-of-$k$-class decision has to be taken based on the $k$ binary LDA classifier outputs. 

The standard approach is to compute an LDA classifier by pooling all target and all non-target data, thus neglecting the stimulus identity. Alternatives are STS and MTS:  we compute a binary classifier for each stimulus, using the mean over the distinct stimulus classes as a shrinkage target (STS) or each mean of each distinct stimulus class as a separate shrinkage target (MTS).
In ERP, the covariance can be considered as general background activity which is independent of the 
stimulus. Hence, for all approaches we take the pooled covariance.

\begin{figure} 
\begin{center}
\includegraphics[width= 0.95 \linewidth]{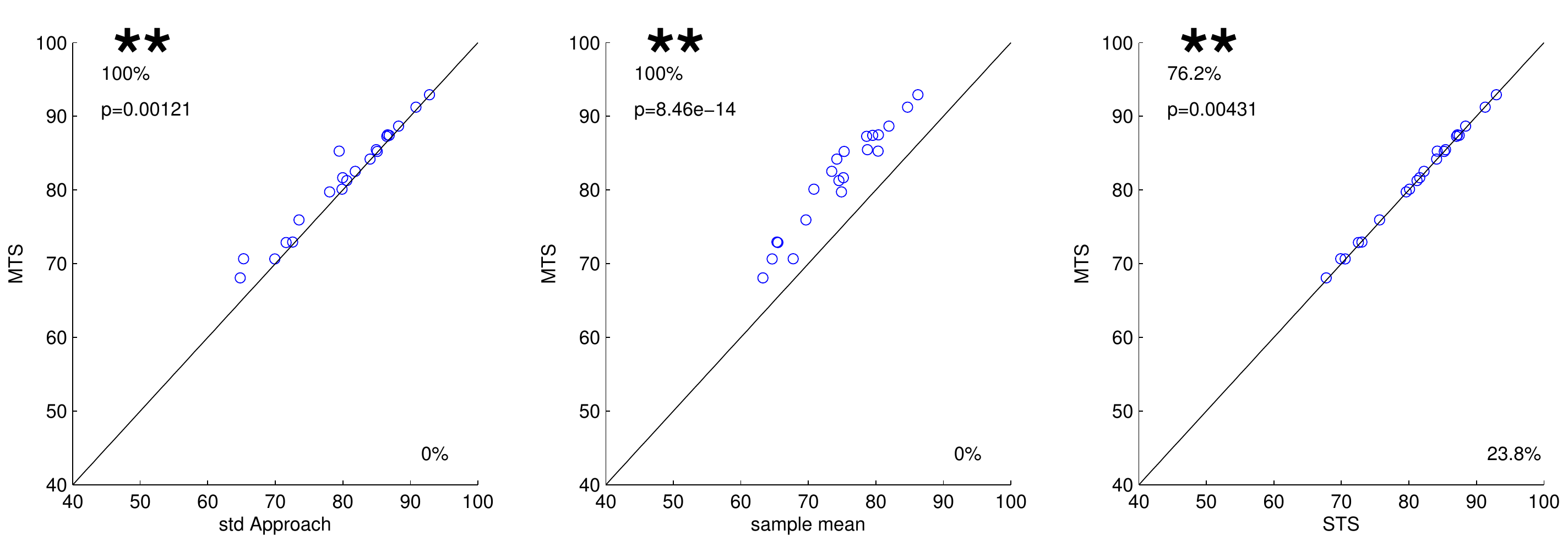}
\caption{classification accuracy of the ERP data using several estimates of the mean. A subject is marked with a circle. It should be noted that all three plots show the same data on the y-axis, being the classification accuracy obtained with the MTS mean estimate.} \label{fig:ERP_MTS}	
\end{center}
\end{figure}

One data set comprising of 21 subjects was reanalyzed \citep{SchRosTan11}. Figure~\ref{fig:ERP_MTS} shows the classification accuracies when computing the MTS mean, comparing against classification accuracies obtained with other estimates for the mean. Next to the MTS estimator, the pooled sample mean (standard approach), sample estimate of the stimulus specific mean and the STS mean estimate was analyzed. For the STS mean estimator, the pooled mean of the remaining classes was considered as target. The analysis shows the MTS estimator of the mean to be superior to all other approaches.

\subsection{MTS of the covariance matrix for motor imagery data}
We reanalyzed a data set from a Brain Computer Interface based on motor imagery. In the experiment, subjects had to imagine two different movements while brain activity was measured via EEG ($p = 55$ channels, 80 subjects, 150 trials per subject, each trial with $n_\text{trial} = 390$ measurements~\citep{blankertz2010neurophysiological}). For each subject the frequency band was optimized. Common Spatial Patterns (CSP) was applied on the class-wise covariance matrices  for feature extraction. 1-3 filters per class were chosen by a heuristic \citep{blankertz2008optimizing} and Linear Discriminant Analysis was applied on log-variance features.

As training is expensive, we are interested in exploiting training data from other subjects.
We compare two approaches: STS to the covariance of all other subjects and Multi-Target Shrinkage to all 80 subjects.
Directions of high variance dominate shrinkage estimators \citep{BarMue13} and the BCI data contains pronounced directions of high variance, the spectrum is heavily tilted. To reduce the impact of the first eigendirections without giving to much importance to low variance noise directions we applied a special form of whitening: we rescaled, only for the calculation of the shrinkage intensities,  the first five principal components to have the same variance as the sixth principal component. Shrinkage is corrected for auto-correlation  \citep{BarMue14}.

\begin{figure} 
\begin{center}
\includegraphics[width= 0.95 \linewidth]{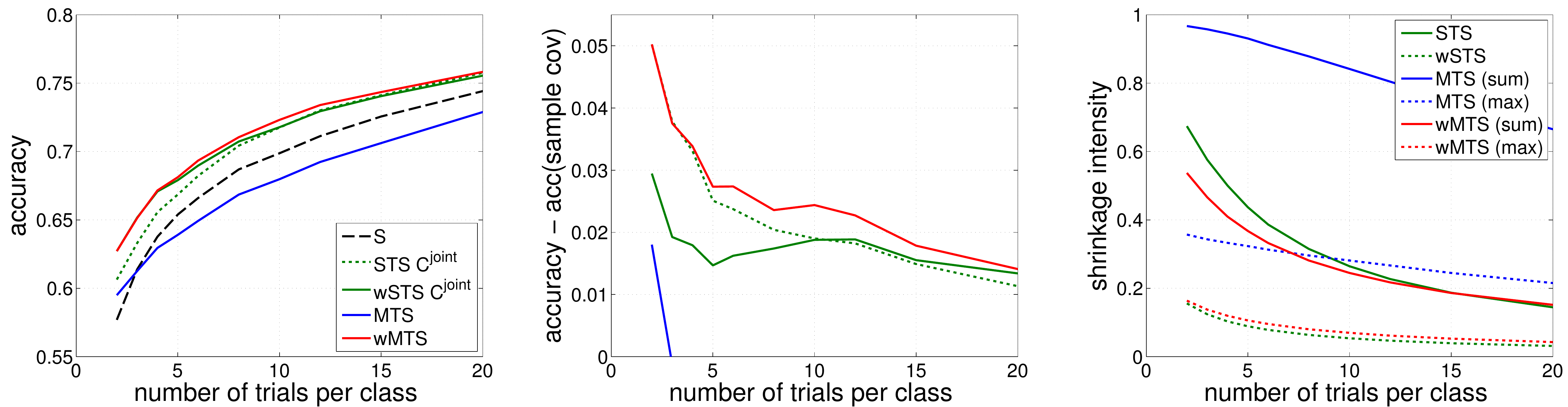}
\caption{dependency on the number of training trials of motor imagery  BCI. Average obtained over $R = 100$ runs.} \label{fig:MI_CSP_tt}
\end{center}
\end{figure}

Figure \ref{fig:MI_CSP_tt}   (left, middle) shows accuracies for different number of training trials per class. One can see that STS outperforms sample covariance matrices, while it is not possible to estimate the high number of parameters for MTS. For few training trials, wSTS outperforms STS, as the averaging over additional dimensions reduces variance. wMTS yields very good accuracies.

Figure \ref{fig:MI_CSP_tt}  (right) shows shrinkage intensities. One can see why MTS fails: when shrinkage is dominated by the first eigendirections, there are targets which appear too good and receive very large shrinkage intensities. Whitened MTS takes more directions into account and yields lower shrinkage intensities.

\begin{figure} 
\begin{center}
\includegraphics[width= 0.95 \linewidth]{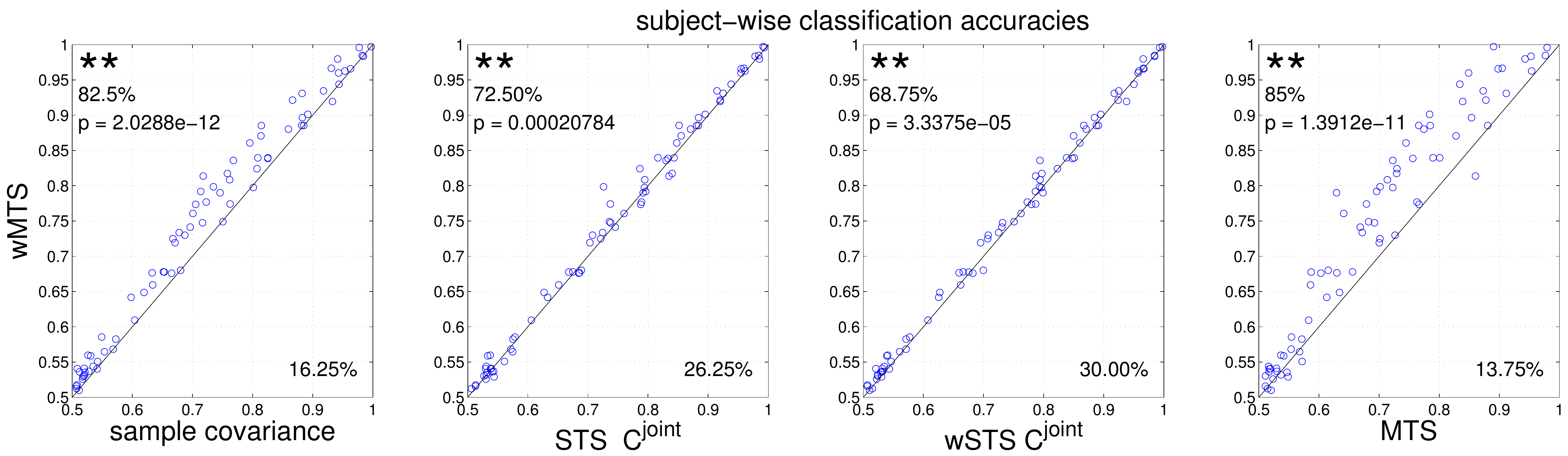}
\caption{subject-wise classification accuracies for motor imagery BCI. 10 training trials. Average obtained over $R = 100$ runs. $^{**}/^{*} :=$ significant at $p\leq0.01$ or $p\leq0.05$, respectively.} \label{fig:MI_CSP_scatter}
\end{center}
\end{figure}

Figure \ref{fig:MI_CSP_scatter} shows subject-wise accuracies for the different covariance matrix approaches for ten training trials. Our proposed wMTS estimator significantly outperforms all other approaches.

\section{Discussion}
\label{sec:discussion}
Shrinkage is a widely applied estimation technique. In the last years the analytic formula for covariance shrinkage of Ledoit and Wolf \citep{LedWol04} has become very  popular: it is a fast and  accurate alternative to cross-validation. 

In this paper, we pointed out several use cases in which a single shrinkage target is not sufficient. This motivates the usage of multiple shrinkage targets (MTS). 
We have derived formulas for optimal Multi-Target Shrinkage and we have shown in theory and simulations that MTS yields improvements over standard shrinkage in several situations. As a practical trick, we proposed whitening as a preprocessing step which increases the robustness of MTS.

On two real world data sets from the  neuroscience domain, our proposed method yields  a significant performance enhancement over standard shrinkage.

Future work will explore  connections to random matrix theory, consider the transfer of domain specific prior knowledge into the proposed  framework, application of MTS to other estimators and the analysis of new real world data sets.  In addition we are interested in incorporating label information into the weighting of the different dimensions and into adaptively whitening only to an extent which sufficiently reduces the variance of the shrinkage estimates.

\newpage 

\acks{Klaus-Robert M\"uller gratefully acknowledges funding by BMBF Big Data Centre (01 IS 14013 A) and the National Research Foundation grant (No.\ 2012-005741) funded by the Korean government. We thank Pieter-Jan Kindermans, Sebastian Bach, Shinichi Nakajima and  Duncan Blythe for valuable discussions and comments.
}

\begin{appendix}

\section{Proofs}

\subsection{Proof of Theorem~\ref{th:solutionMTS} (MTS quadratic program)}

\begin{proof}
We decompose the EMSE into bias and variance
\begin{align}
\label{eq:lw-loss}
\Delta^{\mathrm{MTS}} ( \boldsymbol{\lambda} ) 
&  ={\mathbb E}  \left\|  \bTheta - \hbTheta^{\mathrm{MTS}}(  \blambda ) \right\|^2
=   \bbE \left[  \sum_{i=1}^q \left( \hTheta^{\mathrm{MTS}}(  \blambda )_i - \theta_i \right)^2 \right]  
\\
\notag 
& = \bbE \left[  \sum_{i=1}^q \left( \left(1-\sum_{k=1}^K \lambda_k \right) \hTheta_i + \sum_{k=1}^K \lambda_k \hT_i^k - \theta_i \right)^2 \right] 
\\
\notag 
&  =  \sum_{i=1}^q \Bigg\{  \left(1 - \sum_{k=1}^{K}  \lambda_k \right)^2 \var ( \hTheta_i ) 
 + \sum_{j,k=1}^{K}  \lambda_j \lambda_k  \cov (\hT^j_i, \hT^k_i) 
 \\
 \notag
& {\color{white} abcdefghi}  + \sum_{j=1}^{K}  2  \lambda_j  \left(1 - \sum_{k=1}^{K} \lambda_k \right)  \cov ( \hT^j_i, \hTheta_{i}) 
 \\
  \notag
& {\color{white} abcdefghi}  + \left\{ \sum_{k=1}^{K} \lambda_k \bbE \left[\hT^k_i - \hTheta_{i}   \right]  \right\} 
\left\{ \sum_{j=1}^{K} \lambda_j \bbE \left[ \hT^j_i - \hTheta_{i}   \right]  \right\} \Bigg\}.
\end{align}
This can be simplified to 
\begin{align}
\Delta^{\mathrm{MTS}}  ( \blambda ) 
& =  \sum_{i=1}^q \Bigg\{ \sum_{j,k=1}^{K}  \lambda_j \lambda_k    \bbE \left[ \left( \hT^j_i - \hTheta_{i} \right) \left( \hT^k_i - \hTheta_{i} \right)  \right] \notag \\
& {\color{white} abcdefghi}   +  2 \sum_{k=1}^{K}  \lambda_k \big( \cov ( \hT^k_i, \hTheta_{i} ) - \var (\hTheta_{i})  \big)  
 + \var (\hTheta_{i} ) \Bigg\} \notag \\
& =  \boldsymbol{\lambda}\transpose \mathbf{A} \boldsymbol{\lambda}  - 2 \bb\transpose \boldsymbol{\lambda}  + \sum_{i=1}^q \var (\hTheta_{i}) 
= 2 \Delta^{\mathrm{MTS}}_{\mathrm{qp}}  ( \blambda )  + const. \label{eq:MTSopt}
\end{align}
Therefore the sets of $\blambda$ minimizing $\Delta^{\mathrm{MTS}} ( \blambda )$ and $\Delta^{\mathrm{MTS}}_{\mathrm{qp}} ( \blambda )$ 
are identical.
\end{proof}

\subsection{Proof of Theorem \ref{the:qpconsistency} (consistency of MTS)} 
\begin{proof}
From the constraints, it follows directly that 
\begin{align}
\label{eq:orderlambda}
\| \blambda^\star \| = \mathcal O (1)
\end{align}
 and from the definition of $\bA$ and $\bb$ follows $\forall k: \tau_A^k \geq \tau_{\hTheta}$.
We first prove (i).
We have $\forall k:$
\begin{align}
 {\blambda^{\star}} \transpose \bA \blambda^{\star} 
& = 
\sum_{{k'},l=1}^K \lambda_{k'}^\star \lambda_l^\star {\sum_{i=1}^{q} {\mathbb E} \left[ \left( \hT^k_i - \hTheta_{i} \right) \left( \hT^l_i - \hTheta_{i} \right)  \right] }, 
\notag
\\
& \geq 
{\lambda_k^\star}^2
\min_{\substack{%
        \balpha \in \mathbb{R}^K_{\geq0} \\
        \alpha_k = 1}}
\sum_{i=1}^{q}  \bbE \left(   \sum_{l=1}^K  \alpha_l    (\hT^l_i - \hTheta_{i} ) \right) ^2
\stackrel{\eqref{eq:G3'}}{=} 
{\lambda_k^\star}^2 \Theta \left( p^{\tau_A^k} \right) 
\label{eq:lAl-lim}
 \\
\bb \transpose {\blambda^{\star}}
& \stackrel{\eqref{eq:G2},\eqref{eq:orderlambda}}{=} \mathcal O( p^{\tau_{\hTheta}}) ,
\label{eq:bl-lim}
\end{align}
%
%
We then have $\forall k:$
\begin{align*}
\Theta( p^{\tau_{\hTheta}}) 
& \stackrel{\eqref{eq:G1}}{=} 
\Delta^{\hTheta}
\geq 
\Delta^\mathrm{MTS}(\blambda^\star) 
\stackrel{\eqref{eq:MTSopt}}{=}
 {\blambda^{\star}} \transpose \bA \blambda^{\star} 
-  2 \bb \transpose {\blambda^{\star}}
+ \sum_i \var(\hTheta_i) \\
& \stackrel{\eqref{eq:orderlambda},\eqref{eq:lAl-lim},\eqref{eq:bl-lim}}{\geq} 
{\lambda_k^\star}^2 \Theta ( p^{\tau_A^k} ) + \mathcal O( p^{\tau_{\hTheta}}).
\end{align*}
Rearranging yields $\lambda^\star_k = {\mathcal O}(p^{0.5(\tau_{\hTheta} - \tau_A^k)})$. 
To prove statement (i) for $\hat \lambda_k$, we first define
\begin{align*}
\hDelta^\mathrm{MTS}(\blambda) 
:=  {\blambda} \transpose \hbA  \blambda -  2 \hbb \transpose \blambda + \sum_{i=1}^p \var (\hTheta_{i}).
\end{align*}
Using the result on the limit behaviour of $\blambda^\star$, we obtain
\begin{align}
{\blambda^{\star}} \transpose ( \bA - \hbA ) \blambda^{\star} 
& = 
\sum_{k,l=1}^K \lambda_k^\star \lambda_l^\star ( A_{kl} - \hA_{kl})
\stackrel{\eqref{eq:G3}}{=}
\sum_{k,l=1}^K \lambda_k^\star \lambda_l^\star o \left( p^{0.5 (\tau_A^k + \tau_A^l) }\right)
 = o( p^{\tau_{\hTheta}})
 \label{eq:lAAhl-limit}
\end{align}
This allows us to calculate
\begin{align}
\Delta^\mathrm{MTS}(\blambda^\star) -  \hDelta^\mathrm{MTS}(\blambda^\star)
=    {\blambda^{\star}} \transpose ( \bA - \hbA ) \blambda^{\star} 
- 2 ( \bb - \hbb ) \transpose {\blambda^{\star}}
\stackrel{\eqref{eq:orderlambda},\eqref{eq:lAAhl-limit}}{=}
  o( p^{\tau_{\hTheta}}).
  \label{eq:D-D-1}
\end{align}
In addition, we calculate
\begin{align}
\hDelta^\mathrm{MTS}(\hblambda)  -  \Delta^\mathrm{MTS}(\hblambda) 
=
{\hblambda} \transpose ( \hbA - \bA ) \hblambda
- 2( \hbb - \bb ) \transpose {\hblambda} 
 = 
  \sum_{k} \hat \lambda_k^2 o( p^{ \tau_A^k }) + o( p^{\tau_{\hTheta}}).
    \label{eq:D-D-2}
\end{align}
Using these equations, we obtain
\begin{align*}
\Theta( p^{\tau_{\hTheta}}) 
& \geq \Delta^\mathrm{MTS}(\blambda^\star) 
\stackrel{\eqref{eq:D-D-1}}{=}
\hDelta^\mathrm{MTS}(\blambda^\star)   + o( p^{\tau_{\hTheta}}) 
\geq \hDelta^\mathrm{MTS}(\hblambda)   + o( p^{\tau_{\hTheta}})  
\\
& \stackrel{\eqref{eq:D-D-2}}{=}
\Delta^\mathrm{MTS}(\hblambda)   + o( p^{\tau_{\hTheta}})    + \sum_{k} \hat \lambda_k^2 o( p^{ \tau_A^k }) 
\\
& \stackrel{\eqref{eq:lAl-lim},\eqref{eq:bl-lim}}{\geq} 
{\hat \lambda_k}^2 \Theta ( p^{\tau_A^k} ) + \mathcal O( p^{\tau_{\hTheta}}) 
+ o( p^{\tau_{\hTheta}})    + \sum_{k} \hat \lambda_k^2 o( p^{ \tau_A^k }).
\end{align*}
Rearranging yields $\hat \lambda_k = {\mathcal O}(p^{0.5(\tau_{\hTheta} - \tau_A^k)})$ which concludes (i).
%
%
To prove statement  (ii) we have to relate the difference in ESE to the difference in the estimate of the ESE:
\begin{align*}
& \left( \Delta^\mathrm{MTS}(\hblambda) -   \Delta^\mathrm{MTS}(\blambda^\star)  \right)  - 
\Big(  \hDelta^\mathrm{MTS}(\hblambda) - \hDelta^\mathrm{MTS}(\blambda^\star)   \Big) \\
& = \left( \Delta^\mathrm{MTS}(\hblambda) -     \hDelta^\mathrm{MTS}(\hblambda)^\mathrm{MTS} \right)  - 
\Big( \Delta^\mathrm{MTS}(\blambda^\star)  - \hDelta^\mathrm{MTS}(\blambda^\star)   \Big) 
\stackrel{\eqref{eq:D-D-1},\eqref{eq:D-D-2},(i)}{=}  o(p^{\tau_{\hTheta}})
\end{align*}
Using this and the optimalities of  
$\blambda^\star$ for $ \Delta^\mathrm{MTS}(\blambda)$ and
$\hblambda$ for $ \hDelta^\mathrm{MTS}(\blambda)$, we obtain
\begin{align*}
0 
& \leq ({\Delta^{\hTheta}})^{-1}  
\left( \Delta^\mathrm{MTS}(\hblambda) -   \Delta^\mathrm{MTS}(\blambda^\star)  \right) \\
&   = \Theta (  p^{- \tau_{\hTheta}}) 
\Big(  \hDelta^\mathrm{MTS}(\hblambda) - \hDelta^\mathrm{MTS}(\blambda^\star)   +  o(p^{\tau_{\hTheta}})  \Big)
\\
%
& \leq  0 +  o(1)
\end{align*}
which concludes the proof of (ii).

The proof of part (iii)  is similar to the one of Theorem 2.1 from \citep{Dan73}. On the convex set we have 
\begin{align}
0  & \leq
 (\hblambda - \blambda^\star)  \transpose  \nabla \Delta^\mathrm{MTS}(\blambda^\star)
 \label{eq:Dan1} \\
0 & \leq
(\blambda^\star - \hblambda) \transpose  \nabla \hDelta^\mathrm{MTS}(\hblambda)^\mathrm{MTS}
\label{eq:Dan2}
\end{align}
where the gradients are  $\nabla \Delta^\mathrm{MTS}(\blambda) =  \left( \bA \blambda + \bb \right)$ and $\nabla \hDelta^\mathrm{MTS}(\blambda) = \left( \hbA \blambda + \hbb \right)$.
Multiplying eq.~\eqref{eq:Dan2} by minus one and combining the two equations, we obtain
\begin{align}
\notag
(\hblambda - \blambda^\star)  \nabla \hDelta^\mathrm{MTS}(\hblambda)  \leq (\hblambda - \blambda^\star)  \transpose  \nabla \Delta^\mathrm{MTS}(\blambda^\star) .
\end{align}
Subtracting $ (\hblambda - \blambda^\star)  \nabla \hDelta^\mathrm{MTS}(\blambda^\star)$ from both sides, we obtain
\begin{align*}
 (\hblambda - \blambda^\star)  \transpose  \left( \nabla \hDelta^\mathrm{MTS}(\hblambda) -  \nabla \hDelta^\mathrm{MTS}(\blambda^\star) \right)
 \leq 
  (\hblambda - \blambda^\star)  \transpose  \left( \nabla  \Delta^\mathrm{MTS}(\blambda^\star) -  \nabla \hDelta^\mathrm{MTS}(\blambda^\star) \right).
\end{align*}
The left hand side is 
\begin{align*}
   (\hblambda - \blambda^\star)  \transpose  \hbA  (\hblambda - \blambda^\star)  
& \geq
   \| \hblambda - \blambda^\star\|^2 \min_{\| \balpha \| = 1} \balpha \transpose \bA \balpha  
   + ( \hblambda - \blambda^\star)  \transpose ( \hbA  - \bA ) (\hblambda - \blambda^\star)  \\
&   \stackrel{\eqref{eq:G3'},  \alpha \in \mathbb R^K}{=}
 \| \hblambda - \blambda^\star\|^2 \cdot \Theta ( p^{\tau_{\hTheta}} ).
\end{align*}
The right hand side is
\begin{align*}
\Big(    (\hblambda - \blambda^\star)  \transpose  (\hbA - \bA) \blambda^\star  +   (\hblambda - \blambda^\star)  \transpose  (\bb - \hbb) \Big) 
  = o ( p^{\tau_{\hTheta}}) .
\end{align*}
by \eqref{eq:G1}, \eqref{eq:G2}, \eqref{eq:G3} and the rates of the $\lambda_k$ given by (i).
Therefore, rearranging yields $\| \hblambda - \blambda^\star\|^2 = o ( 1 ) $.
\end{proof}

\subsection{Proof of Theorem \ref{th:LDLconsistency} (LDL consistency of MTS of the mean)}
 \begin{proof}
Without loss of generality, we assume $\bmu = \mathbf{0}$. We start by analysing the asymptotic behaviour of the $\Delta^{\hTheta}$,  $A_{kk}$ and $b$, then we prove the consistency of $\hA_{kl}$ and $\hb$.
\paragraph{\eqref{eq:G1} \& \eqref{eq:G2}: Asymptotic behaviour of $\Delta^{\hTheta}$, $A_{kk}$ and $b$}
We start with the asymptotic behaviour of $\Delta^{\hTheta} = b$. We have
\begin{align}
 \label{eq:LDL_b_asymptotics}
\Delta^{\hTheta}
& = b
 =   \sum_{i=1}^p \var (\hat \mu_{i})   
 =   n^{-1}  \sum_{i=1}^p \var ( x_{is})   
 =   n^{-1}  \sum_{i=1}^p \gamma_i
 \overset{\eqref{ass:sum_sigma}}{=} \Theta(1) 
 \stackrel{!}{= } \Theta(p^{\tau_{\hTheta}})
 \\
 \notag
& \Longleftrightarrow
\tau_{\hTheta} = 0
\end{align}
Using this result, we obtain the asymptotic behaviour of $A_{kk}$:
\begin{align}
   \label{eq:LDL_A_asymptotics}
A_{kk}
& = {\sum_{i=1}^p \bbE \left[ \left( \hat \mu^k_i - \hat \mu_{i} \right)^2 \right] }
 = \sum_{i=1}^p \bbE \left[ 
 (\hat \mu^k_{i})^2  - 2 \hat \mu^k_{i} \hat \mu_{i}  - \hat \mu_i^2
  \right]  
   \\
& = \sum_{i=1}^p \Big\{ \bbE \left[ (\hat \mu^k_{i})^2 \right]+ {\mathbb E} \left[ \hat  \mu_{i}^2
  \right]  \Big\} 
 = \sum_{i=1}^p \Big\{ (\mu^k_{i})^2  + \var \left( {{\hat  \mu}^k_{i}}   \right)  +  \var \left( \hat  \mu_{i}   \right)  \Big\} 
 \notag \\
& = \Theta (p^{\tau^k_\mu}) + \Theta( 1 ) 
\stackrel{!}{= }\Theta ( p^{\tau_A^k } )
\notag 
\\
\notag
\Longleftrightarrow 
& \,
\tau_A^k 
= \max(\tau^k_\mu,0)
%
\end{align}

\paragraph{\eqref{eq:G3}, part I: Consistency of $\hA_{kl}$}
As $\hA_{kl}$ is unbiased, we have to show that 
\begin{align}
\label{eq:dominance}
\var ( \hA_{kl} ) 
= o (p^{\tau_A^k + \tau_A^l} )
= o (p^{\max(\tau^k_\mu,0)+ \max(\tau^l_\mu,0)} )
\end{align}
We introduce the notation 
\begin{align}
\notag
\check x_{is}^{(k)} & = x_{is}^{(k)} - \mu^{(k )}_i,\\
\notag
\check \mu^{(k)}_i & = n^{-1} \sum_s \check x^{(k)}_{is}.
\end{align}
We then have
\begin{align}
\notag
\var( \hA_{kl} )  
& = \var \left( \sum_{i=1}^p  \left( \hat \mu^{k}_i - \hat \mu_{i} \right) \left( \hat  \mu^{l}_i - \hat \mu_{i} \right)  \right) \\
\label{eq:_hA_decomp1}
& = \var \left( \sum_{i=1}^p  
\left(  \check \mu^{k}_i - \check \mu_{i}  + \mu^k_i \right) 
\left( \check  \mu^{l}_i - \check \mu_{i}  + \mu^l_i \right)  \right) 
\end{align}
To show eq.~\eqref{eq:dominance}, it is sufficient to show the that the variance of each combination of terms in eq.~\eqref{eq:_hA_decomp1} is $o (p^{\tau_A^k + \tau_A^l} )$. There are three non-constant types of combinations: First, there is the product of a mean and a sample mean:
\begin{align*}
\var \left( \sum_{i=1}^p   \mu^k_i  \check \mu^{l}_i \right)
& = n_l^{-2} \sum_{ij} \cov(  \mu^k_i  \sum_s \check x^{l}_{is},  \mu^k_j  \sum_t \check  x^{l}_{jt}  )
= n_l^{-1} {\bmu^k} \transpose \bC^l\bmu^k \\
& \stackrel{\eqref{eq:LDLweakness}}{=}
 \frac{ {\bmu^k} \transpose \bC^l\bmu^k}{\| \bmu^k \|^2} \Theta( p^{\tau_\mu^k - 1}) 
= \max_i \gamma_i^l \Theta( p^{\tau_\mu^k - 1})  \\
& \stackrel{\eqref{eq:LDLmeancovrestrictions},\eqref{ass:sum_sigma2}}{=}
  o( p^{\tau_\mu^l + 1}) \Theta( p^{\tau_\mu^k - 1})  
= o ( p^{\tau^k_\mu + \tau^k_\mu})
= o (p^{\tau_A^k + \tau_A^l} )
\end{align*}
Second, there are products of two different sample means:
\begin{align*}
\var \left( \sum_{i=1}^p   \check \mu_i  \check \mu^{k}_i \right)
& = n^{-2} n_k^{-2}\var \left( \sum_{i=1}^p \sum_{s,t}  \check x_{is}  \check x^{k}_{it} \right)
=  n^{-1} n_k^{-1}  \sum_{i,j=1}^p \cov \left(   x_{i1} ,   x_{j1} \right) \cov \left(   x^{k}_{i1} ,   x^{k}_{j1} \right) \\
& =  n^{-1} n_k^{-1}  \sum_{i,j=1}^p \cov \left(   y_{i1} ,  y_{j1} \right) \cov \left(   z^{k}_{i1} ,   z^{k}_{j1} \right) 
 =  n^{-1} n_k^{-1}  \sum_{i=1}^p \gamma_i \bbE   [  (z^{k}_{i1})^2 ]  \\
& \leq  \frac{1}{n n_k}  \sum_{i=1}^p \gamma_i \gamma^{k}_i   
\leq \frac{p}{n n_k} \sqrt{p^{-1} \sum_{i=1}^p \gamma_i ^2 } \sqrt{p^{-1} \sum_{i=1}^p (\gamma_i^k )^2 } \\
& = \Theta \left( p^{ 0.5(\tau_\gamma^k + \tau_\gamma^k) - 1 } \right)
 \overset{\eqref{eq:LDLmeancovrestrictions}}{=} o \left( p^{ \max(0,\tau_\mu^k) +  \max(0,\tau_\mu^l)  } \right)
 = o (p^{\tau_A^k + \tau_A^l} )
\end{align*}
The third combination has two sample means:
\begin{align}
\notag
\var \left( \sum_{i=1}^p   \check \mu_i^2  \right) 
& = n^{-4} \var \left( \sum_{i=1}^p \sum_{s,t}  y_{is}  y_{it} \right) 
 = n^{-4} \sum_{i,j=1}^p  \sum_{s,t,s',t'} \cov \left(  y_{is} y_{it} , y_{js'} y_{jt'} \right) \\
 \notag
& = n^{-4} \sum_{i,j=1}^p   \left\{ 
\sum_{s} \cov \left(  y_{is}^2, y_{js}^2 \right) 
+  \sum_{s,t\neq s} \cov \left(  y_{is} y_{it} , y_{js} y_{jt} \right)
\right\} 
\\
\notag
& \leq n^{-3} \sum_{i,j=1}^p   \cov \left(  y_{i1}^2, y_{j1}^2 \right) 
+ n^{-2} \sum_{i,j=1}^p    \cov^2 \left(  y_{i1} , y_{j1}\right)  \\
\notag
& \leq
p^2 n^{-3} \left( p^{-1} \sum_{i=1}^p   \sqrt{ \bbE  \left[  y_{i1}^4 \right] } \right)^2
+ p n^{-2} \left( p^{-1} \sum_{i=1}^p    \gamma_i^2  \right) \\
\notag 
& \overset{\eqref{ass:sum_sigma},\eqref{ass:fourth_moms}}{=} 
\mathcal O (p^{-1})
+ \Theta \left( p^{\tau_\gamma -1)} \right)
\overset{\eqref{eq:LDLmeancovrestrictions}}{=} 
\mathcal O (p^{-1})
+ o \left( p^{2 \max(0,\min_k \tau_\mu^k)} \right) \\
\label{eq:varmuhsq}
& = o( p^{\tau_A^k + \tau_A^l} ) \qquad  \forall k,l
\end{align}
We have shown that the variance of all terms and hence $\var ( \hA_{kl})$ is  $o( p^{\tau_A^k + \tau_A^l} )$.

\paragraph{\eqref{eq:G3}, part II: Consistency of $\hat b$}The estimator $\hb$ is also unbiased, hence we have to show 
\begin{align*}
\var ( \hb ) = o(p^{\tau_{\hTheta}}) = o(1).
\end{align*}
In a first step, we reformulate the variance:
\begin{align*}
\var ( \hat b ) 
& = \var \left(  \sum _{i=1}^p  \varh (\hat \mu_{i})   \right)
 = \var \left(   n^{-1}(n-1)^{-1} \sum _{i=1}^p \sum_{t=1}^n{ (x_{it} - \hat \mu_i})^2  \right) \\
& = n^{-2}(n-1)^{-2} \var \left(    \sum _{i=1}^p \sum_{t=1}^n x_{it}^2 - n^{-1} \sum _{i=1}^p \sum_{s,t=1}^n x_{is} x_{it} \right).
\end{align*}
The variance is $o(1)$ if the variances of both terms in the sum are $o(p^4)$. We start with 
\begin{align*}
\var \left(    \sum _{i=1}^p\sum_{t=1}^n x_{it}^2  \right) 
& = n \var \left(    \sum _{i=1}^p x_{it}^2  \right) 
 = n \var \left(    \sum _{i=1}^p y_{it}^2  \right) \\
& = n  \sum _{i,j=1}^p \cov \left(   y_{it}^2 , y_{jt}^2  \right) 
 \leq n  \sum _{i,j=1}^p \sqrt{\bbE \left[ y_{it}^4  \right] } \sqrt{\bbE \left[ y_{jt}^4  \right] } 
 \\
& \stackrel{\eqref{ass:fourth_moms}}{\leq} 
p^2 n (1+\alpha_4)  \left( p^{-1} \sum _{i=1}^p \gamma_i  \right)^2 
 \stackrel{\eqref{ass:sum_sigma}}{=}
  \mathcal O(p^3) = o(p^4).
\end{align*}
The variance of the second term in the sum is, following the steps in eq.~\eqref{eq:varmuhsq}, 
\begin{align*}
 \var \left(   n^{-1} \sum _{i=1}^p\sum_{s,t}^n x_{is} x_{it}  \right)  
= \mathcal O (p^{3})
+ o \left( p^{2 \max(0,\min_k \tau^k_\mu) +2} \right)
\stackrel{\eqref{eq:LDLweakness}}{=} o(p^4).
\end{align*}
This concludes the proof the $\var ( \hb )$ is  $o(p^{\tau_{\hTheta}}) = o(1)$.
\paragraph{\eqref{eq:G3'}: Restriction on linear combinations} 
Let $\mathbb L$ be $\mathbb R^p$ or $\mathbb R^p_{\geq0}$. We have
\begin{align}
\label{eq:LDL_lin_comb_proof}
\Theta \left( p^{\tau_A^k} \right) 
& \stackrel{!}{=}
\min_{\substack{\balpha \in \mathbb L  \\ \alpha_k = 1}}
\sum_{i=1}^{q}   \bbE 
\left[ \left(   
\sum_{l=1}^K  \alpha_l    (\hT^l_i - \hTheta_{i} ) 
\right) ^2  \right]
 = \min_{\substack{\balpha \in \mathbb L  \\ \alpha_k = 1}}
\sum_{i=1}^{q}  \bbE 
\left[ \left(   
\sum_{l=1}^K  \alpha_l    (\hmu^l_i - \hmu_{i} ) 
\right) ^2 \right] 
\\
\notag
& = \min_{\substack{\balpha \in \mathbb L  \\ \alpha_k = 1}}
\sum_{i=1}^{q} 
\left\{ \left(   
\sum_{l=1}^K  \alpha_l    (\mu^l_i - \mu_{i} ) \right) ^2  
+ \sum_{l=1}^K  | \alpha_l | \var \left(      \hmu_i  \right)  
+ \sum_{l=1}^K  | \alpha_l  |  \var \left(       \hmu^l_i  \right)  
\right\}
\\
\notag
& \stackrel{\eqref{eq:LDLmean_lin_comb}}{\geq }
\Theta \left( p^{\tau_\mu^k} \right) + \sum_{i=1}^q \var (\hmu_i^k) 
= \Theta \left( p^{\tau_\mu^k} \right) + \Theta(1) 
= \Theta \left( p^{\max(0,\tau_\mu^k)} \right) 
\end{align}
This concludes the proof of Theorem~\ref{th:LDLconsistency}.
 \end{proof}


\subsection{Proof of Theorem~\ref{th:FOLDLconsistency}
(FOLDL consistency of MTS of the mean)}
 \begin{proof}
As above, without loss of generality, we assume $\bmu = \mathbf{0}$. We again start by analysing the asymptotic behaviour of $\Delta^{\hTheta}$, $A_{kk}$ and $b$, then we prove consistency of $\hA_{kl}$ and $\hat b$.

\paragraph{\eqref{eq:G1} \& \eqref{eq:G2}: Asymptotic behaviour of $\Delta^{\hTheta}$, $A_{kk}$ and $b$}
From  equations  
\eqref{eq:LDL_b_asymptotics}  and\eqref{eq:LDL_A_asymptotics}
we directly obtain
\begin{align*}
\tau_{\hTheta} &   = 1
\qquad \text{and} \qquad
\forall k: \tau_A^k   = 1.
\end{align*}

\paragraph{\eqref{eq:G3}, part I: Consistency of $\hA_{kl}$} 
As for the LDL, we show that all types of  terms in eq.~\eqref{eq:_hA_decomp1} are $o(p^{\tau_A^k + \tau_A^l})$. For the FOLDL, this means they have to be  $o(p^2)$. Following similar steps as above, we obtain
\begin{align}
\notag
\var \left( \sum_{i=1}^p   \mu^k_i  \check \mu^{l}_i \right)
& = n_l^{-1} {\bmu^k} \transpose \bC^l\bmu^k
= \max_i \gamma_i^l \Theta( p^{\tau_\mu^k })  
\stackrel{\eqref{eq:FOLDLmeancovrestrictions}}{=} o( p^{1 + \tau^k_\mu }) 
= o(p^2) 
\\
\notag
\var \left( \sum_{i=1}^p   \check \mu_i  \check \mu^{k}_i \right)
& \leq \frac{p}{n n_k} \sqrt{p^{-1} \sum_{i=1}^p \gamma_i ^2 } \sqrt{p^{-1} \sum_{i=1}^p (\gamma_i^k )^2 } 
= o \left( p^{ \max(1,\tau_\mu^k) +  \max(1,\tau_\mu^l)  } \right)
= o(p^2)
\\
 \label{eq:covsquares2}
\var \left( \sum_{i=1}^p   \check \mu_i^2  \right) 
& = n^{-4} \sum_{i,j=1}^p   \left\{ 
\sum_{s} \cov \left(  y_{is}^2, y_{js}^2 \right) 
+  \sum_{s,t\neq s} \cov \left(  y_{is} y_{it} , y_{is} y_{it} \right)
\right\} 
\\
\notag
& \leq
\frac{1}{ n^{3}} \sum_{i,j\neq i}    \cov \left(  (y_{is}^k)^2, (y_{js}^k)^2 \right)
+ \frac{(1+\alpha_4) p}{ n^3} \left( p^{-1} \sum_{i=1}^p    \gamma_i^2  \right) 
+ \frac{p}{ n^2}  \left( p^{-1} \sum_{i=1}^p    \gamma_i^2  \right) 
\\
\notag
& \overset{\eqref{eq:FOLDLmeancovrestrictions},\eqref{eq:avdims}}{=} 
o \left( p^2 \right)
+ o \left( p^2 \right)
\end{align}
We have shown that the variance of all terms and hence $\var ( \hA_{kl})$ is  $o( p^{\tau_A^k + \tau_A^l} )$.

\paragraph{\eqref{eq:G3}, part II: Consistency of $\hat b$}
We have to show that $\var(\hb)$ is $o( p^{2 \tau_{\hTheta} } ) = o( p^2 )$:
\begin{align*}
\var ( \hat b ) 
& = \var \left(   \sum _{i=1}^p  \varh (\hat \mu_{i})   \right) 
 = \var \left(   \frac{1}{n(n-1)} \sum _{i=1}^p \sum_{t=1}^n{ (x_{it} - \hat \mu_i})^2  \right) \\
& =  \frac{1}{n^2(n-1)^2} \var \left(    \sum _{i=1}^p\sum_{t=1}^n x_{it}^2 
- n^{-1} \sum _{i=1}^p \sum_{s=1}^n x_{is} x_{is} 
- n^{-1} \sum _{i=1}^p \sum_{s,t\neq s=1}^n x_{is} x_{it} \right).
\end{align*}
This variance expression is $o(p^2)$ if  the variance of each of the three sums is $o(p^2)$. For the first sum, we use eq.~\eqref{eq:covsquares2} and obtain
\begin{align*}
\var \left(    \sum _{i=1}^p x_{it}^2  \right) 
& = \var \left(    \sum _{i=1}^p y_{it}^2  \right) 
= \sum_{ij} \cov ( y_{it}^2, y_{jt}^2 ) \\
& \leq
\frac{1}{ n^{3}} \sum_{i,j\neq i}    \cov \left(  (y_{is}^k)^2, (y_{js}^k)^2 \right)
+ \frac{(1+\alpha_4) p}{ n^3} \left( p^{-1} \sum_{i=1}^p    \gamma_i^2  \right)  \\
& = o(p^2) + \mathcal O( p ).
\end{align*}
The second sum is proportional to the first sum. For the third sum we obtain, by using eq.~\eqref{eq:covsquares2},
\begin{align*}
& \var \left(  \sum _{i=1}^p\sum_{s,t\neq s}^n x_{is} x_{it}  \right) 
 =    \sum_{ij}  \sum_{s,t,s',t'} \cov \left(  x_{is} x_{it}, x_{js'} x_{jt'} \right) 
= o \left( p^2 \right).
\end{align*}
This concludes the proof that $\var ( \hb )$ is  $o(p^{\tau_{\hTheta}}) = o(p^2)$.
\paragraph{\eqref{eq:G3'}: Restriction on linear combinations} 
Similar to eq.~\eqref{eq:LDL_lin_comb_proof}, we have
\begin{align*}
\Theta \left( p^{\tau_A^k} \right) 
\stackrel{!}{=}
\min_{\substack{\balpha \in \mathbb R \\ \alpha_k = 1}}
\sum_{i=1}^{q} &  \bbE 
\left[ \left(   \sum_{l=1}^K  \alpha_l    (\hT^l_i - \hTheta_{i} ) \right) ^2 
\right]
 \geq  \sum_i \var (\hmu_i^k) 
= \Theta \left( p  \right). 
\end{align*}
This concludes the proof of Theorem~\ref{th:FOLDLconsistency}.
 \end{proof}

\subsection{Proof of Theorem \ref{th:LDLconsistencyCov} (LDL consistency of MTS of the covariance)}
\begin{proof}
The estimators  $\hbA$ and $\bb$ depend on the choice of target. We restrict the proof on targets given by sample covariance matrices of additional data sets. The biased  estimators in \citet{SchStr05} and \citet{LedWol03} have smaller variance, consistency can be shown following similar steps.

\paragraph{\eqref{eq:G1} \& \eqref{eq:G2}: Asymptotic behaviour of $\Delta^{\hTheta}$, $b_k$ and $A_{kk}$}
We first show the asymptotic behaviour 
\begin{align}
\label{eq:LDLcov_b}
\Delta^{\hTheta} =  b_{k} 
= \left( \sum_{ij}  \var  \big( S_{ij} \big) \right) 
= \Theta \left(  p  \right) 
\stackrel{!}{=} \Theta \left(  p^{\tau_{\hTheta}}  \right) 
\qquad \Longleftrightarrow \tau_{\hTheta} = 1.
\end{align}
Rotation invariance allows us to analyse in the eigenbasis. The upper bound follows from 
\begin{align}
\label{eq:LDLcovbupper}
\sum_{i,j}  \var  \big( S'_{ij} \big) 
& \leq  \frac{1}{n} \sum_{i,j} \left\{  \sqrt{ \var (y_{i1}^2) \var ( y_{j1}^2 ) } + \mathbb{E} \left[   y_{i1}^2  \right] \mathbb{E} \left[   y_{j1}^2  \right] - \mathbb{E}^2 \left[   y_{i1} y_{j1}  \right] \right\} \\
\notag
& \leq  \frac{1}{n} \sum_{i,j} \left\{  \sqrt{ \mathbb{E} [ y_{i1}^4 ]  \mathbb{E} [ y_{j1}^4  ] } + \mathbb{E} \left[   y_{i1}^2  \right] \mathbb{E} \left[   y_{j1}^2  \right] - \mathbb{E}^2 \left[   y_{i1} y_{j1}  \right] \right\} \\
\notag 
&  \leq  \frac{2}{n} \sum_{i,j}  \sqrt{ \mathbb{E} [ y_{i1}^4 ]  \mathbb{E} [ y_{j1}^4  ] } 
\leq  \frac{2 p^2}{n} (1 + \alpha_4) \left( \frac{1}{p} \sum_{i}  \mathbb{E} [ y_{i1}^2 ]  \right)^2  
 = \Theta(p).
\end{align}
For the lower bound, we distinguish two cases: for $\tau_\gamma = 1$, we have
\begin{align}
\label{eq:LDLcovblower1}
\sum_{i,j}  \var  \big( S'_{ij} \big) 
& \geq \sum_{i}  \var  \big( S'_{ii} \big) \\
\notag
& =  \frac{1}{n} \sum_{i} \left\{ \mathbb{E} \left[  y_{i1}^4\right] - \mathbb{E}^2 \left[   y_{i1}^2  \right] \right\} \\
\notag
& \geq \frac{1}{n} \sum_{i} \beta_4 \mathbb{E}^2 \left[   y_{i1}^2  \right] 
= \frac{\beta_4 p}{n} \frac{1}{p} \sum_{i}  \gamma_i^2   \\
\notag
& = \Theta \left(  p^{\tau_\gamma}  \right) = \Theta ( p ).
\end{align}
For the case $\tau_\gamma < 1$, we have 
\begin{align}
\label{eq:LDLcovblower2}
\sum_{i,j}  \var  \big( S'_{ij} \big) 
& =  \frac{1}{n} \sum_{i,j} \left\{ \mathbb{E} \left[  y_{i1}^2 y_{j1}^2 \right] - \mathbb{E}^2 \left[   y_{i1} y_{j1}  \right] \right\} \\
\notag
& \geq \frac{1}{n} \sum_{i,j} \left\{  \mathbb{E} \left[   y_{i1}^2  \right] \mathbb{E} \left[   y_{j1}^2  \right] - \mathbb{E}^2 \left[   y_{i1} y_{j1}  \right] \right\} \\
\notag
& \geq \frac{1}{n} \left( \sum_{i}  \mathbb{E} \left[   y_{i1}^2  \right] \right)^2  - \frac{1}{n} \sum_i \mathbb{E}^2 \left[   y_{i1}^2  \right] \\
\notag
& \geq \frac{p^2}{n} \left( \frac{1}{p} \sum_{i}  \mathbb{E} \left[   x_{i1}^2  \right] \right)^2  - \frac{p}{n}  \frac{1}{p} \sum_i \gamma_i^2   \\
\notag
& = \Theta(p) - \Theta( p^{\tau_\gamma} )) = \Theta ( p ).
\end{align}
The asymptotic behaviour of $A_{kk}$ depends on the relationship between the original data $\bX$ and the additional data set $\bX^k$:
\begin{align}
\notag
A_{kk} 
& =  {\sum_{i,j=1}^p \bbE \left[ \left( S^k_{ij} - S_{ij} \right) \left( S^k_{ij} -  S_{ij} \right)  \right] }, \\
\notag
& = \sum_{i,j=1}^p ( C_{ij} - C^k_{ij} )^2 + \var ( S^k_{ij} )  + \var  (S_{ij}) 
\stackrel{\eqref{eq:covsimilarity},\eqref{eq:LDLcov_b}}{=}
 \Theta( p^{\tau_C^k} ) + \Theta( p )
 \stackrel{!}{=} 
\Theta( p^{ \tau_A^k })
\\
\notag
\Longleftrightarrow \tau_A^k 
& = \max(1,\tau_C^k).
\end{align}

\paragraph{\eqref{eq:G3}: Consistency of $\hA_{kl}$} 
As the estimator  $\hA_{kl}$ is unbiased \citep{BarMue13}, we have to  show that 
\begin{align}
 \notag
 \var \left( \hA_{kl} \right)
 & =  \var \left(  \sum_{i,j=1}^p  \left( S^{k}_{ij} -  S_{ij} \right) \left( S^{l}_{i,j} - S_{i,j} \right)   \right) 
 \\ 
 \label{eq:varSdecomp}
 &  =  \var \left(  \sum_{i,j=1}^p  S^{k}_{ij} S^l_{ij} - S^{k}_{ij} S_{ij} - S^{l}_{ij} S_{ij} +  S_{ij}^2  \right), 
 \\
\notag
& = o(p^{\tau_A^k + \tau_A^l} ) 
= o( p^{\max(1,\tau_C^k)  + \max(1,\tau_C^l)}).
 \end{align}
It suffices to show that the variances of all terms in the sum in eq.~\eqref{eq:varSdecomp} are $o(p^{\tau_A^k + \tau_A^l} ) $.

\paragraph{Variance of $\sum_{ij} S_{ij}^2$} We start with the product of two identical sample covariances:
\begin{align}
\notag 
\sum_{ij} S_{ij}^2
& = \sum_{ij} \left( \frac{1}{n}  \sum_s y_{is} y_{jt} \right)^2 
 = \frac{p^2}{ n^2 } \sum_{st} \left( \frac{1}{p}  \sum_i y_{is} y_{it} \right)^2 
 \\
 \label{eq:Ssq_decomp}
& = \frac{p^2}{ n^2 } \sum_{s } \left( \frac{1}{p}  \sum_i y_{is}^2 \right)^2  
+ \frac{1}{ n^2 } \sum_{s,t\neq s} \left(   \sum_i y_{is} y_{it} \right)^2 .
\end{align}
Again, it is sufficient to show that the variance of both terms separately is $ o(p^{\tau_A^k + \tau_A^l} )$. For the first term, we have
\begin{align*}
 \var & \left( \frac{p^2}{ n^2 } \sum_{s } \left( \frac{1}{p}  \sum_i y_{is}^2 \right)^2  \right) 
 \leq  \frac{p^4}{ n^3 } \mathbb{E} \left[ \left( \frac{1}{p} \sum_i y_{i1}^2 \right)^4  \right]  \\
&  \leq  \frac{p^4(1+\alpha_8)}{ n^3 } \mathbb{E} \left[ \frac{1}{p} \sum_i y_{i1}^2 \  \right] 
= \mathcal O (p) 
= o(p^{\tau_A^k + \tau_A^l} ).
\end{align*}
Let us now look at the second term in  eq.~\eqref{eq:Ssq_decomp}:
\begin{align}
\notag 
\var & \left( \frac{1}{ n^2 } \sum_{s,t\neq s } 
\left(   \sum_i y_{is} y_{it} \right)^2  \right) 
\\
\notag 
& = \frac{1}{ n^4 } \sum_{s,t\neq s } \sum_{s',t'\neq s' } 
\cov \left( \left(  \sum_i y_{is} y_{it} \right)^2  , \left(  \sum_i y_{is'} y_{it'} \right)^2  \right). 
\end{align}
The covariance expression only depends on the cardinal of the intersection, which we denote by $\left( \{s,t\}\cup \{s',t'\} \right)^{\#} $ and which can take the values of 0, 1 and 2. When this cardinality is zero,
$$\left( \{s,t\}\cup \{s',t'\} \right)^{\#}  = 0,$$
there is independence and the covariance is zero as well.
For
$$\left( \{s,t\}\cup \{s',t'\} \right)^{\#}  = 1,$$
we have $4n(n-1)(n-2)$ expressions of the form
\begin{align*}
 & \cov \left( \left(   \sum_i y_{i1} y_{i2} \right)^2  , \left(   \sum_i y_{i1} y_{i3} \right)^2  \right) \\
& =  \mathbb{E} \left[ \left(   \sum_i y_{i1} y_{i2} \right)^2  \left(   \sum_i y_{i1} y_{i3} \right)^2  \right] 
- \mathbb{E} \left[ \left(   \sum_i y_{i1} y_{i2} \right)^2 \right] \mathbb{E} \left[  \left(   \sum_i y_{i1} y_{i3} \right)^2  \right] \\
& \leq \max \left(   \mathbb{E} \left[ \left(   \sum_i y_{i1} y_{i2} \right)^2  \left(   \sum_i y_{i1} y_{i3} \right)^2  \right] , 
 \mathbb{E}^2 \left[ \left(   \sum_i y_{i1} y_{i2} \right)^2 \right]  \right),
 \end{align*}
as both terms are positive.  For the first term, we have
 \begin{align*}
 \mathbb{E}  & \left[ \left(   \sum_i y_{i1} y_{i2} \right)^2  \left(   \sum_i y_{i1} y_{i3} \right)^2  \right] 
 =   \sum_{i,j,i',j'} \mathbb{E} \left[  y_{i1}  y_{i'1}  y_{j1}  y_{j'1} \right] \mathbb{E} \left[ y_{i2} y_{i'2}  \right] \mathbb{E} \left[  y_{j3}  y_{j'3}   \right] \\
& =    \sum_{i,j} \mathbb{E} \left[  y_{i1}^2y_{j1}^2 \right] \mathbb{E} \left[ y_{i2}^2  \right] \mathbb{E} \left[  y_{j3}^2   \right] 
 \leq    p^2 \left( \frac{1}{p} \sum_{i} \sqrt{ \mathbb{E} \left[  y_{i1}^4 \right] } \mathbb{E} \left[ y_{i2}^2  \right]  \right)^2 \\
& \stackrel{\text{A6}}{\leq} p^2(1 + \alpha_4) \left( \frac{1}{p} \sum_{i} \mathbb{E}^2 \left[ y_{i2}^2  \right]  \right)^2 
=\mathcal O \left( p^{2 \tau_\gamma  + 2 }\right).
\end{align*}
For the second term, we have
 \begin{align*}
 \mathbb{E}^2 \left[ \left(   \sum_i y_{i1} y_{i2} \right)^2 \right]  
 =  \left(   \sum_{i,j}  \mathbb{E}^2 \left[  y_{i1} y_{j1}  \right]   \right)^2 
 =  p^2 \left( \frac{1}{p}  \sum_{i}  \mathbb{E}^2 \left[  y_{i1}^2  \right]   \right)^2
 =  \mathcal O \left( p^{2 \tau_\gamma + 2 }\right)
\end{align*}
Therefore, we have, combined with the prefactors, 
\begin{align*}
\frac{4n(n-1)(n-2) }{n^4 } 
& \left| \cov  \left( \left( \frac{1}{p}  \sum_i y_{is} y_{it} \right)^2  , \left( \frac{1}{p}  \sum_i y_{is} y_{it} \right)^2  \right) \right| \\
& = \frac{1 }{ n }  \mathcal{O} \left( p^{2 \tau_\gamma + 2} \right) 
= \mathcal{O} \left( p^{2 \tau_\gamma + 1} \right)
\stackrel{\eqref{eq:EVdispgrowthrate}}{=}
o(p^{\tau_A^k + \tau_A^l} ), 
\end{align*}
therefore we have shown that the terms with $\left( \{s,t\}\cup \{s',t'\} \right)^{\#}  = 1$ are 
$ o(p^{\tau_A^k + \tau_A^l} )$.

For 
\begin{align*}
 \left( \{s,t\}\cup \{s',t'\} \right)^{\#} =  2,
 \end{align*}
 we get $2n(n-1)$ expressions of the form
\begin{align*}
&  \left| \cov   \left( \left(   \sum_i y_{is} y_{it} \right)^2  , \left(   \sum_i y_{is} y_{it} \right)^2  \right) \right| 
 = \left| \cov \left( \left(   \sum_i y_{i1} y_{i2} \right)^2  , \left(   \sum_i y_{i1} y_{i2} \right)^2  \right) \right| \\
& \quad \leq   \sum_{i,j,i',j'} \left| \cov \left(  y_{i1}  y_{i2}  y_{i'1}  y_{i'2} , y_{j1} y_{j2}   y_{j'1}  y_{j'2}   \right) \right|.
\end{align*}
We decompose the set of integers into two disjoint subsets: $\{1,\dots,p\}^4 = Q \cup R$, where $Q$ is the set of distinct integers and $R$ is the remainder:
\begin{align*}
& =  \sum_{i,j,i',j'  \in Q} \left| \cov \left(  y_{i1}  y_{i2}  y_{i'1}  y_{i'2} , y_{j1} y_{j2}   y_{j'1}  y_{j'2}   \right) \right| 
 + \sum_{i,j,i',j'  \in R}  \left| \cov \left(  y_{i1}  y_{i2}  y_{i'1}  y_{i'2} , y_{j1} y_{j2}   y_{j'1}  y_{j'2}   \right) \right| .
\end{align*}
The sum over $Q$ we can bring into a form which is dominated as a consequence of  \eqref{eq:LDLcovcons}:
\changed{
\begin{align}
& \left| \cov \left(  y_{i1}  y_{i2}  y_{i'1}  y_{i'2} , y_{j1} y_{j2}   y_{j'1}  y_{j'2}   \right) \right|  \notag \\
& \qquad =  \left| \mathbb{E}^2 \left[  y_{i1}   y_{i'1}  y_{j1}  y_{j'1} \right] 
- \mathbb{E}^2 \left[  y_{i1}   y_{i'1} \right] \mathbb{E}^2 \left[ y_{i2}  y_{i'2} \right]   \right|  
=   \mathbb{E}^2 \left[  y_{i1}   y_{i'1}  y_{j1}  y_{j'1} \right] \notag \\
& \quad =   \left(   \cov \left(  y_{i1}   y_{i'1} , y_{j1}  y_{j'1} \right) +  \mathbb{E} \left[  y_{i1}   y_{i'1} \right] \mathbb{E} \left[ y_{j1}  y_{j'1} \right] \right)^2 
 =    \left(   \cov \left(  y_{i1}   y_{i'1} , y_{j1}  y_{j'1} \right) \right)^2. \label{eq:simpleQ}
\end{align}
}{
\begin{align*}
& \left| \cov \left(  y_{i1}  y_{i2}  y_{i'1}  y_{i'2} , y_{j1} y_{j2}   y_{j'1}  y_{j'2}   \right) \right|  \\
& = \left| \mathbb{E} \left[  y_{i1}  y_{i2}  y_{i'1}  y_{i'2}  y_{j1} y_{j2}   y_{j'1}  y_{j'2}   \right] 
- \mathbb{E} \left[  y_{i1}  y_{i2}  y_{i'1}  y_{i'2} \right] \mathbb{E} \left[ y_{j1} y_{j2}   y_{j'1}  y_{j'2}   \right]  \right| \\
& =  \left| \mathbb{E}^2 \left[  y_{i1}   y_{i'1}  y_{j1}  y_{j'1} \right] 
- \mathbb{E}^2 \left[  y_{i1}   y_{i'1} \right] \mathbb{E}^2 \left[ y_{i2}  y_{i'2} \right]   \right|  \\
& =   \mathbb{E}^2 \left[  y_{i1}   y_{i'1}  y_{j1}  y_{j'1} \right] \\
& =   \left(   \cov \left(  y_{i1}   y_{i'1} , y_{j1}  y_{j'1} \right) +  \mathbb{E} \left[  y_{i1}   y_{i'1} \right] \mathbb{E} \left[ y_{j1}  y_{j'1} \right] \right)^2 \\
& =    \left(   \cov \left(  y_{i1}   y_{i'1} , y_{j1}  y_{j'1} \right) \right)^2.
\end{align*}
}
Taking the prefactors into account, we get
\begin{align*}
 \frac{2n(n-1) }{ n^4 }  & \sum_{(i,j,i',j')  \in Q} \left| \cov \left(  y_{i1}  y_{i2}  y_{i'1}  y_{i'2} , y_{j1} y_{j2}   y_{j'1}  y_{j'2}   \right) \right|  \\
& \leq 48  \frac{p^4   }{ n^2 }    \sum_{(i,j,i',j')  \in Q}  \frac{ \left(   \cov \left(  y_{i1}   y_{i'1} , y_{j1}  y_{j'1} \right) \right)^2 } {  | Q_p | } \\
& \overset{\eqref{eq:LDLcovcons}}{=} \mathcal{O} (p^2 ) o(1) = o(p^2) = o(p^{\tau_A^k + \tau_A^l} )
\end{align*}
%
For the sum over $R$, we have
\begin{align}
 \sum_{(i,j,i',j')  \in R} 
 & \left| \cov \left(  y_{i1}  y_{i2}  y_{i'1}  y_{i'2} , y_{j1} y_{j2}   y_{j'1}  y_{j'2}   \right) \right| \notag \\
& \leq \sum_{i,j,j'} 2 \left| \cov \left(  y_{i1}^2  y_{i2}^2, y_{j1} y_{j2}   y_{j'1}  y_{j'2}   \right) \right| 
+ 4 \left| \cov \left(  y_{i1}  y_{i2} y_{i'1}  y_{i'2}, y_{i1}  y_{i2} y_{j1} y_{j2}      \right) \right|  \notag  \\
& \leq  \sum_{i,j,j'} 2 \sqrt{ \mathbb{E} \left[  y_{i1}^4  y_{i2}^4 \right] \mathbb{E} \left[ y_{j1}^2 y_{j2}^2   y_{j'1}^2  y_{j'2}^2 \right]  } 
+ 4 \sqrt{ \mathbb{E} \left[ y_{i1}^2 y_{i2}^2   y_{j1}^2  y_{j2}^2 \right]  \mathbb{E} \left[ y_{i1}^2 y_{i2}^2   y_{j'1}^2  y_{j'2}^2 \right]  } \notag  \\
& \leq 6 \sum_{i,j,j'} \mathbb{E} \left[  y_{i1}^4  \right] \sqrt{ \mathbb{E} \left[ y_{j1}^4  \right] } \sqrt{ \mathbb{E} \left[  y_{j'1}^4  \right] }
%
  \leq 6 p^3 (1 + \alpha_4) \left( \frac{1}{p} \sum_{i} \mathbb{E}^2 \left[  y_{i1}^2  \right] \right) \left( \frac{1}{p} \sum_{j} \mathbb{E} \left[ y_{j1}^2    \right] \right)^2 \notag  \\
& =   \mathcal O \left(  p^{2 \tau_\gamma + 3} \right). \label{eq:Rrep}
\end{align}
Together with the prefactors, we obtain
\begin{align*}
\frac{2n(n-1)}{ n^4 } & 
\sum_{(i,j,i',j')  \in R} \left| \cov \left(  y_{i1}  y_{i2}  y_{i'1}  y_{i'2} , y_{j1} y_{j2}   y_{j'1}  y_{j'2}   \right) \right| 
 = \frac{1}{n^2}   \mathcal{O} \left( p^{2\tau_\gamma + 3} \right) 
 \\
& = \mathcal{O} \left( p^{2\tau_\gamma +1 } \right)
 \stackrel{\eqref{eq:EVdispgrowthrate}}{=}
o(p^{\tau_A^k + \tau_A^l} ).
\end{align*}
This finishes the proof for  the terms with $\left( \{s,t\}\cup \{s',t'\} \right)^{\#}  = 2$ and in total we have shown that 
$\var ( \sum_{ij} S_{ij}^2)$ is $o(p^{\tau_A^k + \tau_A^l} )$. For $\var ( S_{ij}^k S_{ij}^k), k=l$, an analogue proof holds.

\paragraph{Variance of $\sum_{ij} S^k_{ij}S_{ij}$}
Let us now analyse the products of different sample covariances in  eq.~\eqref{eq:varSdecomp}.
\begin{align*}
\var \left(   \sum_{ij} S^{k}_{ij} S_{ij}  \right) 
& = \var \left(   \sum_{ij}\sum_{st} x^k_{is} x^k_{js} x_{it} x_{jt}  \right) 
 = \frac{1}{n n_k} \sum_{ijgh} \cov \left(   x^k_{i1} x^k_{j1} x_{i2} x_{j2}, x^k_{g1} x^k_{h1} x_{g2} x_{h2}  \right) \\
 & = \frac{1}{n n_k}\sum_{ijgh}\cov \left(  x^k_i  x^k_j , x^k_g x^k_h  \right)  \cov \left(  x_i  x_j , x_g x_h  \right) \\
&  - C_{ij}^k  C^k_{gh} \cov \left(  x_i  x_j , x_g x_h  \right)   
 - C_{ij}  C_{gh} \cov \left(  x^k_i  x^k_j , x^k_g x^k_h  \right) 
 \end{align*}
 The first term  can be separated into the contributions from the two different data sets:
 \begin{align*}
 \frac{1}{n n_k} \sum_{ijgh}\cov \left(  x^k_i  x^k_j , x^k_g x^k_h  \right)  \cov \left(  x_i  x_j , x_g x_h  \right) 
\leq  \frac{1}{n n_k} \sum_{ijgh}\cov^2 \left(  x^k_i  x^k_j , x^k_g x^k_h  \right) +  \cov^2 \left(  x_i  x_j , x_g x_h  \right)
 \end{align*}
These terms are rotation invariant, therefore we analyse
 \begin{align*}
 \frac{1}{n n_k} \sum_{ijgh}  \cov^2 \left(  y_i  y_j , y_g y_h  \right).
 \end{align*}
 For $i,j,g,h$ distinct, this leads directly to assumption \eqref{eq:LDLcovcons}. Otherwise, we have
 \begin{align*}
  \frac{1}{n n_k} \sum_{ijgh}  \cov^2 \left(  y_i  y_j , y_g y_h  \right)
& \leq \frac{4}{n_k n}\sum_{ijg} \cov^2 \left(  y_i  y_g , y_j y_g  \right) + \cov^2 \left(  y_i  y_j , y_g y_g  \right) \\
& \leq \frac{8}{n_k n} \sum_{ijg}  \bbE  \left[  (y_{i1})^2 y_{j1}^2\right]   \bbE \left[ (y_{g1})^4 \right]  \\
& \leq \frac{8 }{n_k n} \sum_{ijg} 
\sqrt{ \bbE  \left[  (y_{i1})^4  \right]  }
\sqrt{ \bbE  \left[  (y_{j1})^4  \right]  }
\bbE  \left[  (y_{g1})^4  \right]  \\
& \leq \frac{8p^3}{n_k n} 
\left ( \frac{1}{p} \sum _i  \gamma_i \right)^2
\left ( \frac{1}{p} \sum _g \gamma_g^2 \right) \\
& = \mathcal O \left( p^{\tau_\gamma + 1} \right) 
= o(p^{\tau_A^k + \tau_A^l} ).
 \end{align*}
Next we consider the second term,
\begin{align*}
 \frac{1}{n n_k} \sum_{ijgh} & C_{ij}^k  C^k_{gh} \cov \left(  x_i  x_j , x_g x_h  \right)   
 =  \frac{1}{n n_k} \sum_{ijgh} \Sigma_{ij}^k  \Sigma^k_{gh} \cov \left(  z_i  z_j , z_g z_h  \right)  \\
& \leq   \frac{1}{n n_k} \sum_{ig} \gamma_{i}^k  \gamma^k_{g} \sqrt{ \bbE \left[ z_i^4  \right] \bbE \left[ z_g^4  \right] } 
 =   \frac{1}{n n_k}  \left( \sum_{i} \gamma_{i}^k   \sqrt{ \bbE \left[ z_i^4  \right]} \right) \\
& \leq   \frac{p^2}{n n_k}  \left( \frac{1}{p} \sum_{i} (\gamma_{i}^k)^2   + \bbE \left[ z_i^4  \right]  \right)^2 
  \leq   \frac{p^2(1+\alpha_4)}{n n_k}  \left( \frac{1}{p} \sum_{i} (\gamma_{i}^k)^2   + \gamma^2_i  \right)^2 \\
&  =    \mathcal O ( p^{2\max(\tau_\gamma,\tau_\gamma^k)})
= o(p^{\tau_A^k + \tau_A^l} ).
\end{align*}
With this we have shown that all terms in $\var \left(   \sum_{ij} S^{k}_{ij} S_{ij}  \right) $ and hence $\var (\hA_{kl})$ is $o(p^{\tau_A^k + \tau_A^l} )$.
\paragraph{\eqref{eq:G3}, part II: Consistency of $\hat b_{k}$} By reformulation we obtain
\begin{align}
\notag
\sum_{ij} \varh S_{ij} 
& =   \sum_{ij} \Bigg( \frac{1}{(n-1)n}  \sum_s \Big( y_{is}y_{js} - \frac{1}{n} \sum_{s'} y_{is'} y_{js'} \Big)^2  \Bigg)   
\\
\notag
& = \frac{1}{(n-1)n}  \sum_{ij} \Bigg(  \sum_s  y_{is}^2y_{js}^2 - \frac{1}{n} \sum_{ss'} y_{is} y_{js} y_{is'} y_{js'}   \Bigg)  
\\
\label{eq:var_b_k}
& = \frac{p^2}{(n-1)n} \sum_s   \left(  \frac{1}{p} \sum_{i}  y_{is}^2 \right)^2  
- \frac{1}{(n-1)} \sum_{ij} S_{ij}^2   .
\end{align}
Both terms, with different prefactors, have been analysed above. The variance of first term is $\mathcal O(p)$ and the bound on the variance of the second term is $n^{-2} o(p^{2\max(1,\tau_C)}) = o( p^2 )$. Hence $\var (\hb)$ is $o(p^{2 \tau_{\hTheta}} )$.

\paragraph{\eqref{eq:G3'}: Restriction on linear combinations} Let $\mathbb L$ be $\mathbb R^p$ or $\mathbb R^p_{\geq0}$. We have
\begin{align}
\label{eq:LDL_cov_lin_comb_proof}
\Theta \left( p^{\tau_A^k} \right) 
& \stackrel{!}{=}
\min_{\substack{\balpha \in \mathbb L  \\ \alpha_k = 1}}
\sum_{i=1}^{q}   \bbE 
\left[ \left(   
\sum_{l=1}^K  \alpha_l    (\hT^l_i - \hTheta_{i} ) 
\right) ^2 \right]
 = \min_{\substack{\balpha \in \mathbb L  \\ \alpha_k = 1}}
\sum_{i=1}^{q}  \bbE 
\left[ \left(   \sum_{l=1}^K  \alpha_l    (S^l_i - S_{i} ) 
\right) ^2 
\right]
\\
\notag
& = \min_{\substack{\balpha \in \mathbb L  \\ \alpha_k = 1}}
\sum_{i=1}^{q}  
\left\{ \left(   
\sum_{l=1}^K  \alpha_l    (S^l_i - S_{i} ) 
\right) ^2  
+ \var \left(   \sum_{l=1}^K  \alpha_l    S_i  \right)  
+ \var \left(   \sum_{l=1}^K  \alpha_l    S^l_i  \right)  
\right\}
\\
\notag
& \geq \Theta \left( p^{\tau_C^k} \right) + \sum_i \var (S_i^k) 
= \Theta \left( p^{\tau_C^k} \right) + \Theta(p) 
= \Theta \left( p^{\max(1,\tau_C^k)} \right) 
\end{align}
This concludes the proof of Theorem~\ref{th:LDLconsistencyCov}.
\end{proof}

\subsection{Proof of Theorem \ref{th:FOLDLconsistencyCov} (FOLDL consistency of MTS of the covariance)}
\paragraph{\eqref{eq:G1} \& \eqref{eq:G2}: Asymptotic behaviour of $\Delta^{\hTheta}$, $b_{k}$ and $A_{kk}$}
\begin{proof}
We first show the asymptotic behaviour 
\begin{align}
\label{eq:FOLDLcov_b}
\Delta^{\hTheta} 
= b_{k}
 = \left( \sum_{ij}  \var  \big( S_{ij} \big) \right)
 = \Theta \left(  p^2  \right) 
 \stackrel{!}{=}
  \Theta \left(  p^{ \tau_{\hTheta}}  \right) 
 \qquad \Longleftrightarrow 
 \tau_{\hTheta} = 2
\end{align}
The upper bound follows from (compare to~eq.~\eqref{eq:LDLcovbupper})
\begin{align*}
b_k = \sum_{i,j}  \var  \big( S'_{ij} \big) 
\leq  \frac{2 p^2}{n} (1 + \alpha_4) \left( \frac{1}{p} \sum_{i}  \mathbb{E} [ y_{i1}^2 ]  \right)^2  
 = \Theta(p^2).
\end{align*}
For the lower bound, we again distinguish two cases: for $\tau_\gamma = 1$, we have (compare to~eq.~\eqref{eq:LDLcovblower1})
\begin{align*}
\sum_{i,j}  \var  \big( S'_{ij} \big) 
= \frac{\beta_4 p}{n} \frac{1}{p} \sum_{i}  \gamma_i^2    = \Theta \left(  p^{1 + \tau_\gamma}  \right) = \Theta ( p^2 ).
\end{align*}
For the case $\tau_\gamma < 1$, we have  (compare to~eq.~\eqref{eq:LDLcovblower2})
\begin{align*}
\sum_{i,j}  \var  \big( S'_{ij} \big) 
& \geq \frac{p^2}{n} \left( \frac{1}{p} \sum_{i}  
\mathbb{E} \left[   x_{i1}^2  \right] \right)^2  - \frac{p}{n}  \frac{1}{p} \sum_i \gamma_i^2
 = \Theta(p^2) - \Theta( p^{1+ \tau_\gamma} ) 
 = \Theta ( p^2 ).
\end{align*}
For the asymptotic behaviour of $A_{kk}$ we then have
\begin{align}
A_{kk} 
& = \sum_{i,j=1}^p ( C_{ij} - C^k_{ij} )^2 + \var ( S^k_{ij} )  + \var  (S_{ij}) 
= \Theta( p^{\tau^k_C} ) + \Theta( p^2 )
\stackrel{!}{=}
\Theta( p^{\tau_A^k}),
\\
\notag
& \Longleftrightarrow 
\forall k: \tau_A^k
 = 2
\end{align}
where used the fact that  $\sum _{ij} \var(  S^k_{ij}  )$ has the same limit behaviour as $\sum _{ij} \var(  S_{ij}  )$. 

\paragraph{\eqref{eq:G3}, part I: Consistency of $\hA_{kl}$} 
The proof is analogue to the proof in Theorem~\ref{th:LDLconsistencyCov}. We only show that $\var \left (\sum_{ij} S_{ij}^2 \right)$, the expression with the highest variance, is $o(p^{\tau_A^k + \tau_A^l} ) = o(p^4)$.
We use the same decomposition as above:
 \begin{align}
 \sum_{ij} S_{ij}^2
& = \frac{1}{ n^2 } \sum_{s } \left(  \sum_i y_{is}^2 \right)^2  
+ \frac{1}{ n^2 } \sum_{s,t\neq s} \left(   \sum_i y_{is} y_{it} \right)^2. \label{eq:Ssq_decompnfix}
\end{align}
This asymptotic setting is easier, because the sums over $s$ and $t$ are finite sums. We have a finite number of terms in the first sum in eq.~\eqref{eq:Ssq_decompnfix}: 
 \begin{align}
\var & \left(   \left(  \sum_i y_{is}^2 \right)^2  \right)
= \sum_{i,j,i',j'} \cov \left(  y_{i1}^2  y_{j1}^2, y_{i'1}^2  y_{j'1}^2  \right) \notag \\
&  =   \sum_{i,j,i',j'  \in Q} \cov \left(  y_{i1}^2  y_{j1}^2, y_{i'1}^2  y_{j'1}^2  \right) 
  +  \sum_{i,j,i',j'  \in R}  \cov \left(  y_{i1}^2  y_{j1}^2, y_{i'1}^2  y_{j'1}^2  \right). \label{eq:k1sumRb}
\end{align}
For the sum over $Q$, we need assumption \eqref{eq:LDL_cov_lin_combs}:
\begin{align*}
& \sum_{i,j,i',j'  \in Q} \cov \left(  y_{i1}^2  y_{j1}^2, y_{i'1}^2  y_{j'1}^2  \right) 
\leq p^4 24 \frac{ \sum_{i,j,i',j'  \in Q} \cov \left(  y_{i1}^2  y_{j1}^2, y_{i'1}^2  y_{j'1}^2  \right) } { | Q_p | } 
\stackrel{\eqref{eq:LDL_cov_lin_combs}}{=}
 o( p^4 ).
\end{align*}
For the sum over $R$, we have,
\begin{align}
\notag 
\sum_{(i,i',j,j') \in R} &  \cov \left(   y_{i1}^2 y_{j1}^2,  y_{i'1}^2 y_{j'1}^2 \right) 
\\
\notag 
& \leq 6\sum_{i,j,i'}  \cov \left(   y_{i1}^2 y_{j1}^2,  y_{i'1}^2 y_{i1}^2 \right)  
+ \cov \left(   y_{i1}^2 y_{j1}^2,  y_{i'1}^4 \right) 
 \\
 \notag
& \leq 6 \sum_{i,j,i'}  \sqrt{ \mathbb{E} [y_{i1}^4 y_{j1}^4 ]} \sqrt { \mathbb{E}[  y_{i'1}^4 y_{i1}^4 ] } 
+ \sqrt{ \mathbb{E} [y_{i1}^4 y_{j1}^4 ]} \sqrt { \mathbb{E}[  y_{i'1}^8 ]}  
 \\
 \notag
& \leq 6 \sum_{i,j,i'}  \sqrt[4]{ \mathbb{E} [y_{i1}^8 ] \mathbb{E} [ y_{j1}^8 ]} \sqrt[4] { \mathbb{E}[  y_{i'1}^8 ] \mathbb{E} [ y_{i1}^8 ] } 
+ \sqrt[4]{ \mathbb{E} [y_{i1}^8 ] \mathbb{E} [ y_{j1}^8 ]} \sqrt { \mathbb{E}[  y_{i'1}^8 ]   } 
\\
\notag 
& \leq 12  (1+\alpha_8) \sum_{i,j,i'}  \mathbb{E} [y_{i1}^2 ] \mathbb{E} [ y_{j1}^2]  \mathbb{E}^2[  y_{i'1}^2 ] 
 =  \mathcal{O}\left(  p^{3 + \tau_\gamma}  \right)
\stackrel{\eqref{eq:EVdispgrowthrateFOLDL}}{=} 
 o(p^4).
\end{align}
For the terms in the second sum in eq.~\eqref{eq:Ssq_decompnfix}, we have
 \begin{align}
\notag 
\var  \left(   \left(   \sum_i y_{i1} y_{i2} \right)^2 \right) 
& = \sum_{i,j,i',j'} \cov \left(     y_{i1} y_{i2}y_{j1} y_{j2}  ,   y_{i'1} y_{i'2}y_{j'1} y_{j'2}  \right) 
\\
\notag
&  \leq    \sum_{i,j,i',j'  \in Q \cup R } \left| \cov \left(  y_{i1}  y_{i2}  y_{i'1}  y_{i'2} , y_{j1} y_{j2}   y_{j'1}  y_{j'2}   \right) \right|.    
\end{align}
For the sum over $Q$, we simplify using eq.~\eqref{eq:simpleQ} and obtain
\begin{align*}
  \sum_{i,j,i',j'  \in Q}  & \left| \cov \left(  y_{i1}  y_{i2}  y_{i'1}  y_{i'2} , y_{j1} y_{j2}   y_{j'1}  y_{j'2}   \right) \right|  
  = \sum_{i,j,i',j'  \in Q}    \left(   \cov \left(  y_{i1}   y_{i'1} , y_{j1}  y_{j'1} \right) \right)^2 \\
& \ \leq  24 p^4   \sum_{(i,j,i',j')  \in Q}  \frac{ \left(   \cov \left(  y_{i1}   y_{i'1} , y_{j1}  y_{j'1} \right) \right)^2 } { |Q_p| } 
\overset{\eqref{eq:LDL_cov_lin_combs}}{=}  o(p^4)
\end{align*}
For the sum over $R$, we have, as in eq.~\eqref{eq:Rrep},
 \begin{align*}
&   \sum_{i,j,i',j'  \in R}  \left| \cov \left(  y_{i1}  y_{i2}  y_{i'1}  y_{i'2} , y_{j1} y_{j2}   y_{j'1}  y_{j'2}   \right) \right| 
=   \Theta \left(  p^{3 + \gamma_\tau} \right)
\stackrel{\eqref{eq:EVdispgrowthrateFOLDL}}{=} 
o(p^4).
\end{align*}
With this we have shown that all terms and hence $\var (\hA_{kl})$ is $o(p^{\tau_A^k + \tau_A^l} )$.

\paragraph{\eqref{eq:G3}, part II: Consistency of $\hat b_{k}$} 
As in eq.~\eqref{eq:var_b_k} we have
\begin{align*}
\sum_{ij} \varh ( S_{ij}  )
& = \frac{p^2}{(n-1)n} \sum_s   \left(  \frac{1}{p} \sum_{i}  y_{is}^2 \right)^2  
- \frac{1}{(n-1)} \sum_{ij} S_{ij}^2   .
\end{align*}
The first term is equal to the first term in eq.~\eqref{eq:Ssq_decompnfix} and hence its variance $o(p^4)$.
The second term is proportional to the left hand side of eq.~\eqref{eq:Ssq_decompnfix} and its variance therefore also $o(p^4)$.
In total, $\var (\hb)$ is $o(p^{2 \tau_{\hTheta}} )$.

\paragraph{\eqref{eq:G3'}: Restriction on linear combinations} Following the same steps as in eq.~\eqref{eq:LDL_cov_lin_comb_proof}, we obtain
\begin{align}
\notag
\Theta \left( p^{\tau_A^k} \right) 
\stackrel{!}{=}
\min_{\substack{\balpha \in \mathbb R^p  \\ \alpha_k = 1}}
\sum_{i=1}^{q} &  \bbE 
\left[ \left(   \sum_{l=1}^K  \alpha_l    (\hT^l_i - \hTheta_{i} ) 
\right) ^2 \right]
 \geq  \sum_i \var (S_i^k) 
=  \Theta(p^2) 
\end{align}
This concludes the proof of Theorem~\ref{th:FOLDLconsistencyCov}.
\end{proof}

\end{appendix}

\bibliographystyle{apalike2}
\bibliography{ida,machineLearning,finance,dbML,bbci_extern,bbci_neurophysics}{}
\end{document}